\documentclass[aps,prb,twocolumn,showpacs,superscriptaddress,longbibliography]{revtex4-1} 

\usepackage{graphicx}  % needed for figures
\usepackage{dcolumn}   % needed for some tables				
\usepackage{bm}        % for math
\usepackage{amssymb}   % for math
\usepackage{float}

\usepackage{amsmath,mathrsfs}
\usepackage{amsthm}
\usepackage{epsfig}

\usepackage{color}
\usepackage{soul}
\newcommand{\ket}[1]{\left|#1\right\rangle}
\newcommand{\bra}[1]{\left\langle#1\right|}

\def\bra#1{\left\langle#1\right|}
\def\ket#1{\left|#1\right\rangle}

\begin{document}

\title{Geometric Construction of Quantum Hall Clustering Hamiltonians}

\author{Ching Hua Lee}
\affiliation{Department of Physics, Stanford University, Stanford, CA 94305, USA}
\author{Zlatko Papi\'c}
\affiliation{School of Physics and Astronomy, University of Leeds, Leeds, LS2 9JT, United Kingdom}
\affiliation{Perimeter Institute for Theoretical Physics, Waterloo, Ontario, Canada N2L 2Y5}
\affiliation{Institute for Quantum Computing, Waterloo, ON N2L 3G1, Canada}
\author{Ronny Thomale}
\affiliation{Institute for Theoretical Physics, University of W\"urzburg, D-97074, Germany}

\date{\today}

\begin{abstract}

Many fractional quantum Hall wave functions are known to be unique highest-density zero modes of certain ``pseudopotential'' Hamiltonians. 
While a systematic method to construct such parent Hamiltonians has been available for the infinite plane and sphere geometries, the generalization to manifolds where relative angular momentum is not an exact quantum number, i.e. the cylinder or torus, has remained an open problem. This is particularly true for non-Abelian states, such as the Read-Rezayi series (in particular, the Moore-Read and Read-Rezayi $\mathbb{Z}_3$ states) and more exotic non-unitary (Haldane-Rezayi, Gaffnian) or irrational (Haffnian) states, whose
parent Hamiltonians involve complicated many-body interactions. Here we develop a universal geometric approach for constructing pseudopotential Hamiltonians that is applicable to all geometries. Our method straightforwardly generalizes to the multicomponent SU($n$) cases with a combination of spin or pseudospin (layer, subband, valley) degrees of freedom. We demonstrate the utility of our approach through several examples, some of which involve non-Abelian multicomponent states whose parent Hamiltonians were previously unknown, and verify the results by numerically computing their entanglement properties. 
\end{abstract}

\maketitle

%\tableofcontents

\section{Introduction}

Among the most striking emergent phenomena in condensed matter are the incompressible quantum fluids in the regime of the fractional quantum Hall effect~\cite{Tsui-PhysRevLett.48.1559}. A key theoretical insight to understanding the many-body nature of such phases of matter was provided by Laughlin's wave function~\cite{Laughlin-PhysRevLett.50.1395}. Shortly thereafter, Haldane~\cite{Haldane1983} realized that the $\nu=1/3$ Laughlin state also occurs as an exact zero-energy ground state of a certain positive semi-definite short-range parent Hamiltonian which annihilates any electron pair in a relative angular momentum $l>1$ state, but which assigns an energy penalty for each  state with $l=1$.  

Haldane's construction of the parent Hamiltonian for $\nu=1/3$ is just one aspect of a more general ``pseudopotential'' formalism that applies to quantum Hall systems in the infinite plane or on the surface of a sphere~\cite{Haldane1983}. In those cases, the system is invariant under rotations around at least a single axis, and by virtue of the Wigner-Eckart theorem, any long range interaction (such as Coulomb interactions projected to Landau level) decomposes into a discrete sum of components $V_l$.
The different components are quantized according to the relative angular momentum $l$, which is odd for spin-polarized fermions and even for spin-polarized bosons. The unique zero mode of the $V_1$ pseudopotential at $\nu=1/3$ is precisely the Laughlin state. Furthermore, it is the \emph{densest} such mode because all the other states, at $\nu=1/3$ or any filling $\nu'>\nu$, are separated by a finite excitation gap as indicated by overwhelming numerical evidence. In numerical simulations, it is furthermore possible to selectively turn on the magnitude of longer range pseudopotentials ($V_3$,$V_5$, etc.), and verify that the Laughlin state adiabatically evolves to the exact ground state of the Coulomb interaction~\cite{prangegirvin}. The corrections induced to the Laughlin state in this way are notably small (below 1\% in finite systems containing about 10 particles), and the gap is maintained during the process~\cite{FanoPhysRevB.34.2670}. These findings constitute an important support of Laughlin's theory.

Remarkably, the existence of pseudopotential Hamiltonians is not limited to the Laughlin states. Generically, more complicated parent Hamiltonians arise in the case of quantum trial states with non-Abelian quasiparticles, which are one important route towards topological quantum computation~\cite{RevModPhys.80.1083, pachos2012introduction}. For example, the celebrated Moore-Read ``Pfaffian" state~\cite{Moore1991362}, believed to describe the quantum Hall plateau at $\nu=5/2$ filling fraction~\cite{Greiter91}, was shown to possess a parent Hamiltonian which is the shortest-range repulsive potential acting on 3 particles at a time~\cite{Greiter91,Greiter92,Read-PhysRevB.54.16864,nick2001}. This state is a member of a family of states -- the parafermion Read-Rezayi (RR) sequence -- where the parent Hamiltonian of the $k$th member is the shortest-range $k$-body pseudopotential~\cite{Read-PhysRevB.59.8084}. Similar parent Hamiltonians have been formulated for non-Abelian $S=k/2$ chiral spin liquid states~\cite{Ronny-CSLPhysRevLett.102.207203}. More recent analysis has allowed to resolve the structure of quantum Hall trial states interpreted as null spaces of pseudopotential Hamiltonians~\cite{nick2013,PhysRevB.88.165303,PhysRevB.91.085103,PhysRevB.91.085115}. 

The knowledge of the parent Hamiltonian class is crucial for the complete characterization of a quantum Hall state. In certain cases, when the state can be represented as a correlator of a conformal field theory (CFT), the charged (quasihole) excitations can be constructed using the tools of the CFT~\cite{Moore1991362}. (Note that such approaches become rather cumbersome for quasielectron excitations~\cite{PhysRevB.80.165330}, and even more so in the torus geometry~\cite{schnells}.) The tools of CFT relating wave functions to conformal blocks certainly become less useful when information about \emph{neutral} excitations is needed. 
The knowledge of the parent Hamiltonian is thus indispensible, e.g. for estimating the neutral excitation gap of the system.  In some cases, the neutral gap can be computed by the single-mode approximation~\cite{GMP85, GMP86}, which is microscopically accurate for Abelian states~\cite{SMA}, but requires non-trivial generalizations for the non-Abelian states~\cite{projective}. Similarly, entanglement spectra may help to characterize the elementary excitations solely deduced from the ground state wave function~\cite{LiPhysRevLett.101.010504,ThomalePhysRevLett.104.180502,PES}, but only unfold their full strength as a complementary tool to the spectral analysis of the associated parent Hamiltonian. While a CFT underlies the structure of entanglement, conformal blocks, and clustering properties within most quantum Hall states of interest~\cite{nick2013}, it is desirable to have a general framework that complements it with a pseudopotential parent Hamiltonian.

An appealing ``added value" of pseudopotentials as building blocks of parent Hamiltonians is the possibility of discovering new states by varying the considered pseudopotential terms and searching for new zero-mode ground states.  This strategy is exemplified by the 
spin-singlet Haldane-Rezayi state~\cite{Haldane-PhysRevLett.60.956}, initially discovered as a zero mode of the ``hollow core" interaction between spinful fermions at half filling. Another example is the 3-body interaction of a slightly longer range than the one that gives rise to the Moore-Read state. It was found~\cite{greiterGaff} that such an interaction has the densest zero-energy ground state at filling factor $\nu=2/5$, subsequently named the ``Gaffnian"~\cite{Simon-PhysRevB.75.075318}. These examples illustrate that a systematic description of an operator space of parent Hamiltonians holds the promise of the discovery of previously unknown quantum Hall trial states.

All appreciable aspects of parent Hamiltonians discussed so far are independent of the geometry of the manifold in which the quantum Hall state is embedded. For several purposes, however, knowing the parent Hamiltonian on geometries without rotation symmetry, such as the torus or cylinder, is particularly desirable. For example, quantum Hall states only exhibit topological ground state degeneracy on higher genus manifolds such as the torus~\cite{WenPhysRevB.41.9377}. Accessing the set of topologically degenerate ground states further allows to extract the modular $\mathcal{S}, \mathcal{T}$ matrices which encode all topological information about the quasiparticles~\cite{Wen-modular,Zhang-modular}. Furthermore, parent Hamiltonians on a cylinder or torus can be used to derive solvable models for quantum Hall states when one of the spatial dimensions becomes comparable to the magnetic length~\cite{LeePhysRevLett.92.096401,seidel2005, Jansen-2012JMP....53l3306J, Wang-PhysRevB.87.245119, Soule-PhysRevB.85.155116, Papic14}. It has been shown that such models can be used to construct ``matrix-product state" representations for quantum Hall states, and in some cases can be used to study the physics of ``non-unitary" states and classify their gapless excitations~\cite{Seidel-PhysRevB.84.085122, Papic14, Weerasinghe14}.

A systematic construction of many-body parent Hamiltonians for quantum Hall states was first undertaken by Simon \emph{et al.} for the infinite plane or sphere geometry~\cite{simon2007,davenport2012}. This approach relies on the relative angular momentum which in this case is an exact quantum number. As such, it cannot be applied to the cylinder, torus, or any quantum Hall lattice model such as fractional Chern insulators, for which the analogue of pseudopotentials has been developed recently~\cite{lee2013,YangleWu,lee2014,claassen2015position}.

Ref.~\onlinecite{lee2013} has introduced the closed-form expressions for all two-body Haldane pseudopotentials on the torus and cylinder.
In this work, inspired by Refs.~\onlinecite{simon2007,davenport2012,lee2013} as the starting point of our analysis, we provide a complete framework for constructing general quantum Hall parent Hamiltonians involving $N$-body pseudopotentials, for fermions as well as bosons, in  cylindric and toroidal geometries. This advance proves particularly important for the non-Abelian states, most of which necessitate many-body pseudopotentials in their parent Hamiltonian class. From the construction scheme laid out in this work, all topological properties of the non-Abelian states such as their modular matrices and topological ground state degeneracy can now be conveniently studied from their associated toroidal parent Hamiltonian. Complementing previous results for the sphere and infinite plane, our formalism furthermore directly generalizes to multicomponent systems with an arbitrary number of ``spin types" or ``colors". Therefore, our construction of many-body clustered Hamiltonians not only applies to arbitrary geometries, but also crucially simplifies previous approaches. We illustrate this by numerical examples, including non-Abelian multicomponent states whose parent Hamiltonians were previously unknown.

The article is organized as follows. In Sec.~\ref{sec_haldane}, we start with a brief overview of Haldane's pseudopotential formalism from the viewpoint of clustering conditions, and define our notation of generalized many-body multicomponent pseudopotentials. The clustering conditions, as well as the polynomial constraints following from them, form the core of our systematic geometric construction of many-body parent Hamiltonians described in Sec.~\ref{sec:construction}. The main result of that section is the appropriately chosen integral measure which allows us to generate all pseudopotentials through a direct Gram-Schmidt orthogonalization. The extension to many-body parent Hamiltonians for spinful states is described in Sec.~\ref{sec:spinful}. Technical details of the construction are delegated to the appendices. In Section~\ref{sec:examples}, explicit examples of parent Hamiltonian studies are worked out in detail. We discuss spin polarized states such as the Gaffnian, the $1/q$ Pfaffian and the Haffnian states, as well as spinful states such as the spin-singlet Gaffnian, the NASS and the Halperin-permanent states. Apart from deriving the second-quantized parent Hamiltonians, we also provide extensive numerical checks of our construction using exact diagonalization and analysis of entanglement spectra. In Sec.~\ref{sec:conclusions}, we conclude that our geometric construction of parent Hamiltonians promises ubiquitous use in the analysis of fractional quantum Hall states and outline a few immediate future directions.

\section{Haldane pseudopotentials}\label{sec_haldane}

\begin{figure}[thb]
\includegraphics[width=\linewidth]{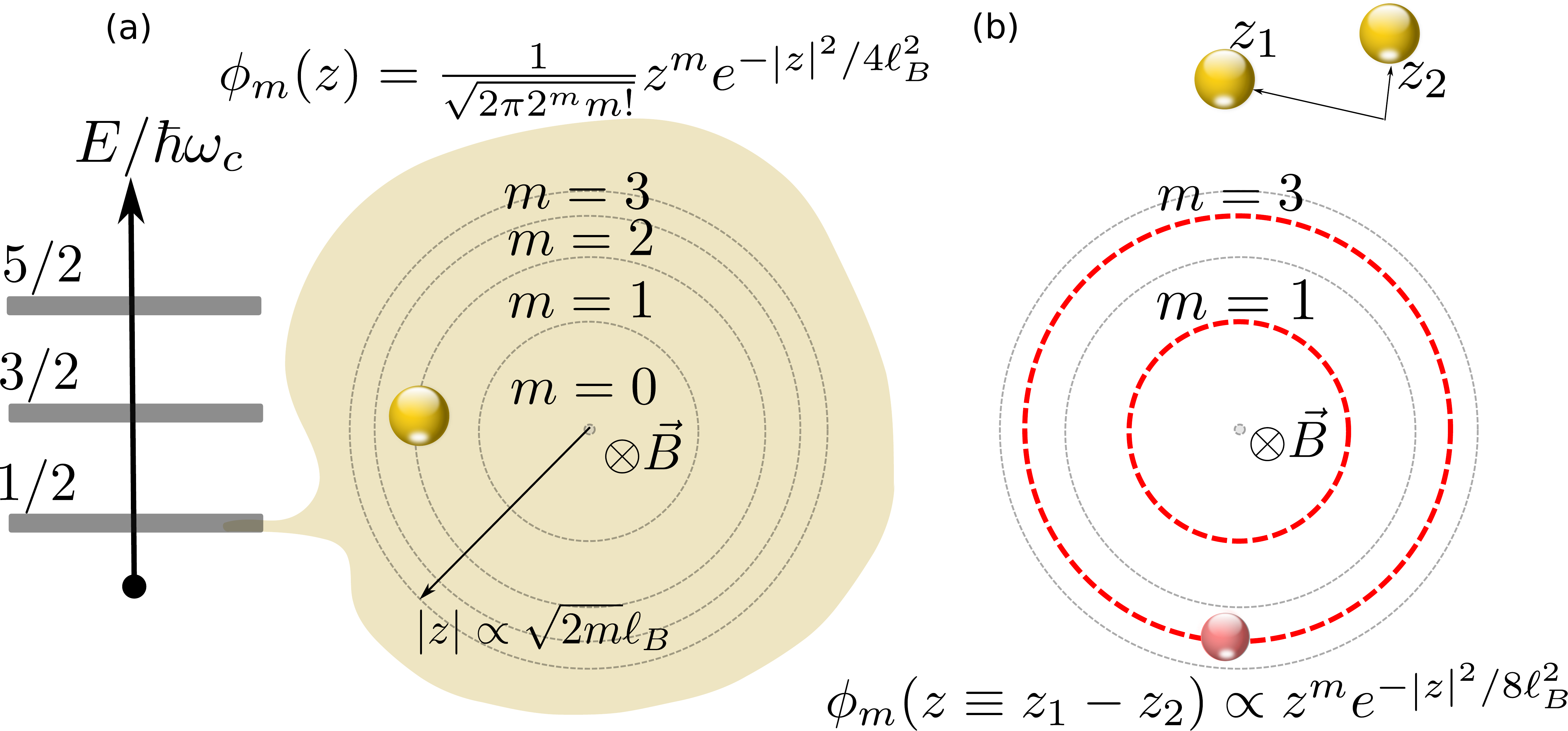}
\caption{(Color online) (a) Landau levels and the single-particle states on the disk. The lowest LL is separated from the next lowest level by the cyclotron gap $\hbar \omega_c$. The wave functions of the LLL states are $\phi_m$,  labelled by an integer $m=0,1,2\ldots$, and Gaussian-localized along the ring of radius $\sim \sqrt{m} \ell_B$. (b) Any two-particle state separates into a product of two wave functions, one depending on the center-of-mass and the other depending on the relative coordinate $z=z_1-z_2$. The relative wave function $\psi_m$ has identical form to the single-particle wave function, with two important differences: its effective magnetic length is rescaled ($\ell_B \to \ell_B \sqrt{2}$), and the integer value of $m$ is constrained by particle statistics (assuming the spin is fully polarized): for fermions, the $m$'s entering the relative wave function are odd (red dashed lines).}
\label{fig:disk}
\end{figure}

Within a given Landau level, the kinetic energy term in the quantum Hall (QH) Hamiltonian is ``quenched", i.e. is effectively a constant. Hence the remaining effective Hamiltonian only depends on the interaction between particles (e.g., the Coulomb potential) projected to the given Landau level~\cite{prangegirvin}. In the infinite plane, the lowest Landau level (LLL) projection amounts to evaluating the matrix elements of the Coulomb interaction between the single-particle states of the form
\begin{equation}\label{spwf}
\phi_m(z)=z^m \exp(-|z|^2/4\ell_B^2),
\end{equation}
where $z$ is a complex parametrization of 2D electron coordinates and $\ell_B=\sqrt{\hbar/eB}$ the magnetic length (Fig.~\ref{fig:disk}a). For simplicity, we consider a QH system in the background of a fixed (isotropic) metric~\cite{HaldaneGeometry}, which allows us to write $z=x+i y$. The states in Eq.~\ref{spwf} are mutually orthonormal, and span the basis of the LLL. There are $N_\phi$ of these states, which is also the number of magnetic flux quanta through the system. The above will be assumed throughout this paper, which is appropriate in strong magnetic fields when the particle-hole excitations to other Landau levels are suppressed by the large cyclotron energy gap $\hbar\omega_c$.

Restricting to a single LL, a large class of QH states can be classified by their \emph{clustering} properties. (See Refs.~\onlinecite{jack,PhysRevB.77.235108} for classification schemes based on clustering.)  These are a set of rules which describe how the wave function vanishes as particles are brought together in space. To define the clustering rules, it is essential to first consider the problem of two particles restricted to the LLL (Fig.~\ref{fig:disk}b). As usual, the solution of the two-body problem proceeds by transforming from coordinates $z_1,z_2$ into the center of mass (COM) $Z=(z_1+z_2)/2$ and the relative coordinate $z\equiv z_1-z_2$ frame. In the new coordinates, the two-particle wave function decouples. As we are interested in translationally-invariant problems, only the relative wave function (which depends on $z$) will play a fundamental role in the following analysis. For any two particles, the relative wave function turns out to have an identical form to the single-particle wave function~\eqref{spwf}
\begin{equation}\label{2pwf}
\psi_m(z \equiv z_1-z_2) \propto z^m \exp(-|z|^2/8\ell_B^2),
\end{equation}
up to the rescaling of the magnetic length $\ell_B$. An important difference between Eqs.~(\ref{2pwf}) and (\ref{spwf}) is the new meaning of $m$: since $z$ now represents the \emph{relative} separation between two particles, $m$ in Eq.~(\ref{2pwf}) is related to particle statistics, and therefore encodes the clustering properties. For spinless electrons, $m$ in Eq.~(\ref{2pwf}) is only allowed to take odd integer values since the wave function must be antisymmetric with respect to $z\to -z$, while for spinless bosons $m$ can be only be an even integer. Finally, $m$ is also the eigenvalue of the relative angular momentum for two particles ($\hbar=1$), as we can directly confirm from
\begin{equation}
L^z = \sum_i z_i \frac{\partial}{\partial z_i}.
\end{equation}

After this two-particle analysis (summarized in Fig.~\ref{fig:disk}), we are in position to introduce the notion of clustering properties for $N$-particle states. Let us pick a pair of coordinates $z_i$ and $z_j$ of indistiguishable particles in a many-particle wave function. We say that these particles are in a state $\psi^{ij}_m$ which obeys the clustering property with the power $m$
if $\psi^{ij}_m$ vanishes as a polynomial of total power $m$ as the coordinates of the two particles approach each other:
\begin{equation}\label{vanishing}
\psi_m^{ij} (z_i \to z_j) \sim  (z_i-z_j)^{m}.
\end{equation} 
Similarly as before, we can relate the exponent $m$ to the angular momentum $L^z$ if the latter is a conserved quantity.
Clustering conditions like this directly generalize to cases where more than $2$ particles approach each other, with the polynomial decay also specified by a power. For example, we say that an $N$-tuple of particles $z_1, z_2, \dots z_N$ is in a state $\psi^{N}_m$ with total relative angular momentum $m$ if $\psi_m^N$ vanishes as a polynomial of total degree $m$ as the coordinates of $N$ particles approach each other. Fixing an arbitrarily chosen reference particle of the $N$-tuple, i.e. $z_1=:z_r$, we have:
\begin{equation}\label{vanishing2}
\psi_m^{N} (z_2,\dots, z_N \to z_r) \sim \prod_{j=2}^N (z_r-z_j)^{m_j},
\end{equation} 
 with $\sum_{j=2}^N m_j =m$ as all remaining $N-1$ particles approach the reference particle $z_r$.
If the system is rotationally invariant about at least a single axis (such as for a disk or a sphere) it directly follows that the state in Eq.~\ref{vanishing2} is also an eigenstate of the corresponding $L_z$ relative angular momentum operator with eigenvalue $m$. 

The simplest illustration of the clustering condition is the fully filled Landau level. The wave function for such a state is the single Slater determinant of states $\phi_m(z_i)$ in Eq.~\ref{spwf}. Due to the Vandermonde identity, this wave function can be expressed as
\begin{equation}
\psi_{\nu=1} = \prod_{i<j} (z_i-z_j) \exp(-\sum_k |z_k|^2/4\ell_B^2).
\end{equation}
We see that when any pair of particles $z_i$ and $z_j$ is isolated, the relevant part of the wave function is $(z_i-z_j)$. Therefore, the wave function of the filled Landau level vanishes with the exponent $m=1$ as particles are brought together. This is the minimal clustering constraint that any spinless fermionic wave function must satisfy. (As we will see below, interesting many-body physics results from \emph{stronger} clustering conditions on the wave function.)

When properly orthogonalized, the states $\{\psi_m^{N}\}$ form an orthonormal basis in the space of magnetic translation invariant QH states. They allow to define the $N$-body \emph{Haldane pseudopotentials}\cite{Haldane1983} (PPs)
\begin{equation}
U^{N}_{m}= \ket{\psi^{N}_m}\bra{\psi^{N}_m},
\end{equation}   
which obey the null space condition $U^N_{m'} \psi_m^N=0$ for $m' \neq m$. Since they are positive-definite, the PPs $U^N_m$ give \emph{energy penalties} to $N$-body states with total relative angular momentum $m$. With a given many-body wave function, the Hamiltonian representation of $U^N_m$ will involve the sum over all $N$-tuple subsets of particles. Concrete examples of this will be given in Secs.~\ref{sec:construction} and \ref{sec:examples}.

For a given filling fraction, it can occur that a certain QH state is the unique and \emph{densest} ground state lying in the null space of a certain linear combination of PPs, i.e. it is annihilated by a certain number of PPs. (The requirement of being the densest state is necessary to render the finding non-trivial, because it is simple in principle to construct additional zero modes of a given PP Hamiltonian by increasing the magnetic flux, i.e. by nucleating quasihole excitations.)  
The most elementary examples are the Laughlin states at $1/m$ filling, which lie in the null space of $U^2_{m'}$ for all $m'<m$. As it also represents the densest configuration that is annihilated by the PP, the Laughlin state emerges as the unique ground state of a Hamiltonian at filling $1/m$, $H=\sum_{m'<m} c_{m'} U^2_{m'}$, where the coefficients are arbitrary as long as $c_{m'} > 0$. That is, the fermionic $1/3$ Laughlin state is the unique ground state of $U^2_1$, while the fermionic $1/5$ state is the unique groudstate of any linear combination of $U^2_1$ and $U^2_3$ with positive weights. Note that the PPs of even $m'$ are precluded by fermionic antisymmetry. For short-ranged two-body interactions on the disk where the degree of the clustering polynomial $m$ coincide with the exact relative angular momentum $l$, the traditional notation by Haldane~\cite{Haldane1983} relates to ours via $V_l \equiv U^2_m$.  
As we elaborate below, many more exotic states can be realized as the highest density ground states of combinations of PPs involving $N\geq 2$ bodies. 

In view of other geometries than disk or sphere, one question immediately arises: Is there any hope of defining PPs {\it in the absence of continous rotation symmetry} and hence no exact relative angular momentum quantum number? This question is natural because one of the popular choices for the gauge of the magnetic field -- the Landau gauge -- is only compatible with periodic boundary conditions along one or both directions in the plane, which breaks continuous rotational symmetry.

The answer to the above question is {\it affirmative}, which is the central message of this paper. This is mainly because the clustering conditions, i.e. the polynomial exponent $m$, are more fundamental than their interpretation as the relative angular momentum. The clustering conditions are short-distance properties:  their support is the area associated with the fundamental droplet of $N$ particles $\sim N \times 2\pi\ell_B^2$. Assuming a one-to-one correspondence between a clustering power and a pseudopotential, the PPs should be independent of the specific geometry as long as they act in a manifold that is homogeneous and much larger than the fundamental $N$-particle droplet. This viewpoint has been confirmed by the explicit constructions of quantum Hall trial wave functions. For instance, the successful generalization of the Laughlin wave function to the torus, given in the classic paper by Haldane and Rezayi~\cite{HalRezTorusPhysRevB.31.2529}, has demonstrated that its short-distance properties are identical to its original version defined on the disk. The states in both geometries are uniquely characterized by their clustering properties and are locally indistinguishable; their main difference lies in the global properties, e.g., the fact that the torus Laughlin state is $m$-fold degenerate due to its invariance under the center of mass (COM) translation. This degeneracy is of intrinsically topological origin~\cite{WenPhysRevB.41.9377}. At filling $\nu=p/q$, where $p,q$ are relatively co-prime integers, the quantum Hall state on a torus is invariant under the translation that moves every particle by $q$ orbitals (see Fig.~\ref{fig:hopping}c below). This symmetry guarantees an exact $m$-fold degeneracy of the Laughlin $\nu=1/m$ state on the torus~\cite{Haldane-PhysRevLett.55.2095}. Additional topological degeneracies can arise for more complicated (non-Abelian) states~\cite{Moore1991362,Greiter92}.

In the following, we assume there exists, in general, a well-defined deformation of the null space specified by a planar QH trial wave function to the multi-dimensional null space specified by the associated set of topologically degenerate QH ground states on the torus. What we are then interested in is to find a suitable deformation of the planar Laplacian, whose bilinear form is the known planar parent Hamiltonian composed of the spherical PPs, to the toroidal Laplacian, whose bilinear form is the toroidal Hamiltonian. In the following sections, we solve this problem by what we refer to as the ``geometric construction" of pseudopotentials.

\section{Geometric construction of pseudopotentials: spinless case}
\label{sec:construction}

We describe the construction of generic QH pseudopotential Hamiltonians with a perpendicular magnetic field applied to the 2D electron gas. We assume the field to be sufficiently strong such that the spin is fully polarized, and that there are no further internal degrees of freedom for the particles. We first introduce a suitable single-particle basis for generic $N$-body interactions that obeys the magnetic translation symmetry and conserves the center-of-mass (COM) momentum. Next, we show how the explicit functional form of Haldane PPs in Sec.~\ref{sec_haldane} can be easily obtained from geometric principles and symmetry. We constrain ourselves to single-component PPs in this section, and  generalize our construction to multicomponent (spinful) PPs in Sec.~\ref{sec:spinful}.

\subsection{Basis choice}

\begin{figure}[htb]
\includegraphics[width=\linewidth]{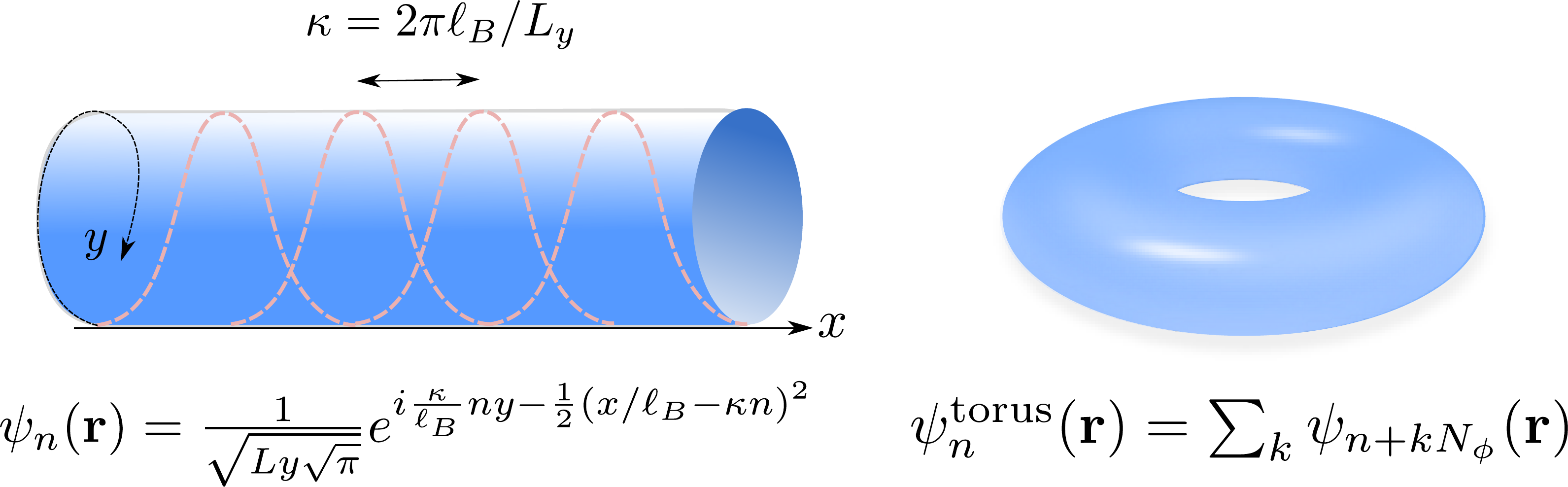}
%\captionsetup{justification=centerlast}
\caption{(Color online) Boundary conditions considered in this paper, compatible with Landau gauge: cylinder (left) and torus (right). The associated single-particle states $\psi_n(\mathbf{r})$ are arranged along a 1D chain, with the parameter $\kappa=2\pi\ell_B/L_y$ controlling the distance between nearest neighbor orbitals.}
\label{fig:torus}
\end{figure}

A well-developed pseudopotential formalism is available in the literature~\cite{Haldane1983, Simon-PhysRevB.75.075318} for the infinite plane or the sphere, where the $z$-component of angular momentum is conserved. A different approach to pseudopotential Hamiltonian construction, however, is needed when the system is no longer invariant under continuous rotations around the $z$-axis. This can occur when periodic boundary conditions are imposed [Fig.\ref{fig:torus}], either along one direction (cylinder geometry) or both directions (torus).
Under periodic boundary conditions (PBCs) in one direction (say $\hat y$), the single-particle Hilbert space is spanned by the Landau gauge basis wave function labelled by $n$:
\begin{equation}
\psi_n(x,y) \equiv  \langle \vec{r} | c_n^\dagger \ket{0} \propto e^{i\frac{\kappa}{\ell_B}ny}e^{-\frac{1}{2}\left(\frac{x}{\ell_B}-\kappa n \right)^2},
\label{LLL}
\end{equation}
where $\ell_B=\sqrt{\hbar/eB}$ is the magnetic length and $c_n^\dagger$ is the second-quantized operator that creates a particle in the state $|n\rangle$. The parameter $\kappa=2\pi \ell_B/L_y$ sets the effective separation between the one-body states in the $x$-direction, as each one-body state is a Gaussian packet approximately localized around $\kappa n$ in $x$-direction (Fig.~\ref{fig:torus}). 

\begin{figure}[htb]
\includegraphics[width=\linewidth]{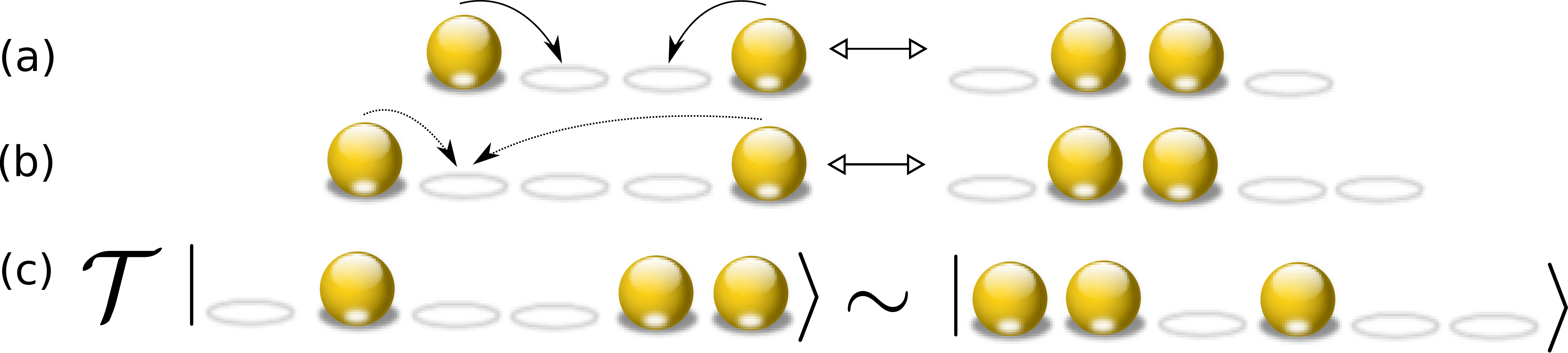}
%\captionsetup{justification=centerlast}
\caption{(Color online) Scattering schemes and symmetries in a Landau level. Preservation of COM allows a cluster of particles to scatter according to (a), but forbids the scattering according to (b).  (c) The many-particle translation operator $\hat{T}$ acts on a given configuration by moving every particle $q$ orbitals to the right (in this example, $q=2$). Many-body states at filling $\nu=p/q$ are invariant under this symmetry, which gives rise to the topological degeneracy equal to $q$. }
\label{fig:hopping}
\end{figure}

Due to this simple one-parameter labeling of the one-body states, an $N$-body interaction matrix element is labeled by $2N$ indices. Additional constraints on its functional form arise when we consider interactions projected to a given Landau level (Fig.~\ref{fig:hopping}). Due to the magnetic translation invariance, the interactions projected to a Landau level only allow for scattering that conserves total momentum, or equivalently leaves the COM of the particles fixed. For example, the process in Fig.~\ref{fig:hopping}a is allowed, but the one in Fig.~\ref{fig:hopping}b is forbidden. This special structure in the interaction Hamiltonian gives rise to the symmetry under many-body translations that shifts every particle by $q$ orbitals at filling $\nu=p/q$ (Fig.~\ref{fig:hopping}c). This is the symmetry that underlies the topological ground state degeneracy. 

The matrix elements of any interaction $V({\mathbf r}_1,...,{\mathbf r}_N)$ projected to a Landau level are given by
\begin{eqnarray}
V_{\{ n_j \}, \{ n'_j \}} &=& \int d{\mathbf r}_1 \ldots d{\mathbf r}_N \notag \\ 
 && \left( \prod_j \psi^*_{n'_j}( {\mathbf r}_j)   \psi_{n_j}( {\mathbf r}_j) \right) V( {\mathbf r}_1,..., {\mathbf r}_N) , 
\label{Vint}
\end{eqnarray}
which corresponds to the second-quantized Hamiltonian
\begin{eqnarray}
\notag && H = \sum_\alpha \sum_{n} b^{\alpha\dagger}_n b^\alpha_n, \\
&& \notag b^\alpha_n \propto \sum_{ \sum_j  n_j = n}p_\alpha(\kappa \bar n_1,...,\kappa \bar n_N)e^{-\frac{\kappa^2}{2}\sum_{j=1}^N \bar n_j^2}c_{n_1} \ldots c_{n_N}.\\
\label{bn}
\end{eqnarray}
Here $\bar n_j=n_j-n/N$ denotes the orbital of particle $j$ with respect to the COM, $n=n_1+...+n_N$, and $p_\alpha$ is a polynomial in $N$ variables $\kappa \bar n_j$.  The index $\alpha$, as will be clear from the explicit construction of $p_\alpha$ below, specifies the degree of the polynomial and, for a multicomponent state discussed in Section~\ref{sec:spinful}, its spin sector. The polynomial $p_\alpha$ can be chosen to be real, as will be evident from its geometric construction to follow. From now on, $n$ will refer to the COM, and not the index of a single particle previously appearing in Eq. \ref{LLL}. $c_{n_j}^\dagger$ is the same operator as in Eq. \ref{LLL}, creating the $j$th particle in state $|n_j\rangle$.  The Hamiltonian thus consists of a product sum over all positions of the COM $n$, as well as all polynomial degrees $\alpha$.

Eqs.~\ref{Vint} and~\ref{bn} represent a general translationally invariant Hamiltonian projected to a Landau level. This Hamiltonian has a rather special form: it decomposes into a linear combination of positive-definite operators $b^{\alpha\dagger}_n b^\alpha_n$, such that the form factor $p_\alpha e^{-\frac{\kappa^2}{2}\sum_{j=1}^N \bar n_j^2}$ in each $b^\alpha_n$ depends only on the relative coordinates of particles. Any short-range Hamiltonian can be explictly expressed in this form~\cite{LeePhysRevLett.92.096401,seidel2005, lee2013}, which reveals its Laplacian structure. 

Physically, Landau-level-projected Hamiltonians can be visualized as a long-range interacting 1D chain [Fig.~\ref{fig:hopping}]. The interaction terms can be interpreted as long-range (though Gaussian suppressed) hopping processes labeled by $\alpha$. For each $\alpha$, $N$ particles ``hop" from sites $n_j$ to sites $n'_j$, $j=1,...,N$ according to a COM independent amplitude given by $p_\alpha(\kappa \bar n_1,...,\kappa \bar n_N) e^{-\frac{\kappa^2}{2}\sum_{j=1}^N \bar n_j^2}$, such that the initial and final COM $n$ remains unchanged (Fig.~\ref{fig:hopping}c). Note that although we use the term ``hopping'', there is no clear distinction between ``hopping'' and ``interaction'' in our case, as opposed to the Hubbard model. Rather, ``hopping'' is designated for any interaction term that is purely quantum (i.e., not of Hartree form). The goal of the remainder of this paper is to show how to systematically construct the polynomial amplitudes $p_\alpha$ that need to be inserted into Eq.~(\ref{bn}) to obtain the parent Hamiltonian for the desired quantum Hall state.

\subsection{Geometric derivation}
\label{geometric}

We now specialize Eq.~\ref{Vint} to $N$-body interactions that are Haldane PPs $U^N_m$. One appealing feature of the second-quantized form of Eq.~\ref{Vint} is that we can construct the  desired PPs from symmetry principles alone, without referring to the first-quantized form of the interaction $V( \mathbf{r}_1,...,\mathbf{r}_N)$. 

The many-body PPs are, by definition, supposed to project onto orthogonal subspaces labeled by $m$, where $m$ denotes the relative angular momentum:
\begin{equation}
U^N_{m'}U^N_m=\ket{\psi^N_{m'}}\bra{ \psi^N_{m'}}\ket{\psi^N_{m}}\bra{ \psi^N_{m}}=U^N_m\delta_{m'm}.
\label{Uortho}
\end{equation}
This requires that
\begin{eqnarray}
\bra{0}b^{m'\dagger}_nb^m_n\ket{0} &\propto& \sum_{\bar n_1+...+\bar n_N=0}p_{m'}(\kappa \bar n_1,...,\kappa \bar n_N) \notag \\
&\times& p_{m}(\kappa \bar n_1,...,\kappa \bar n_N) \exp{\left(-\kappa^2\sum_{j=1}^N \bar n_j^2\right)} \notag \\
&=& \delta_{m'm}.
\end{eqnarray}
If $\psi^N_m$ is to vanish with $m$th total power as $N$ particles approach each other, the polynomial $p_m$ must be of degree $m$. Hence the PPs will be \emph{completely} determined once we find a set of polynomials $\{p_m\}$ such that: (1)  $p_m$ is of total degree $m$;
(2) $p_m$ has the correct symmetry property under exchange of particles, i.e. is totally (anti)symmetric for bosonic (fermionic) particles; 
(3) the $p_m$'s are orthonormal under the inner product measure
\begin{eqnarray}\label{innerproductdef}
\langle p_{m'}, p_m\rangle &=& \sum_{\bar n_1+...+\bar n_N=0} p_{m'}(\kappa \bar n_1,...,\kappa \bar n_N) \notag \\ 
&\times& p_{m}(\kappa \bar n_1,..., \kappa\bar n_N)\exp\left(-\kappa^2\sum_{j=1}^N \bar n_j^2 \right).
\end{eqnarray}
Using the barycentric coordinates (Appendix~\ref{sec:barycentric}) to represent the tuple $(\bar n_1,...,\bar n_N)$
\begin{eqnarray}
\langle p_{m'}, p_m\rangle  &\approx & \int_{\mathbb{R}^{N-1}} dW d\Omega \; \; p_{m'}(W ,\Omega) p_m(W ,\Omega) \notag \\
&\times& \exp{(-\frac{N-1}{N}W^2 )}J_{N-1}(W,\Omega)  \notag\\
&=& \delta_{m'm},
\label{innerproduct}
\end{eqnarray}
where $W$ and $\Omega$ are the radial and angular coordinates of the vector $\vec x\in \mathbb{R}^{N-1}$ representing the tuple $(\bar n_1,...,\bar n_N)$, and $J_{N-1}$ is the Jacobian for the spherical coordinates in $\mathbb{R}^{N-1}$. We have exploited the magnetic translation symmetry of the problem in quotienting out the COM coordinate $n$. (It is desirable to quotient out $n=\sum_j^N n_j$, since $n$ takes values on an infinite set when the particles lie on the 2D infinite plane, and that complicates the definition of the inner product measure.) 
Each quotient space is most elegantly represented as an $N-1$-simplex in barycentric coordinates, where particle permutation symmetry (or subgroups of it) is manifest. 
Explicitly, the set of $\bar n_j$ can be encoded in the vector  
\begin{equation}
\vec x =\kappa \sum_{j=1}^N \bar n_j \vec\beta_j,
\end{equation}
where the $N$ basis vectors $\{\vec \beta_j\}$ form a set that spans $\mathbb{R}^{N-1}$. A configuration $(\bar n_1,...,\bar n_N)$ is uniquely represented by a point $\vec x\in \mathbb{R}^{N-1}$ that is independent of $n=\sum_j^Nn_j$. Since $\vec x$ should not favor any particular $\bar n_j$, any pair of vectors in the basis $\{\vec \beta_j\}$ must form the same angle with each other. Specifically,
\begin{equation}
\vec \beta_j \cdot \vec \beta_k = \frac{N}{N-1}\delta_{jk} - \frac{1}{N-1},
\end{equation}
so that each vector points at the angle of $\pi-\cos^{-1}\frac1{N-1}$ from another. 

With this parametrization, the Gaussian factor reduces to the simple form 
\begin{equation}
W^2=|\vec x|^2=\frac{N}{N-1}\kappa^2\sum_{j=1}^N\bar n_j^2.
\end{equation}
Further mathematical details can be found in the examples that follow, as well as in Appendix \ref{sec:barycentric}.
 
The integral approximation in Eq.~\ref{innerproduct} becomes exact in the infinite plane limit, and is still very accurate for values of $m$ where the characteristic inter-particle separation is smaller than the smallest of the two linear dimensions of the QH system. In the following, we will assume this to be the case; otherwise, there can be significant effects from the interaction of a particle with its periodic images. This was systematically studied in the appendix of Ref. \onlinecite{lee2013} for $N=2$. We note that the approximation in Eq.~\ref{innerproduct} does not affect the exact zero mode property of the trial Hamiltonians constructed below.

\subsection{Orthogonalization}
\label{PPortho}

We are now ready to evaluate the pseudopotentials. To find a second-quantized PPs with relative angular momentum $m$, one needs to follow the rules listed in Table~\ref{table:spinless}.  
\begin{table}[htb]
\centering
\renewcommand{\arraystretch}{2}
\begin{tabular}{|p{8.0cm}|}\hline
(1) Write down the allowed ``primitive" polynomials~\cite{simon2007,Simon-PhysRevB.75.075318, davenport2012} of degree $m'\leq m$ consistent with the symmetry of the particles. \\ \hline
(2) Orthogonalize this set of primitive polynomials according to the inner product measure in Eq.~\ref{innerproduct}. \\ \hline
\end{tabular}
%\captionsetup{justification=centerlast}
\caption{Summary of the PP construction for spinless particles. Examples of this procedure are given in Sec.~\ref{sec:examples}.
}
\label{table:spinless}
\end{table}
In the following, we shall execute this recipe explicitly for $N=2,3$, and, to some extent, $N=4$-body interactions. 

\subsubsection{$2$-body case}
\label{n2}

For two-body interactions, we have 
\begin{equation}
\langle p_{m'},p_m\rangle = \int_{-\infty}^{\infty} p_{m'}( W)p_{m}( W)e^{-\frac1{2}W^2}dW, 
\label{inner2}
\end{equation}
where $W=-2\kappa\bar n_1=2\kappa\bar n_2$. For this $N=2$ case, we have allowed $W$ to take negative values as the angular direction spans the 1D circle, which consists of just two points. The primitive polynomials for bosons are $\{1,W^2,W^4,W^6,...\}$ while those for fermions are $\{ W,W^3,W^5,...\}$. After performing the Gram-Schmidt orthogonalization procedure, the $2$-body PPs are found to be $U^2_m\propto \kappa^3\sum_n b^{m\dagger}_n b^m_n$, where
\begin{equation} b^m_n =\sum_{\bar n_1+\bar n_2=0} p_m(\kappa \bar n_1,\kappa \bar n_2)e^{-\frac{1}{2}\kappa^2 (\bar n_1^2+\bar n_2^2)}c_{n/2+\bar n_1}c_{n/2+\bar n_2}, \label{b2}\end{equation}
$n/2 + \bar n_j$ are integers, and $p_m$ is a $m$th degree Hermite polynomial given in Table~\ref{N2}. In particular, we recover the Laughlin $\nu=1/2$ bosonic or $\nu=1/3$ fermionic state for $m=0$ or $m=1$ respectively.

\begin{table}[htb]
\begin{minipage}{0.99\linewidth}
\centering
\renewcommand{\arraystretch}{2}
\begin{tabular}{|l|l|l|}\hline
$m$ &\ Bosonic $p_m(W)$ &\ Fermionic $p_m(W)$ \\    \hline
0 &\ $1$ &\ 0 \\ \hline 
1 &\ $0$ &\ $W$ \\ \hline  
2 &\ $ \frac{1}{\sqrt{2!}}(-1+W^2)$ &\ 0 \\  \hline
3 &\ $0$ &\ $\frac{1}{\sqrt{3!}}(-3+W^2)W$ \\ \hline  
4 &\ $ \frac{1}{\sqrt{4!}}(3-6 W^2+W^4) $ &\ 0  \\ \hline 
5 &\ $0$ &\ $\frac{1}{\sqrt{5!}}(15-10 W^2+W^4)W$ \\ \hline  
6 &\ $ \frac{1}{\sqrt{6!}}(-15+45 W^2$ &\ $0$ \\ &\  $-15 W^4 +W^6)$ &\ \\ \hline 
7 &\ $0$  &\ $\frac{1}{\sqrt{7!}}(-105+105 W^2-21 W^4$ \\ &\ &\ $+ W^6)W$ \\ \hline  
\end{tabular}
\end{minipage}
%\captionsetup{justification=centerlast}
\caption{Representative polynomials $p_m$ for the first few $N=2$-body PPs for bosons and fermions. $p_m(W)e^{-W^2/4}$ represents the amplitude that two particles $W/\kappa$ sites apart are involved in a two-body hopping on the chain. }
\label{N2}
\end{table}
One can easily check that PPs become more delocalized in $W$-space as $m$ increases. Indeed,
\begin{eqnarray}
\notag \kappa^2\langle\sum_j \bar n_j^2\rangle&=&\frac1{2}\langle W^2\rangle =\frac1{2}\int_{-\infty}^{\infty}p_m(W)W^2e^{-\frac1{2}W^2}dW \notag\\
& =& m+\frac{1}{2}, 
\label{PPm}
\end{eqnarray}
which is reminiscent of the interpretation of $m$ as the angular momentum of a pair of particles in rotationally-invariant geometries (Fig.~\ref{fig:disk}).
Obtaining the pseudopotentials in this second-quantized form is highly advantageous. In particular, note that this construction is free from ambiguities in the choice of $V(\mathbf{r}_1,...,\mathbf{r}_N)$, since several possible real-space interactions, e.g., those of the Trugman-Kivelson type~\cite{trugman1985}, can all be grouped into the same $m$ sector. This is discussed in more detail in Appendix~\ref{sec:TK}.

\subsubsection{$3$-body case}
\label{n3}

For $N=3$, the inner product measure takes the form
\begin{equation}
\langle p_{m'},p_m\rangle = \int_0^\infty \int_0^{2\pi}p_{m'}(W ,\theta) p_m(W ,\theta)e^{-\frac{2}{3}W^2}Wd\theta dW
\label{inner3}
\end{equation}
with
\begin{subequations}
\begin{align} \bar n_1 &=  \frac{2W}{3\kappa}\cos\theta \\
\bar  n_2 &=  \frac{W}{3\kappa}(\sqrt{3}\sin\theta-\cos\theta) \\
 \bar n_3 &=  \frac{W}{3\kappa}(-\sqrt{3}\sin\theta-\cos\theta) 
\end{align}\label{n3def}\end{subequations} 
Each of the $\bar n_j$'s are treated on equal footing, as one can easily check graphically. The above expressions are the simplest nontrivial cases of the general expressions for barycentric coordinates found in the Appendix (Eqs.~\ref{barybasis}-\ref{eqg10}).

The bosonic primitive polynomials are made up of elementary symmetric polynomials $S_1,S_2,S_3$ in the variables $\kappa \bar n_j=\kappa (n_j-n/N)$. Since $S_1=\kappa \sum_j \bar n_j=0$, the only two symmetric primitive polynomials are
\begin{equation} 
-S_2=-\kappa^2\sum_{i<j}\bar n_i\bar n_j=\frac{S_1^2-2S_2}{2}=\frac{\kappa^2}{2}\sum_i \bar n_i^2 = \frac{1}{3}W^2, 
\end{equation} 
and
\begin{eqnarray} Y= S_3 &=&  \kappa^3\prod_i\bar n_i\notag\\
&=&\frac{\kappa^3}{27}(2n_1-n_2-n_3)(2n_2-n_3-n_1) \notag \\
&& (2n_3-n_1-n_2)\notag\\
&=& \frac{2}{27}W^3\cos 3 \theta. 
\label{Y}
\end{eqnarray}

The fermionic primitive polynomials are totally antisymmetric, and can always be written\cite{simon2007,davenport2012} as a symmetric polynomial multiplied by the Vandermonde determinant
\begin{eqnarray} 
\nonumber A &=& \kappa^3(n_1-n_2)(n_2-n_3)(n_3-n_1) \\
\nonumber &=& \kappa^3(\bar{n}_1-\bar{n}_2)(\bar{n}_2-\bar{n}_3)(\bar{n}_3-\bar{n}_1)\\
&=& -\frac{2}{3\sqrt{3}}W^3 \sin 3\theta. 
\label{A}
\end{eqnarray}
Note that $W^2$ is of degree 2 while $A$ and $Y$ are of degree 3. All of them are independent of the COM coordinate $n$, as they should be. $N=3$ PPs were derived in Ref.~\onlinecite{lee2013} through explicit integration, and the approach discussed here considerably simplifies those computations by exploiting symmetry.

To generate the fermionic (bosonic) PPs up to $U^3_m$, we need to orthogonalize the basis consisting of all possible (anti)symmetric primitive polynomials up to degree $m$. For instance, the first seven (up to $m=9$) 3-body fermionic PPs are generated from the primitive basis $\{ A, AW^2, AY,AW^4,AYW^2, AY^2, AW^3\}$. Note that the last two basis elements both contribute to the $m=9$ PP sector. 

The 3-body PPs are found to be $U^3_m\propto \kappa^3\sum_n b^{m\dagger}_nb^m_n$, where
\begin{equation} 
b^m_n = \sum_{n_1+n_2+n_3=n} p_m(W,Y,A)e^{-\frac1{3} W^2}c_{n_1}c_{n_2}c_{n_3}, \label{b3}
\end{equation}
with the polynomials $p_m$ listed in Table~\ref{N3}. These results are fully compatible with those from Ref. \onlinecite{simon2007}. As mentioned, there can be more than one (anti)symmetric polynomial of the same degree for sufficiently large $m$. This leads to the degenerate PP subspace, a specific example of which is presented in Sec.~\ref{sec:pf}. 

\begin{table*}[htb]
\centering
\renewcommand{\arraystretch}{2}
\begin{tabular}{|l|l|l|}\hline
$m$ &\ Bosonic $p_m(W ,Y)$ &\ Fermionic $p_m(W ,Y, A)$ \\    \hline
0 &\ $ 1 $ &\ 0 \\ \hline 
1 &\ $ 0$ &\ 0 \\  \hline
2 &\ $  1-\frac{2}{3}  W^2  $ &\ 0  \\ \hline 
3 &\ $ 3\sqrt{2}  Y$  &\ $A$ \\ \hline 
4 &\ $ 1-\frac{4}{3}  W^2+ \frac{2}{9}  W^4$ &\ 0 \\ \hline 
5 &\ $  \sqrt{2}  Y(6-  W^2)$  &\ $  A\left(2-\frac{1}{3}  W^2\right)  $ \\ \hline 
6 &\ \shortstack{(i)  $ 1-2  W^2 +\frac{2}{3} W^4 -\frac{4}{81}  W^6$  \\ (ii) $ \frac1{81}\sqrt{\frac{2}{5}}(2W^6-729Y^2)$ } &\ $ \frac{3}{\sqrt{5}}AY$ \\ \hline 
7 &\ $\frac{2}{3\sqrt{5}}(45-15W^2+W^4)Y$ &\ $ \sqrt{10}A\left(1-\frac{1}{3}  W^2 +\frac{1}{45}  W^4\right)$ \\ \hline
8 &\ (i)  $1+\frac{2}{243}W^2(W^2-6)(54-18W^2+W^4)$ &\ $  3 \sqrt{\frac{7}{5}}\left(1-\frac{2}{21} W^2\right)AY$ \\ &\ (ii) $\frac1{243}\sqrt{\frac{2}{35}}(2W^2-21)(2W^6-729Y^2)$ &\ \\ \hline
9 &\ \begin{tabular}{@{}c@{}}(i) $\sqrt{\frac{6}{155}}Y(-30+15W^2-2W^4+18Y^2)$ \\ (ii) $\frac1{27}\sqrt{\frac{2}{1085}}Y(-5760W^2+756W^4-31W^6 +81(140+9Y^2))$ \end{tabular}&\  \begin{tabular}{@{}c@{}} (i) $\frac{  A}{\sqrt{5}} (-10+5  W^2-\frac{2}{3}  W^4+\frac{2}{81} W^6)$   \\  (ii) $\frac{ A}{81\sqrt{105}}\left(W^6-729 Y^2\right)$ \end{tabular}
\\ \hline
%$\frac{  A}{\sqrt{5}} (-10+5  W^2-\frac{2}{3}  W^4
%+\frac{2}{81} W^6)$ \\ $ \frac{ A}{81\sqrt{105}}\left(W^6-729 Y^2\right)$ 
\end{tabular}
%\captionsetup{justification=centerlast}
\caption{The polynomials $p_m$ for $N=3$-body PPs for bosons and fermions up to $m=9$. 
Note that there is more than one possible PP for larger $m$, since there are multiple ways to build a homogeneous (anti)symmetric polynomial from the primitive and elementary symmetric polynomials. For instance, with $Y$ and $A$ defined in Eqs. \ref{Y} and \ref{A}, there are two possible bosonic PPs for $m=6$, since there are two ways ($W^6$ and $Y^2$) to build a homogeneous 6-degree polynomial from elementary symmetric polynomials. For the fermionic cases, we have explicitly kept only one factor of $A$, since even powers of $A$ can be expressed in terms of $W^2$ and $Y$. For all cases, the mean-square spread of the PPs also increases linearly with $m$, i.e. $\kappa^2\langle\sum_j^3 \bar n_j^2\rangle=m+1 $ (compare with the two-body case in Eq.~\ref{PPm}).  }
\label{N3}
\end{table*}

\subsubsection{$N\ge4$-body case}

For general PPs involving $N$ bodies, the inner product measure takes the form
\begin{eqnarray}
\nonumber \langle p_{m'},p_m\rangle &=& \int_0^\infty e^{-\frac{N-1}{N}W^2}W^{N-2} dW \\
\nonumber &&\times \int_0^{2\pi} d\phi_{N-2}\prod_{k=1}^{N-3}\int_0^\pi \sin^{N-2-k}(\phi_k) d\phi_k  \\
\nonumber && \times p_{m'}(W ,\phi_1,...,\phi_{N-2}) p_m(W ,\phi_1,...,\phi_{N-2}), \\
\label{innerN}
\end{eqnarray}
where the Jacobian determinant from Eq. \ref{innerproduct} has already been explicitly included. One transforms the tuple $\{\bar n_1,...,\bar n_N\}$ into $(N-1)$-dim spherical coordinates via the barycentric coordinates detailed in Appendix~\ref{sec:barycentric}. 

\begin{figure}[ttt]
\includegraphics[width=\linewidth]{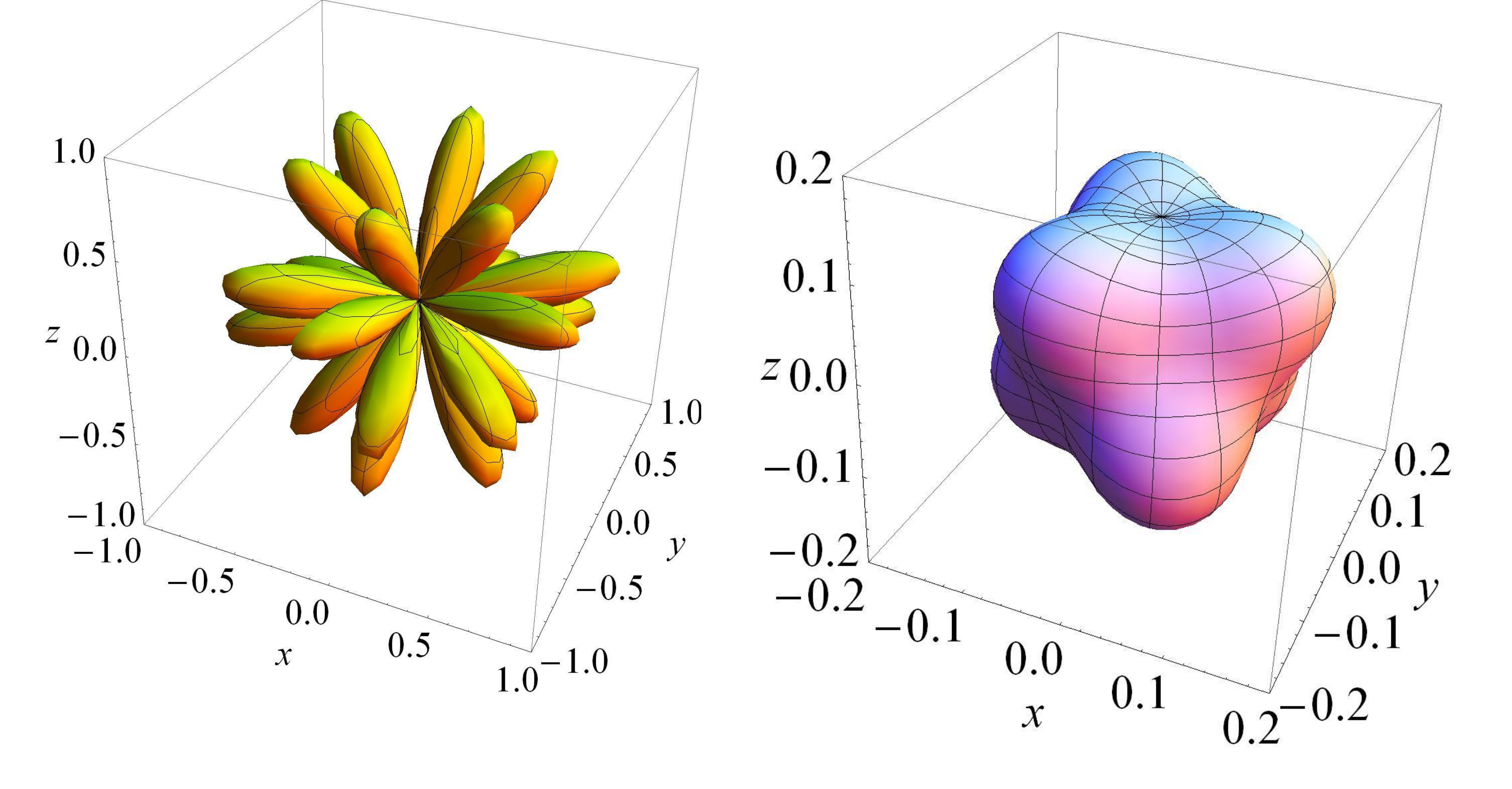}
%\captionsetup{justification=centerlast}
\caption{(Color online) The primitive polynomials have beautiful geometric shapes when plotted in $N-1$-dim spherical coordinates. Shown above are the constant $W$ plots of the antisymmetric polynomial $A=(\bar n_1-\bar n_2)(\bar n_1-\bar n_3)(\bar n_1-\bar n_4)(\bar n_2-\bar n_3)(\bar n_2-\bar n_4)(\bar n_3-\bar n_4)$ (left), and the degree $m=4$ symmetric expression in the orthonormalized space spanned by the elementary symmetric polynomials $1,S_2,S_3,S_4$ and $S^2_2$ (right). On the left, there are $24$ lobes that each maximally avoid the vertices of the tetrahedron (3-simplex). 
On the right, the $6$ lobes lie around the centers of the $6$ edges of the tetrahedron, where two of the $\bar n_i$'s are equal.   
}
\label{symantisymm4}
\end{figure}

The bosonic primitive basis is spanned by the elementary symmetric polynomials $\{S_2,S_3,...,S_N\}=\{\sum_{ij} \bar n_i\bar n_j,\sum_{ijk}\bar n_i\bar n_j\bar n_k,...,\prod_j\bar n_j\}$ and combinations thereof. For instance, with $N=5$ particles at degree $m=6$, there are $3$ possible primitive polynomials: $S_2^3,S_3^2$ and $S_2S_4$. The fermionic primitive basis is spanned by all the symmetric polynomials as above, \emph{times} the degree $\binom N{2}$ Vandermonde determinant $\prod_{i<j}(\bar n_i-\bar n_j)$ shown in Fig. \ref{symantisymm4}. 

From the examples above, one easily deduces the degeneracy of the PPs $U^N_m$ to be $P(m,N)-P(m-1,N)$ for bosons and $P\left(m-\binom N{2},N\right)-P\left(m-1-\binom N{2},N\right)$ for fermions, where $P(m,N)$ is the number of partitions of the integer $N$ into at most $m$ parts\cite{simon2007,davenport2012}. In particular, the degeneracy is always nontrivial ($\geq 2$) whenever $N\geq 4$ and $m\geq 4$. 

\section{Geometric construction of pseudopotentials: spinful case}
\label{sec:spinful}

In the presence of ``internal degrees of freedom" (DOFs) which we also refer to as ``spins" or ``components'' for simplicity, there are considerably more possibilities for the diverse forms of the PPs. This is because the PPs consist of products of spatial and spin parts, and either part can have many possible symmetry types, as long as they conspire to produce an overall (anti)symmetric PP in the case of (fermions) bosons.

A generic multicomponent PP takes the form
\begin{equation} 
U^N_m=\sum_\lambda\sum_\sigma U^N_{m,\lambda} \ket{\sigma^\lambda }\bra{\sigma^\lambda }. 
\end{equation}
The notation here requires some explanation. $\lambda=[\lambda_1,\lambda_2,...]$ defines a partition of all particles into several subgroups $\lambda_i$ each of which imposing a symmetry constraint among particles of this subgroup. Associated with it is $U^m_\lambda$ which is a spatial term (in $(\bar n_1,...,\bar n_N)$-space) exhibiting this symmetry. $\ket{\sigma^\lambda} =\ket{\sigma_1\sigma_2...}$ refers to an (internal) spin basis that is consistent with the symmetry type $\lambda$. Each symmetry type corresponds to a partition of $N$, with $\sum_j \lambda_j=N$. For $N$ bosons, $\lambda$ represents the situation where there is permutation symmetry among the first $\lambda_1$ particles, among the next $\lambda_2$ particles, etc. but \emph{no additional symmetry} between the $\lambda_i$ subsets. This is often represented by the Young Tableau with $\lambda_i$ boxes in the $i^{th}$ row. For fermions, we use the conjugate representation $\bar\lambda$, with the rows replaced by columns and symmetry conditions replaced by antisymmetry ones. For instance, the totally (anti)symmetric types are $[1,1,...,1]$ and $[N]$, respectively.

Hence the space of PPs is specified by three parameters: $N$ -- the number of particles interacting with each other, $m$ -- the total relative angular momentum or the total polynomial degree in $\bar n_i$, and $b$ -- the number of internal DOFs (spin). While the symmetry type $\lambda$ and hence $U^N_{m,\lambda}$ depends only on $N$ and $m$, the set of possible $\ket{\sigma^\lambda }$ also depends on $b$.
To further illustrate our notation, we specify $N,m,b$ parameters for the interactions relevant to some commonly known states: in the archetypical single-layer FQH states, we have $b=1$ components, and the $N=2$ PP interactions for the Laughlin state penalize pairs of particles with relative angular momentum $m$, where $\frac1{m+2}$ is the filling fraction. For bilayer FQH states, we have $b=2$ and $N=2$-body interactions. PPs as energy penalties in the $[2]$ sector with angular momentum $<m$ and $[1,1]$ sectors with angular momentum $<n$ give rise to the Halperin $(mmn)$ states. Here, the $[2]$ sector is also known as the triplet channel, as it is spanned by the following three basis vectors: $\{\ket{\uparrow\uparrow},\ket{\uparrow\downarrow}+\ket{\downarrow\uparrow},\ket{\downarrow\downarrow}\}$. By contrast, the $[1,1]$ sector only contains $\{\ket{\uparrow\downarrow}-\ket{\downarrow\uparrow}\}$ as dictated by antisymmetry.

\subsection{Multicomponent pseudopotentials for a given symmetry}

The construction of multicomponent PPs here parallels that of multicomponent wave functions described in Ref.~\onlinecite{davenport2012}. For completeness, we first review this construction, and proceed to show how an orthonormal multicomponent PP basis, adapted to the cylinder or torus, can be explicitly found through the geometric approach. We describe how to first find the spatial part of the PP $U^N_{m,\lambda}$, and second the spin basis $\ket{\sigma^\lambda}$. 

\subsubsection{Spatial part}

For each symmetry type $\lambda$, we can construct the spatial part $U^N_{m,\lambda}$ with elementary symmetric polynomials in \emph{subsets} of the particle indices $\bar n_i=n_i-n/N$. They are, for instance, $S_{1,12}= \bar n_1+\bar n_2$, $S_{2,234}=\bar n_2\bar n_3+\bar n_3 \bar n_4 +\bar n_2\bar n_4$, etc. Of course, we must have $S_{1,123...N}=\sum_i \bar n_i =0$.

Like in the single-component case, the spatial part $U^N_{m,\lambda}$ consists of a primitive polynomial which enforces the symmetry, and a totally symmetric factor that does not change the symmetry. Here, the main step in the multicomponent generalization is the \emph{replacement} of primitive polynomials $1$ and $A$ by primitive polynomials consistent with the symmetry type $\lambda$. 
As the simplest example, the primitive polynomial in 
$U^{N=3}_{m=1,[2,1]}$ is $S_{1,12}=-\bar n_3$ (and cyclic permutations). It is the only possible degree $m=1$ expression symmetric in two (but not all three) of the indices. 

In general, there can be more than one candidate monomial obeying a symmetry consistent with $\lambda$. 
For instance, for $U^{N=3}_{m=2,[2,1]}$ they are $S_{1,12}^2=(\bar n_1+\bar n_2)^2$ and $S_{2,12}=\bar n_1\bar n_2$ (and cyclic permutations thereof). To find the primitive polynomials, we will have to construct one or more linear combinations of these terms which do not have any higher symmetry other than $[2,1]$ (i.e. in this case, this higher symmetry channel could be $[3]$). Elementary computation reveals that the only primitive polynomial should be $S_{1,12}^2+2S_{2,12}$, because it is the only linear combination that is manifestly symmetric in indices $1,2$ {\it and} disappears upon symmetrization over all three particles.  

{\it $N=4$, $m=2$, $\lambda=[2,2]$.} As a more involved example, we demonstrate how to find the primitive polynomial corresponding to the symmetry type $[2,2]$. Independent monomials that satisfy this symmetry include $S_{1,12}^2$, $S_{2,12}$ and $S_{2,34}$. The primitive polynomial is then given by the linear combination 
\begin{equation}
S_{1,12}^2+\mu_1 S_{2,12}+\mu_2 S_{2,34}
\end{equation}
with $\mu_1$ and $\mu_2$ to be determined by demanding that the linear combination disappears upon symmetrizing over permutations under $[3,1]$ and $[4]$. The symmetrized sums are
\begin{eqnarray}
&&2((n_2+n_3)^2+(n_2+n_4)^2+(n_3+n_4)^2+\notag\\
&&\mu_1(n_1n_2+n_1n_3+n_2n_3)+\mu_2(n_1n_4+n_2n_4+n_3n_4))\notag\\
\end{eqnarray} and 
\begin{eqnarray}
4\left(n_2^2+n_3^2+n_4^2+n_2n_3+n_3n_4+n_2n_4\right)(4-\mu_1-\mu_2).\notag\\
\end{eqnarray}
Setting them both to zero, we find $\mu_1=\mu_2=2$, so the primitive polynomial is $S_{1,12}^2+2 S_{2,12}+2 S_{2,34}$.

\begin{figure}[ttt]
\includegraphics[width=0.8\linewidth]{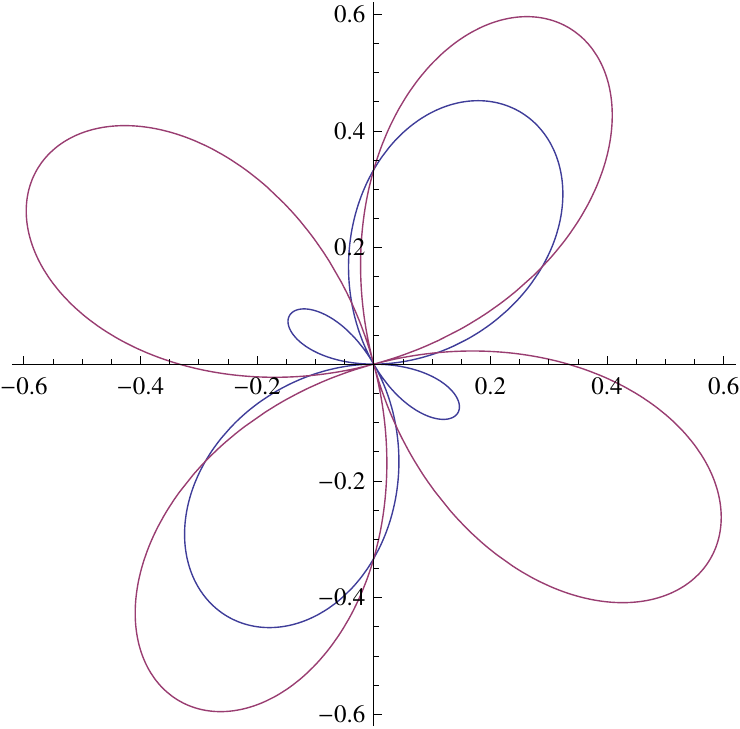}
%\captionsetup{justification=centerlast}
\caption{(Color online) Polar plots of $\beta_2(\theta)$ as a function of $\theta$, with $\mu=0.5$ (blue curve) and $\mu=2$ (purple curve). $\beta_2(\theta)+\beta_2(\theta+2\pi/3)+\beta_2(\theta+4\pi/3)$ sum to zero only when the lobes are of equal size, which is the case for $\mu=2$ only.   }
\label{fig:beta2}
\end{figure}

{\it $N=3$, $m=2$, $\lambda=[2,1]$.} The above procedure works for arbitrarily complicated cases, but quickly becomes cumbersome. This is when our geometric approach again becomes useful. %Below, we shall show how one can elegantly read out the the primitive polynomials through graphical inspection. 
We first write down the relevant monomials in barycentric coordinates given by Eq.~\ref{n3def} for $N=3$ bodies (or Eq.~\ref{qs} for general $N$). The coefficients in the primitive polynomial can then be elegantly determined through graphical inspection. 
We demonstrate this explicitly by revisiting the example on $\lambda=[2,1]$. Recall that the primitive polynomial (call it $\beta_{m=2}\equiv \beta_2$) is a linear combination of $S_{1,12}^2$ and $S_{2,12}$, i.e. 
\begin{eqnarray}
\beta_2 &=& S_{1,12}^2+\mu S_{2,12}\notag\\
&=& -\frac{W^2}{9\kappa}\left(\mu-2+(1+\mu)(\cos 2 \theta -\sqrt{3}\sin 2 \theta)\right)\notag\\
\end{eqnarray}
where $\theta=0,2\pi/3,4\pi/3$ points towards the vertices favoring $\bar n_1,\bar n_2,\bar n_3$ respectively. The correct value of $\mu$ will cause $\beta_2$ to disappear under symmetrization of the $3$ particles. Graphically, it means that the lobes of three copies of the plot of $\beta_2$, each rotated an angle $2\pi/3$ from each other, must cancel upon addition. This is illustrated in Fig. \ref{fig:beta2}, where $\mu=2$ is readily identified as the correct value. 

All in all, we have the $m=2$ primitive polynomial for $[2,1]$: 
\begin{equation}\beta_2= (\bar n_3)^2+2 \bar n_1 \bar n_2 =\frac{W^2}{3}\left(-\cos 2\theta+\sqrt{3}\sin 2\theta\right).
\label{beta2}
\end{equation}
{\it $N=3$, $m=1$, $\lambda=[2,1]$.} The reader is invited to also visualize the simpler case for the $[2,1]$ sector, which takes the form
\begin{equation}\beta_1= S_{1,12} =-\bar n_3=\frac{W}{3\kappa}\left(\cos\theta+\sqrt{3}\sin\theta\right).
\label{beta1}
\end{equation}
Compared to the antisymmetric case for fermions, we have two primitive polynomials $\beta_1,\beta_2$, instead of just $A$. The primitive polynomials for a wide variety of cases are also listed in Table II and III of Ref. \onlinecite{davenport2012}, and can alternatively be systematically derived via the theory of Matric units explained in the appendix of the same reference.

\subsubsection{Admissible spin bases}

In general, not all possible $N$-particle spin bases survive under symmetrization with respect to a given symmetry type $\lambda$. Only those that survive should be included, i.e. are admissible, in the set of basis in $\ket{\sigma^\lambda}$. For instance, the basis $\ket{\alpha\alpha\alpha }$ does not appear in $\ket{\sigma^{[2,1]}}$. To see this, try symmetrizing the combination $p(\bar n_1,\bar n_2,\bar n_3)\ket{\alpha\alpha\alpha }$ subject to the condition that $p(\bar n_1,\bar n_2,\bar n_3)$ has no higher symmetry than $\lambda=[2,1]$, i.e.  the totally symmetrized sum $\mathcal{S}\left[ p(\bar n_1,\bar n_2,\bar n_3)\right]=0$. Obviously, $p(\bar n_1,\bar n_2,\bar n_3)\ket{\alpha\alpha\alpha }$ then has to symmetrize to zero and should not be included in $[2,1]$. 

There is a nice way to write down the set of admissible bases by looking at the labelings of semistandard Young Tableaux. From Ref. \onlinecite{davenport2012}, the admissible bases in $\ket{\sigma^\lambda}$, up to permutations, are in one-to-one correspondence with the labelings of semistandard Young Tableaux with numbers $1$ to $N$. (A semistandard Young Tableau has non-decreasing entries along each row and strictly increasing entries along each column.) 
For instance, a symmetry type of $[3,1]$ with $b=3$ corresponds to labelings (with labels $a,b,c$ defined to be in increasing order): 
\begin{eqnarray}
\nonumber && [aaa,b], \; [aab,b], \; [aac,b], \; [abb,b], \; [abc,b], \\
\nonumber && [acc,b], \; [aaa,c], \; [aab,c], \; [aac,c], [abb,c], \\
 \nonumber && [abc,c], \; [acc,c], \; [bbb,c], \; [bbc,c], \; [bcc,c] \; {\rm and} \; [ccc,c],
\end{eqnarray}
where rows are separated by commas. From these, the admissible states for $\lambda=[3,1]$ can be written down by copying the labelings verbatim:
\begin{eqnarray} 
\nonumber && \ket{aaab}, \; \ket{aabb}, \; \ket{aacb}, \; \ket{abbb}, \; \ket{abcb}, \\
\nonumber && \ket{accb}, \; \ket{aaac}, \; \ket{aabc}, \; \ket{aacc}, \; \ket{abbc}, \\
\nonumber && \ket{abcc}, \; \ket{accc}, \; \ket{bbbc}, \; \ket{bbcc}, \; \ket{bccc} \; {\rm and} \; \ket{cccc}.
\end{eqnarray}
In the Young Tableau corresponding to $\lambda$, each row represents a set of particles that are symmetric with each other in the $\lambda$ representation. The requirement that labels increase monotonically within each row defines an ordering, and prevents the repeated listing of basis states related by permutation. The requirement of strictly increasing labelings down each column also prevents that, and also avoids the listing of bases that do not survive under symmetry constraints.

With these preliminary considerations, we are now in position to formulate the general recipe for constructing PPs in the multicomponent case, which is given in Table~\ref{table:summary}. In the following section we apply this recipe to the case of SU(2) spins with 3-body interactions.
\begin{table}[htb]
\centering
\renewcommand{\arraystretch}{2}
\begin{tabular}{|p{8.0cm}|}\hline
(1) Given $N$,$m$ and $b$, specify which symmetry type
the PP is associated with. For example, we specify the $[2]$ or the $[1,1]$ channel when computing PPs realizing the fermionic Halperin states in bilayer QH systems ($b=2,N=2$). \\ \hline
(2) Next, determine the appropriate primitive polynomials by finding the coefficients multiplying the allowed monomials. Then multiply the primitive polynomial by symmetric polynomials, and orthogonalize to obtain the spatial parts $U^N_{m,\lambda}$. \\ \hline
(3) Finally, choose the desired admissible spin channels, and (anti)symmetrize the resultant product of the spatial and spin parts depending on whether we want a bosonic (fermionic) PP. \\ \hline
\end{tabular}
%\captionsetup{justification=centerlast}
\caption{Summary of the PP construction procedure for the multicomponent case. Examples of this procedure are given in Section~\ref{sec:examples}.
}
\label{table:summary}
\end{table}

\subsection{Example: $N=3$-body case for SU(2) spins }

Here we explicitly work out the multi-component case $b=2$ which is also the most common multi-component scenario. We focus on $N=3$-body interactions to illustrate the nontrivial aspects of our approach.

The symmetry type is given by $[\lambda_1,\lambda_2]=[N/2+S,N/2-S]$, where $S$ is the total spin of the particles. Let us denote the spins by $\uparrow,\downarrow$. There are only $N=3$ boxes in the Young Tableaux, with the following possible symmetry types: $[3]$, $[2,1]$, $[1,1,1]$. 

For type $\lambda=[3]$, the possible spin bases are $\ket{\uparrow \uparrow \uparrow }$, $\ket{\uparrow \uparrow \downarrow }$, $\ket{\uparrow \downarrow \downarrow }$ and $\ket{\downarrow \downarrow \downarrow }$. For $\lambda=[2,1]$, the possible spin bases are $\ket{\uparrow \downarrow \downarrow }$, $\ket{\uparrow \uparrow \downarrow }$, corresponding to Tableau labellings $[\uparrow\downarrow,\downarrow]$ and $[\uparrow\uparrow,\downarrow]$ respectively. For $\lambda=[1,1,1]$, there is actually \emph{no} admissible spin basis: total (internal DOF) antisymmetry is impossible for 3 particles, if there are only 2 different spin states to choose from.

Consider bosonic particles in the following. The symmetry type $[3]$ case corresponds to the primitive polynomial $1$, so the resultant PP takes the same form as in the single-component case. After symmetrizing the spin part, we have the following available sets of spin channels for $[3]$: 
\begin{eqnarray}
\nonumber && \{\ket{\uparrow \uparrow \uparrow }\}, \\
\nonumber && \{\ket{\uparrow \uparrow \downarrow },\ket{\uparrow \downarrow \uparrow },\ket{\downarrow \uparrow \uparrow }\}, \\
\nonumber && \{\ket{\downarrow \downarrow \uparrow },\ket{\downarrow \uparrow \downarrow },\ket{\uparrow \downarrow \downarrow }\}, \\
\nonumber && \{\ket{\downarrow \downarrow \downarrow }\}.
\end{eqnarray}
For the PP $U^3_{m,[3]}\propto \kappa^3\sum_n b^{m\dagger}_nb^m_n$, we have
\begin{equation}
b^{m\dagger }_n\ket{0} = \sum_{\sum_j n_j=n}p_m e^{-\frac1{3}W^2}\ket{\uparrow \uparrow \uparrow }\
\label{bS32}
\end{equation}
for $S=3/2$, and
\begin{equation}
b^{\dagger m}_n\ket{0} =\mathcal{S} \left[ \sum_{\sum_j n_j=n}p_m e^{-\frac1{3}W^2}\ket{\uparrow \uparrow \downarrow }\right]
\label{bS12}
\end{equation}
for $S=1/2$. Other contributions with $S\rightarrow -S$ can be obtained via the identification $\ket{\uparrow} \leftrightarrow \ket{\downarrow}$. Here $\ket{\uparrow \uparrow \downarrow }$ is the shorthand for $c^\dagger_{n_1\uparrow}c^\dagger_{n_2\uparrow}c^\dagger_{n_3\downarrow}\ket{0}$, etc, and $p_m \equiv p_{m,[3]}$ refers to the \emph{same} polynomial as in the single-component case.

To construct the orthonormal PP basis for the first few $m$, we orthogonalize the set $\{\beta_1,\beta_2,\beta_1 W^2, \beta_2W^2,\beta_1 Y,\beta_2Y,\beta_1W^4,... \}$. 
The results are shown in Table \ref{N21}. Notice that the first PP for symmetry type $[2,1]$ occurs at $m=1$, whereas that of single-component bosons/fermions occur at $m=0$ and $m=3$ respectively. Indeed, there are more ways of constructing PP polynomials when only a subgroup of the full symmetric/alternating group is involved. The onset of PPs degenerate in $m$ also occurs earlier, at $m=4$.

\begin{table}[ttt]
\centering
\renewcommand{\arraystretch}{2}
\begin{tabular}{|l|l|}\hline
$m$ &\ $p_{m,[2,1]}(\beta_1,\beta_2,W,Y)$ \\    \hline
1 &\ $ \beta_1$ \\  \hline
2 &\ $ \frac{1}{\sqrt{3}}\beta_2 $ \\ \hline 
3 &\ $ \frac{1}{3} \sqrt{2} \beta_1 \left(-3+W^2\right)$  \\ \hline 
4(i) &\ $ \frac1{\sqrt{5}}\left(\beta_2+6 \beta_1 Y\right) $ \\ %\hline 
4(ii) &\ $ \frac1{27 \sqrt{5}}\left(-54 \beta_2+12 \beta_2 W^2+W^4 \left(\cos [4 \theta]+\sqrt{3} \sin [4 \theta]\right)\right)$ \\ \hline 
5(i) &\ $ \frac1{\sqrt{33}}\left(\beta_1 \left(-3+2 W^2\right)+6 \beta_2 Y\right) $ \\ %\hline 
5(ii) &\ $ \frac{1}{81} \sqrt{\frac{2}{55}} (-\sqrt{3} W^5 \cos [5 \theta ]+3 (5 \sqrt{3} \beta_1 (27-18 W^2$ \\ 
&\ $+2 W^4)+W^5 \sin [5 \theta ])) $ \\ \hline 
\end{tabular}
%\captionsetup{justification=centerlast}
\caption{The polynomials $p_m$ for the first few $N=3$-body PPs for bosons in the total spin $|S|=\frac{1}{2}$, i.e. $\lambda=[2,1]$ channel. The primitive polynomials $\beta_1$ and $\beta_2$ are given by Eqs. \ref{beta2} and \ref{beta1}.  $\cos 4\theta,\sin 4\theta, \cos 5\theta, \sin 5\theta$ can all be decomposed into the elementary symmetric polynomials and primitive polynomials $W,Y,\beta_1$, and $\beta_2$.  }
\label{N21}
\end{table}

To illustrate the full procedure for PP construction with internal DOFs, we detail the case of $S=-1/2$ below. For $m=1$, 
\begin{eqnarray}
b^{1\dagger}_n\ket{0} &=& \mathcal{S} \left[\sum_{\sum_j n_j=n}(-\bar n_3) e^{-\frac1{3}W^2}\ket{\uparrow \downarrow \downarrow }\right]\notag\\
\nonumber &=& -e^{-\frac{1}{3}W^2}[(\bar n_2 + \bar n_3) \ket{\uparrow\downarrow\downarrow} \\
&+& (\bar n_1 + \bar n_3) \ket{\downarrow\uparrow\downarrow}+(\bar n_1 + \bar n_2) \ket{\downarrow\downarrow\uparrow}]\notag\\
&=& e^{-\frac{1}{3}W^2}[\bar n_1 \ket{\uparrow\downarrow\downarrow} +\bar n_2 \ket{\downarrow\uparrow\downarrow}+\bar n_3 \ket{\downarrow\downarrow\uparrow}]\notag, \\
\label{b1S12}
\end{eqnarray}
while for $m=2$ we have 
\begin{eqnarray}
b^{2\dagger }_n|0\rangle &=& \mathcal{S} \left[\sum_{\sum_j n_j=n}((\bar n_3)^2+2\bar n_1\bar n_2) e^{-\frac1{3}W^2}\ket{\uparrow \downarrow \downarrow }\right]\notag\\
&= &e^{-\frac1{3}W^2} [((\bar n_2)^2 + (\bar n_3)^2+2\bar n_1 (\bar n_2+\bar n_3)) \ket{\uparrow\downarrow\downarrow}\notag\\ && +((\bar n_1)^2 + (\bar n_2)^2+2\bar n_3 (\bar n_1+\bar n_2)) \ket{\downarrow\downarrow\uparrow}\notag\\
&&+ ((\bar n_1)^2 + (\bar n_3)^2+2\bar n_2 (\bar n_1+\bar n_3)) \ket{\downarrow\uparrow\downarrow}], \notag\\
\label{b2S12}
\end{eqnarray}
Expressions for $S=1/2$ are obtained via $\ket{\uparrow} \leftrightarrow \ket{\downarrow}$.

For the present case of SU$(2)$ spins, i.e. $b=2$, there exists a nice closed-form generating function for the dimension of the spatial basis for each $m$. 
Define the generating function $Z_{[\lambda_1,\lambda_2]}(q)= \sum_m d({[\lambda_1,\lambda_2]};m)q^m $ where $ d({[\lambda_1,\lambda_2]};m)$ is the number of different polynomials with degree $m$ and symmetry type $[\lambda_1,\lambda_2]$. It can be shown that for bosons, %(with $\lambda_1\geq\lambda_2$),
\begin{eqnarray}
 Z_{[\lambda_1,\lambda_2]}(q) &=& \frac{1-q}{ \prod_{m=1}^{\lambda_1}(1-q^m)\prod_{n=1}^{\lambda_2}(1-q^n)} \notag\\
 &-&\frac{1-q}{ \prod_{m=1}^{\lambda_1+1}(1-q^m)\prod_{n=1}^{\lambda_2-1}(1-q^n)}.
\end{eqnarray}
This is obtained\cite{davenport2012} by considering the dimensionality from two symmetry subsets, and then subtracting overlaps from the higher symmetry case $[\lambda_1+1,\lambda_2-1]$. The dimension is related to the q-binomial coefficient.
For $b>2$, however, the situation is much more complicated, involving Kostka coefficients which count the number of semisimple labelings of $\lambda$ with a given alphabet.

\section{Pseudopotential Hamiltonians: case studies}
\label{sec:examples}

In this Section we provide several applications of the general pseudopotential construction developed in the previous sections, with examples arranged in the order of increasing complexity. We start with spin-polarized states (Sec.~\ref{sec:ex_polarized}), whose Hamiltonians have been obtained via alternative methods and are well-known in the literature. As an illustration of the method, we provide a detailed derivation of the fermionic Gaffnian parent Hamiltonian. 
Note that the resulting second-quantized form of the Hamiltonian applies (with minimal modifications) to both cylinder and torus geometries. In addition to pedagogical examples which illustrate our approach, we construct pseudopotential Hamiltonians for some non-Abelian states that have not been available in the literature for any geometry. 

The main sequence of steps is stated as follows:
\begin{enumerate}
\item Assuming that the ground-state wave function is known in the first-quantized form, find its thin torus pattern (this step has been frequently discussed in the literature; for completeness, we provide a brief summary in Appendix~\ref{sec:power}).
\item Using the formalism developed in the previous sections, write down the parent Hamiltonian corresponding to the root pattern in the second-quantized form.
\item Use numerics (exact diagonalization) to verify that the ground state of the proposed Hamiltonian is indeed given by the initial wave function. The verification criteria include testing for the unique zero-energy ground state at the given filling factor, the correct thin-cylinder root patterns as the circumference of the cylinder is taken to zero, and the level counting of entanglement spectra. For the purpose of numerical calculations, we place a finite total number of particles $N$ on the surface of a torus or an open cylinder. The two linear dimensions of the Hall surface ($L$ and $H$) satisfy the relation $LH=2\pi l_B^2 N_{\text{orb}}$, where $N_{\text{orb}}$ is the number of available orbitals (it is equal to $N_\phi$ on the torus, and equal to $N_\phi+1$ on the cylinder). Unless stated otherwise, we also assume $L=H$.
\end{enumerate}

\subsection{Spin-polarized states}\label{sec:ex_polarized}

The construction of two-body as well as the shortest-range three-body pseudopotential Hamiltonians has been discussed in depth in the literature, see e.g., Refs.~\onlinecite{prangegirvin, Greiter91, Greiter92, Read-PhysRevB.54.16864}. Starting from there, as our most elementary example for spin-polarized particles, we consider ground states of longer ranged 3-body potentials: the fermionic Gaffnian state at filling $\nu=2/5$ as well as the Haffnian and the generalized Moore-Read Pfaffian state at filling $\nu=1/q$, for which the $q=4$ case will be discussed in detail.

\subsubsection{Fermionic Gaffnian}

\begin{figure}[htb]
\includegraphics[width=0.98\linewidth]{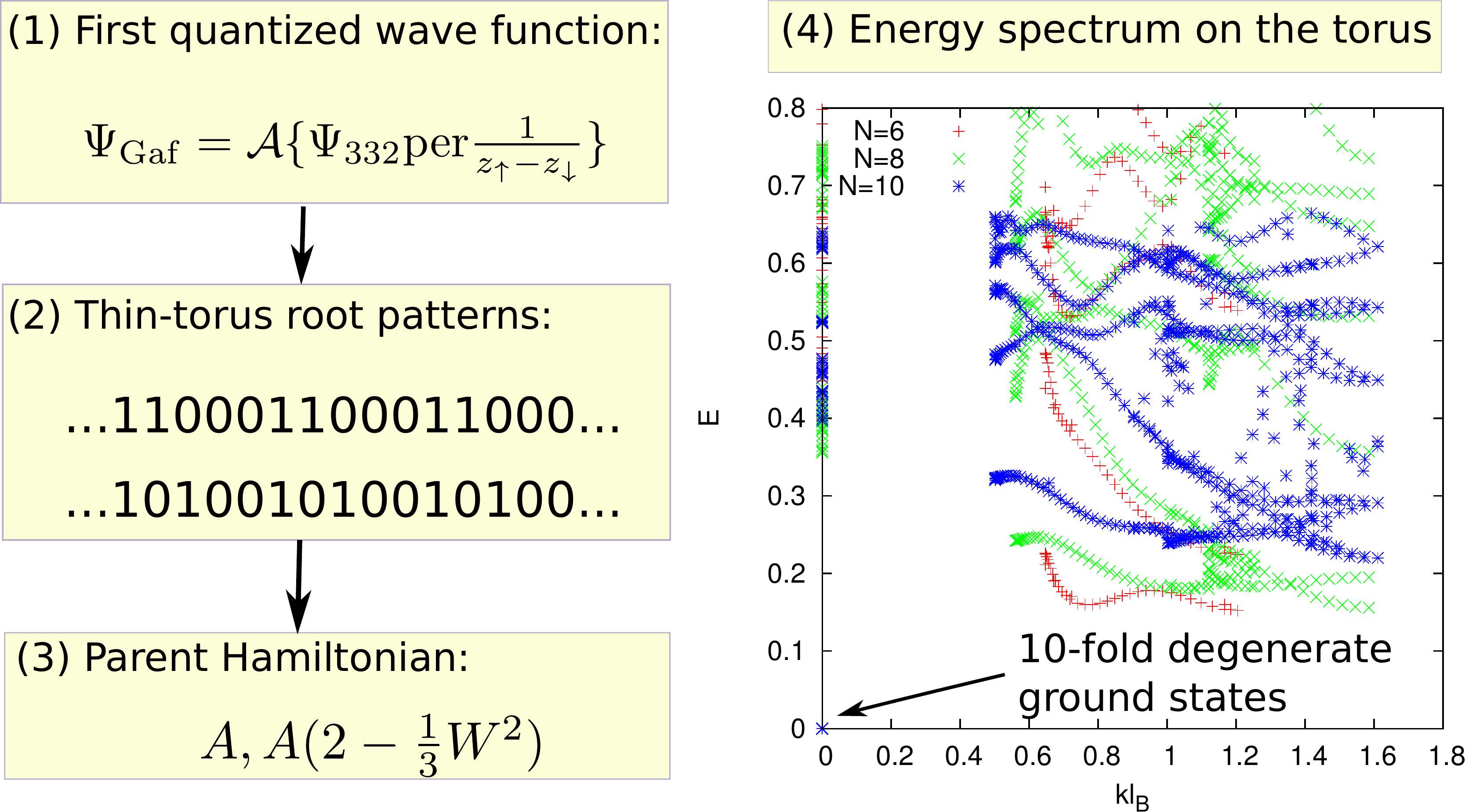}
%\captionsetup{justification=centerlast}
\caption{(Color online) Main steps in constructing the fermionic Gaffnian parent Hamiltonian. (i) Assume the first quantized wave function, in this example given by Eq.~\ref{gaffwf}. (ii) Find the corresponding thin torus root patterns (here $1100011000...$ and $1010010100...$). (iii) Construct the parent Hamiltonian by consulting Eq.~(\ref{b3}) and Table~\ref{N3}. (iv) Verify the construction and compute properties by exact diagonalization. Here, this is performed on a hexagonal torus with $N=6,8,10$ particles, finding a 10-fold ground state with exactly zero energy. The energy scale is set by $\gamma=1$ in Eq.~\ref{gaffnian_ham}. As the torus is stretched along one direction, the ground states evolve to the expected patterns $1100011000...$ and $1010010100...$. In addition, we obtain the entire low-lying neutral excitation spectrum, allowing one, in principle, to extract many other properties of the system.}
\label{fig:gaffnian_summary}
\end{figure}

The derivation of the fermionic Gaffnian parent Hamiltonian is summarized in Fig.~\ref{fig:gaffnian_summary}. The wave-function of the Gaffnian state on the infinite plane is given by~\cite{Simon-PhysRevB.75.075317}
\begin{equation}\label{gaffwf}
\Psi_{\rm Gaf} = \mathcal{A}\Big\{ \Psi_{332} (\{z_\uparrow \}, \{ z_\downarrow\}) {\rm Per}\left( \frac{1}{z_\uparrow - z_\downarrow}\right)\Big\},
\end{equation}
where we have suppressed the usual Gaussian factors. Although physically a one-component (spin-polarized) state, the above wave-function reflects the underlying ``two-component" nature of the Gaffnian state: in order to write down the wave function, we have divided electrons into two groups $\uparrow$ and $\downarrow$ (not to be confused with physical spin) that are correlated through the Jastrow factors within the 332 Halperin state, which is defined as
\begin{eqnarray}
\Psi_{332}(\{z_\uparrow,z_\downarrow\}) &=& \prod_{i<j}(z_{\uparrow,i}-z_{\uparrow,j})^3 \prod_{i<j}(z_{\downarrow,i}-z_{\downarrow,j})^3 \notag \\ 
&& \prod_{i,j} (z_{\uparrow,i}-z_{\downarrow,j})^2, \label{332}
\end{eqnarray}
as well as the ``permanent",
\begin{equation}\label{eq:per}
{\rm Per} \left( \frac{1}{z_\uparrow - z_\downarrow}\right) = \sum_\sigma \frac{1}{\prod_j (z_{\uparrow j}-z_{\downarrow \sigma(j)})}.
\end{equation}
Ultimately, the distinction between $\uparrow$ and $\downarrow$ particles is erased by the overall antisymmetrization $\mathcal{A}$, producing a well-defined single-component wave function. 

Having obtained the first quantized wave function for the Gaffnian, the second step is to determine its thin cylinder root patterns. (Note that a similar, but different analysis can also be executed on the sphere by analyzing the root partitions of the parent Hamiltonian null space~\cite{PhysRevB.86.125316}). The detailed procedure for finding the root patterns of Halperin bilayer states was given in Ref.~\onlinecite{Seidel-PhysRevLett.101.036804}, of which a brief summary is outlined in Appendix~\ref{sec:power}. The Gaffnian wave function vanishes as power $6$ as three particles are brought together, hence its thin torus root patterns are $1100011000...$ and $1010010100...$, accordingly.

From the form of the root patterns, we conclude that we need $U^{N=3}_{m=3}$ and $U^{N=3}_{m=5}$ terms in order to build the parent Hamiltonian. The polynomial amplitudes for these terms can be found in Eq. \ref{b3} and Table \ref{N3}. The Gaffnian parent Hamiltonian therefore reads
\begin{eqnarray}\label{gaffnian_ham}
&& H=\sum_n b_n^\dagger b_n + \gamma \sum_n d_n^\dagger d_n, \\
&& b_n = \sum_{\bar n_1+\bar n_2 + \bar n_3 = 0} \hspace{-10pt} A({\bar n}_1,{\bar n}_2,{\bar n}_3) e^{-\frac{\kappa^2}{2}\sum_{j=1}^3 \bar n_j^2} c_{n_1}c_{n_2} c_{n_3}, \\
\nonumber && d_n = \sum_{\bar n_1+\bar n_2 + \bar n_3 = 0} A({\bar n}_1,{\bar n}_2,{\bar n}_3)\left[ 2-\frac{1}{3}W^2({\bar n}_1,{\bar n}_2,{\bar n}_3) \right] \\
&& \hspace{20pt} \times e^{-\frac{\kappa^2}{2}\sum_{j=1}^3 \bar n_j^2} c_{n_1}c_{n_2} c_{n_3},
\end{eqnarray}
where as before  $\bar n_j=n_j-n/3$. The Hamiltonian (\ref{gaffnian_ham}) assigns positive energies to $m=3,5$ in a cluster of 3 particles in an infinite system, while all the other energies are exactly zero. The constant $\gamma$ tunes the ratio of the two non-zero energies and can be set to any positive number. The precise value of $\gamma$ does not affect the ground state wave function or its energy, but it does have an effect on the energetics of the low-lying excited states. In the following, for the sake of brevity, we refer to the Hamiltonian in Eq.~(\ref{gaffnian_ham}) simply by $A + \gamma A (2-\frac{1}{3}W^2)$.

Finally, having obtained the parent Hamiltonian, we can verify that it yields the correct ground state we started from (Eq.~\ref{gaffwf}). As we see in Fig.~\ref{fig:gaffnian_summary}, exact diagonalization on the torus for $N=6, 8, 10$ electrons consistently finds a zero-energy ground state that is 10-fold degenerate and has zero momentum (i.e., does not break translation symmetry). One can further verify that we have obtained the correct ground state by computing its entanglement properties, as we discuss below. Another non-trivial check is to perform diagonalization on a torus stretched along one axis, and directly identify the root patterns in the thin-torus limit. In doing so we find, as expected, the ground states to evolve to the single Fock states $1100011000...$ and $1010010100...$ (and those obtained from these by COM translation), in agreement with the fermionic Gaffnian state.

\subsubsection{Pfaffians and Haffnians}\label{sec:pf}

We next consider the generalized non-Abelian Pfaffian state at $1/q$ filling~\cite{Moore1991362,Read-PhysRevB.54.16864}. The wave function in the disk geometry reads
\begin{equation}\label{Pfq}
\text{Pf}\left(\frac1{z_i-z_j}\right)\prod_{i<j}(z_i-z_j)^q,
\end{equation} 
where even (odd) $q$ corresponds to a fermionic (bosonic) state. The Pfaffian is defined as
$$
{\rm Pf}(A) = \frac{1}{2^{n/2} (n/2)!}\sum_{\sigma\in S_{n}}\operatorname{sgn}(\sigma)\prod_{i=1}^{n/2}A_{\sigma(2i-1),\sigma(2i)},
$$
where $A$ is the $n\times n$ skew-symmetric matrix $A_{ij}=1/(z_i-z_j)$ for $n$ even.
The case $q=2$ reduces to the familiar Moore-Read state, which is the ground state of purely 3-body interaction $U_3^3$. One would naively expect that $q>2$ states can be obtained by adding 2-body terms to $U_3^3$, but as one can explicitly verify, this is not the case. 

By a power counting procedure~\cite{Read-PhysRevB.54.16864} (further elaborated on in Appendix \ref{sec:power}), the root configuration of the $1/q$-Pfaffian state is given by 
\begin{equation}10^{q-2}10^{q}10^{q-2}10^q......10^{q-2}10^q...
\label{pfaffianroot}\end{equation}
where $0^q$ represents a string of $q$ zeros. It follows that its parent Hamiltonian is given by the 3-body PP
\begin{equation}
U^3_{m=3(q-1)},
\end{equation}
as well as all nonzero spinless 2-body PPs 
\begin{equation}
U^2_{m<q-2}.
\end{equation}
For the more interesting $q=4$ case which we have studied numerically, we require the two-body $U^2_1$ PP and three-body $U^3_{9,(i)}$ and $U^3_{9,(ii)}$ PPs given by Eq. \ref{b3} and Table \ref{N3}. (Again, we emphasize that $U$'s promoted to how they appear in the Hamiltonian should be understood as polynomial amplitudes that enter the definition of operators $b_n^\dagger,b_n$ in Eq.~\ref{bn}.)

Note that labels $(i), (ii)$ stand for two linearly independent 3-body PPs that occur for $m=9$. Thus, the 1/4 Pfaffian state is an example whose parent Hamiltonian contains a degenerate subspace of PPs. The state can alternatively be obtained numerically through Jack polynomials~\cite{jack2} and the root configuration given above.  We have confirmed that the overlap between the ground state of the parent Hamiltonian and the Jack polynomial wave function is equal to one (within machine precision) for $q=4$ and $N=6, 8$ electrons. Note that experiments~\cite{luhman} find  some evidence for an incompressible, potentially non-Abelian, $\nu=1/4$ state. This may be attributed to the Pfaffian $1/4$ state, although theoretical calculations suggest that the Halperin 553 state is also a candidate~\cite{Papic09}. 

For our final spin-polarized example we consider the fermionic Haffnian state~\cite{green-10thesis}, whose bosonic counterpart was recently considered in Ref.~\onlinecite{Papic14}. The fermionic Haffnian occurs at the filling fraction $\nu=1/3$, and, up to COM translation, has the following root patterns~\cite{Hermanns-PhysRevB.83.241302} on the torus: $110000110000...$ and $100100100100...$. The latter pattern  is identical to that of the Laughlin state. The fermionic Haffnian vanishes as power $7$ as three particles are brought together, hence we need to impose an energy penalty for $m=6$ too, and the parent Hamiltonian consists of 
\begin{equation}
U^3_3, \; U^3_5, \; {\rm and} \; U^3_6,
\end{equation}
which are given by Eq. \ref{b3} and Table~\ref{N3}.

\subsection{Spinful states}

We have previously mentioned that if we allow the spin degree of freedom to enter, the number of possible states obviously becomes
much richer with interactions involving three or more particles. A systematic investigation
of these states is left for future work. Here we content ourselves with illustrating our method by formulating
parent Hamiltonians for several states that have been the subject of recent attention: the Spinful Gaffnian~\cite{davenport2013} and the non-Abelian Spin Singlet (NASS) states~\cite{ardonne2001}. The Hamiltonians for these states have previously been written down for the sphere (or disk) geometry, which we now extend to the cylinder and torus. Furthermore, we propose the parent Hamiltonian for a certain type of state involving the permanent (``221 times permanent" state~\cite{ardonne2011}), for which the Hamiltonian was previously unknown. 

\begin{figure}[htb]
\includegraphics[width=0.98\linewidth]{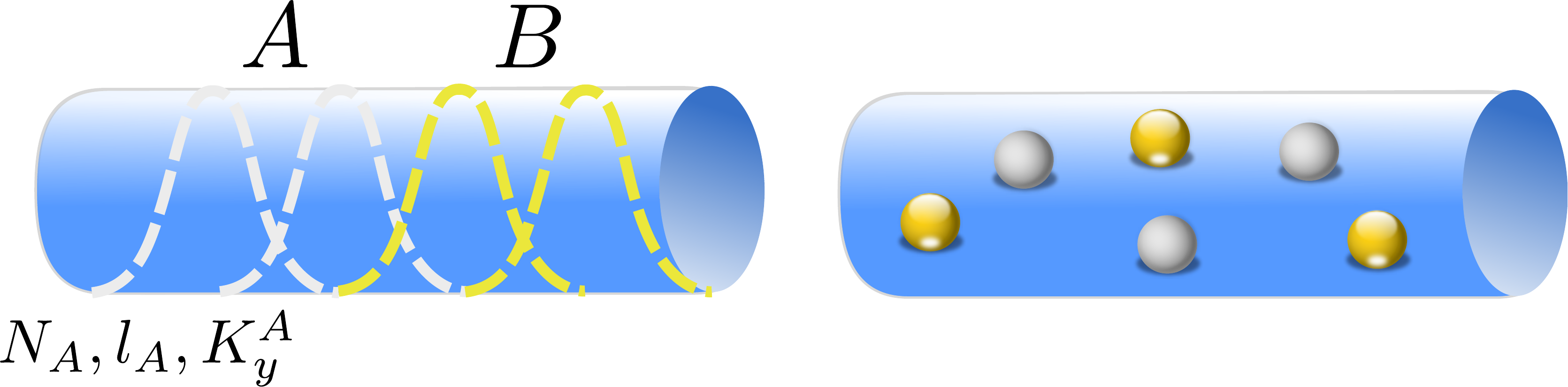}
%\captionsetup{justification=centerlast}
\caption{(Color online) Two choices of partitioning the system for computing the entanglement spectrum. Orbital cut (left) amounts to partitioning the system into two groups of orbitals $A$ and $B$, and tracing out the orbitals in $B$ (marked yellow). The Schmidt levels of $A$ can be classified by the number of particles $N_A$, the number of orbitals $l_A$, and the total momentum $K_y^A$. An alternative choice is particle partitioning (right), where some particles (marked in yellow) are traced out, regardless of their position. }
\label{fig:es_cuts}
\end{figure}
In addition to formulating the parent Hamiltonians, we also computed the orbital (OES) and particle entanglement spectra (PES) for the respective states. Fig.~\ref{fig:es_cuts} illustrates the two types of partitioning in the case of an open cylinder. 
After performing the cut, the entanglement spectrum can still have some remaining symmetry which can be used to classify the Schmidt levels. For example, for the OES, the total number of particles $N_A$ and the total number of orbitals $l_A$ in the left subsystem remain good quantum numbers. On the other hand, for the PES, the translation or rotation symmetry of the full system is preserved. For an open cylinder, as in Fig.~\ref{fig:es_cuts}, this means that the total momentum $K_y^A$ of the left subsystem is also a good quantum number:
\begin{equation}
K_y^A=\sum_{m\in A} m  \hat n_m,
\end{equation}
with $\hat n_m$ the density operator acting on the momentum orbital $m$. This linear momentum on the cylinder corresponds to the $L_z$ projection of angular momentum on the sphere. The PES on the sphere, however, is also invariant under full SU(2) rotation, i.e. it forms multiplets of $L^2$. This symmetry is absent on a cylinder, where we can only use $K_y^A$ to classify the levels of the entanglement spectrum. This is discussed in more detail in the example in Sec.~\ref{sec:gaffniansinglet}. As we perform an orbital partition, the symmetry of the subsystem is reduced, and even on the sphere, the only remaining quantum number is $L_z$. Thus, the OES on the sphere can be directly compared with the OES on the open cylinder.  Finally, for spinful states, the spin quantum number commutes with the reduced density operator of the OES, and thus allows for additional resolution of the ES level counting~\cite{PhysRevB.84.045127}.

\subsubsection{Spin-singlet Gaffnian state}\label{sec:gaffniansinglet}

\begin{figure}[bbb]
\includegraphics[width=0.98\linewidth]{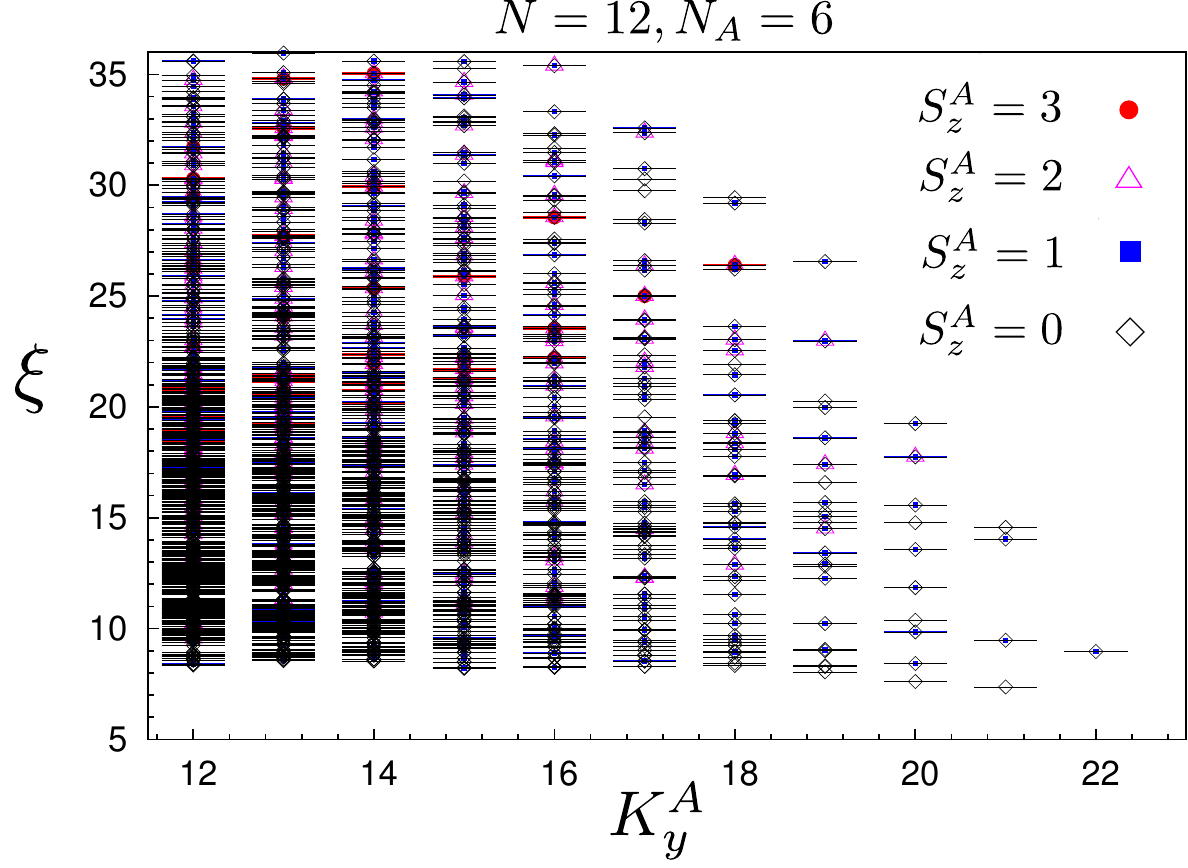}
%\captionsetup{justification=centerlast}
\caption{(Color online) PES of the spin-singlet Gaffnian state for $N=12$ particles and $N_\phi=12$ flux quanta on an open cylinder with aspect ratio equal to 1. The spectrum is obtained by tracing out $N_B=N_A=6$ particles and plotted as a function of momentum $K_y^A$ in part A. Different values of $S_z^A$, the total $z$ component of spin in part $A$, are indicated in the inset. }
\label{fig:gaffnian}
\end{figure}
The spin-singlet Gaffnian state is a nontrivial spinful generalization of the bosonic  spin-polarized Gaffnian state by Davenport \emph{et al.}\cite{davenport2013}. In addition to the two spin polarized 3-body terms with $S=3/2$, its parent Hamiltonian also contains the shortest range ($m=1$) term with $S=1/2$. In total, this encompasses the PPs
\begin{equation}\label{spinfulgaffnian}
U^3_{0,S=3/2}, \; U^3_{2,S=3/2} \; {\rm and} \; U^3_{1,S=1/2}.
\end{equation}
As such, the spin-singlet Gaffnian wave function vanishes as the third power in the $S=3/2$ channel, and the second power in $S=1/2$. These projectors are given by Eqs.~\ref{b3},~\ref{bS32} and Table~\ref{N3} for $U^3_{0,3/2}$ and $U^3_{2,3/2}$, and Eq.~\ref{b1S12} for $U^3_{1,S=1/2}$ (with both spin orientations $\ket{\uparrow}\leftrightarrow \ket{\downarrow}$). In addition to these, we also add the total spin operator $S^2$ to our Hamiltonian, to ensure that the ground state is a spin-singlet. 

By diagonalizing the Hamiltonian (\ref{spinfulgaffnian}) numerically, we find a zero-energy ground state at filling factor $\nu=4/5$ and shift of $-3$ on a finite cylinder for $N=4, 8,$ and  $12$ particles. We have furthermore computed the PES~\cite{PES}  for $N=12$ particles and $N_\phi=12$ flux quanta, shown in Fig.~\ref{fig:gaffnian}. As illustrated in Fig.~\ref{fig:es_cuts} (right), to perform a PES ``partition", we divide the system into parts $A$ and $B$, which both contain $N_\phi$ orbitals, but $N_A$ and $N_B$ particles, respectively, such that $N_A+N_B=N$. To obtain Fig.~\ref{fig:gaffnian}, we have traced out $N_B=6$ particles from the system. The PES obtained in this way provides information about the counting of quasihole excitations of the given state, as shown in Ref.~\onlinecite{PES}. 

We compare the counting in Fig.~\ref{fig:gaffnian} with the corresponding PES obtained on the sphere in Davenport \emph{et al.}~\cite{davenport2013}. Note that our result in Fig.~\ref{fig:gaffnian} superficially looks different from the result in Ref.~\onlinecite{davenport2013}. This is because the PES partition preserves the symmetry of the ground state, causing the PES on the sphere to have exact rotational symmetry which is absent for our open cylinder. In other words, on an open cylinder, the good quantum number after partitioning is only the linear momentum $K_y^A$. This quantum number, in turn, corresponds to the $L_z$ projection of angular momentum on the sphere. This, however, does not exhaust all symmtries of the PES on the sphere, where the full angular momentum $L^2$ is a good quantum number. This additional degeneracy of the PES is factored out in Ref.~\onlinecite{davenport2013}. Furthermore, on both the sphere and cylinder, due to the singlet property of the wave function, we expect the PES levels to be multiplets of the $S^2$ operator~\cite{PhysRevB.84.045127}, as can be verified in Fig.~\ref{fig:gaffnian}. The most important universal information is the counting of PES levels per momentum sector, which we find to be in agreement with Ref.~\onlinecite{davenport2013}. As we restore the $L^2$ degeneracy in the PES given in  Ref.~\onlinecite{davenport2013}, the counting for the first three sectors is a single level with $S=1$ and $L^z=22$, 4 levels at $L_z=21$ (two with $S=1$ and two with $S=0$), and 10 levels at $L_z=20$ (one of them with $S=2$, 6 with $S=1$ and 3 with $S=0$). While the finite-size splitting between these levels is non-universal (and differs between sphere and cylinder), we indeed obtain the identical counting per sector (Fig.~\ref{fig:gaffnian}). 

\begin{figure*}[ttt]
  \begin{minipage}[l]{0.9\linewidth}
\includegraphics[width=0.98\linewidth]{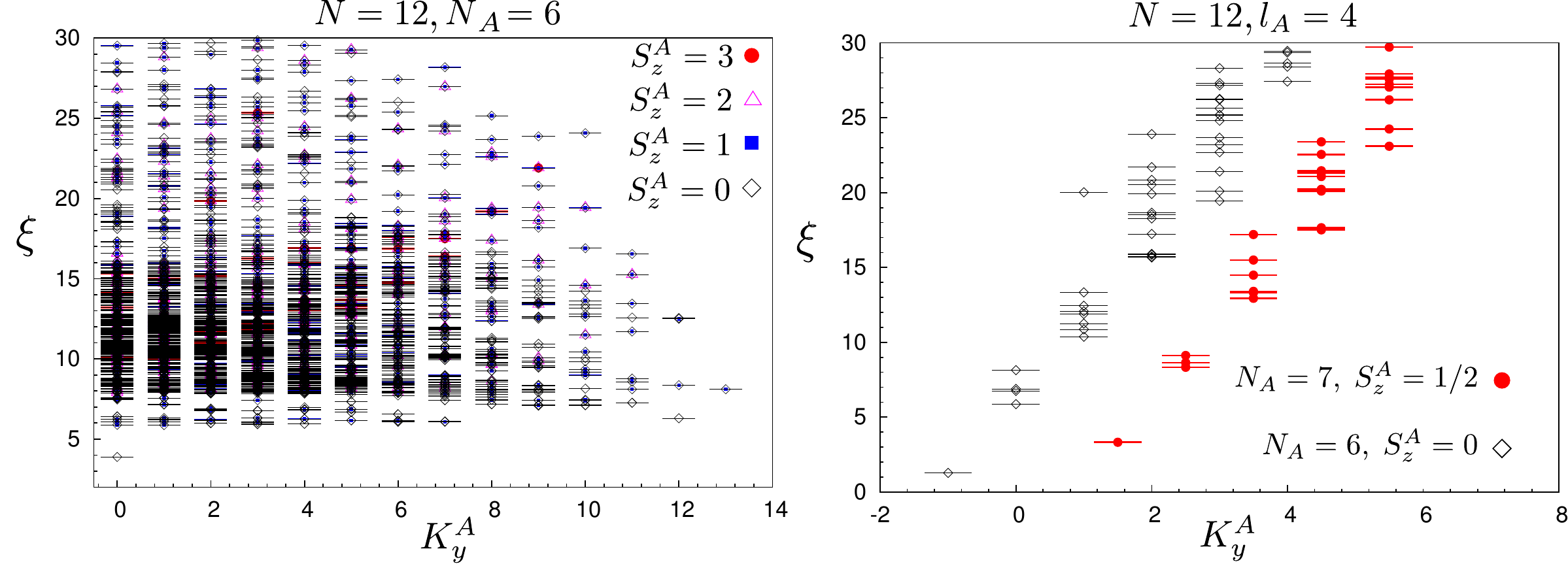}
  \end{minipage}
%	\captionsetup{justification=centerlast}
\caption{(Color Online) PES (left) and OES (right) for the NASS $\nu=4/3$ state.  The system contains $N=12$ bosons on a finite cylinder with aspect ratio 1. For PES, the subsystem $A$ contains $N_A=6$ particles, and PES is plotted as a function of momentum $K_y^A$ of subsystem $A$. The states are labeled by different values of $S_z^A$, the total $z$ component of spin in part $A$, as indicated in the inset.
For OES, the subsystem $A$ contains $l_A=4$ orbitals, where we show data for $N_A=6$ and $N_A=7$. The spectrum is resolved as a function of momentum $K_y^A$ of subsystem $A$.}
\label{fig:NASS43}
\end{figure*}

\subsubsection{The NASS state}

Another class of non-Abelian spin-singlet states has been proposed by Ardonne \emph{et al.}\cite{ardonne2001} under the name ``non-Abelian spin singlet" states (NASS). The bosonic family of such states occurs at filling factors $\nu=2k/3$. According to Ref.~\onlinecite{ardonne2011}, the NASS wave function ought to vanish quadratically when we bring together $k+1$ particles of the same spin, and linearly for $k+1$ particles of different spins. For SU(2) spins, such wave functions only exist in the $[k+1]$ and $[k,1]$ representations. Its parent Hamiltonian should therefore contain the PPs $U^k_{m=0,S=k/2}$ and $U^k_{m=0,S=k/2-1}$, since the simplest nontrivial totally/partially symmetric polynomial are of degrees two and one, respectively. 
Explicitly, the $b_n^{m=0\dagger}$ operators are
\begin{eqnarray}\label{NASSham}
b_{n,S=k/2}^{0\dagger}\ket{0} &=&e^{-\frac{\kappa^2}{2}\sum_i^k \bar n_i^2} \ket{\uparrow \uparrow...\uparrow }, \\
\nonumber \label{NASSham2} b_{n,S=k/2-1}^{0\dagger}\ket{0} &=& e^{-\frac{\kappa^2}{2}\sum_i^k \bar n_i^2}[\ket{\uparrow \uparrow...\uparrow\downarrow } +\ket{\uparrow \uparrow...\downarrow\uparrow } \\
&+&\ket{\uparrow \uparrow...\downarrow\uparrow\uparrow }+...+\ket{\downarrow \uparrow...\uparrow }] ,
\end{eqnarray}
and similarly for the terms $S\rightarrow -S$ obtained by exchanging $\ket{\uparrow}\leftrightarrow \ket{\downarrow}$.

The simplest member $k=2$ of the NASS family has recently been studied from the ``squeezing" perspective~\cite{ardonne2011}. Below we complement these results by independently generating the NASS $k=2$ state using its parent Hamiltonian in Eqs.~\ref{NASSham} and~\ref{NASSham2}. We compute its PES and OES on a finite cylinder, Fig.~\ref{fig:NASS43}. The PES and OES are computed for $N=12$ bosons on a cylinder with aspect ratio 1. Similar to the previous cases, the universal information in the PES spectrum -- the counting of levels per momentum sector -- is in agreement with the independently obtained data in Ref.~\onlinecite{ardonne2011} on the sphere. In this case, the bipartition of the system is obtained by fixing the cut such that there are $l_A=4$ orbitals in part A. We compute the OES for two choices of the total number of particles in A, $N_A=6$ and $N_A=7$. We see that the counting of the OES changes as a function of $N_A$, as it often happens for non-Abelian states where moving the cut probes different topological sectors of the theory, in this case associated with SU$(3)_2$ CFT. 

\subsubsection{Halperin-permanent states}

Finally, we tackle more complex examples of non-Abelian states that involve a product of Halperin multicomponent states~\cite{Halperin83} and the permanent. To be specific, we study states of the following form in the disk geometry: 
\begin{eqnarray} 
\Phi_{lln}&=& \text{Per}\left(\frac{1}{z_i-w_j}\right) \psi_{lln}(\{z\},\{w\}), 
\label{phidef}
\end{eqnarray}
where $z_i,w_i$ refer to the spin up and spin down particle coordinates, respectively, and $\psi_{lln}$ is the Halperin wave function~\cite{Halperin83}, an example of which was given in Eq.~\ref{332} for $l=3$, $n=2$. We have suppressed, as usual, the spinor and Gaussian parts of the wave function. The permanent was defined in Eq.~(\ref{eq:per}).

We are particularly interested in the states $\Phi_{111}$ and $\Phi_{221}$. The former was introduced by Moore and Read~\cite{Moore1991362}, and subsequently analyzed in detail by Read and Rezayi~\cite{Read-PhysRevB.54.16864} (see also Ref.~\onlinecite{green-10thesis}). As a prototypical example of a state that derives from a non-unitary CFT, it was shown to describe a critical point between the ferromagnet and paramagnet. Its thin-torus limit was recently studied in Ref.~\onlinecite{Papic14}. The latter state, $\Phi_{221}$, has been addressed by Ardonne and Regnault~\cite{ardonne2011}, who computed some of its properties by identifying its root configuration on the sphere and deriving its ``squeezing" properties. Here we show how to write down the parent Hamiltonian for these two states for the cylinder and torus geometry. This is of particular interest in the case of $\Phi_{221}$, where such a Hamiltonian has not been reported for any geometry before.

\begin{figure*}[t]
  \begin{minipage}[l]{0.9\linewidth}
\includegraphics[width=0.98\linewidth]{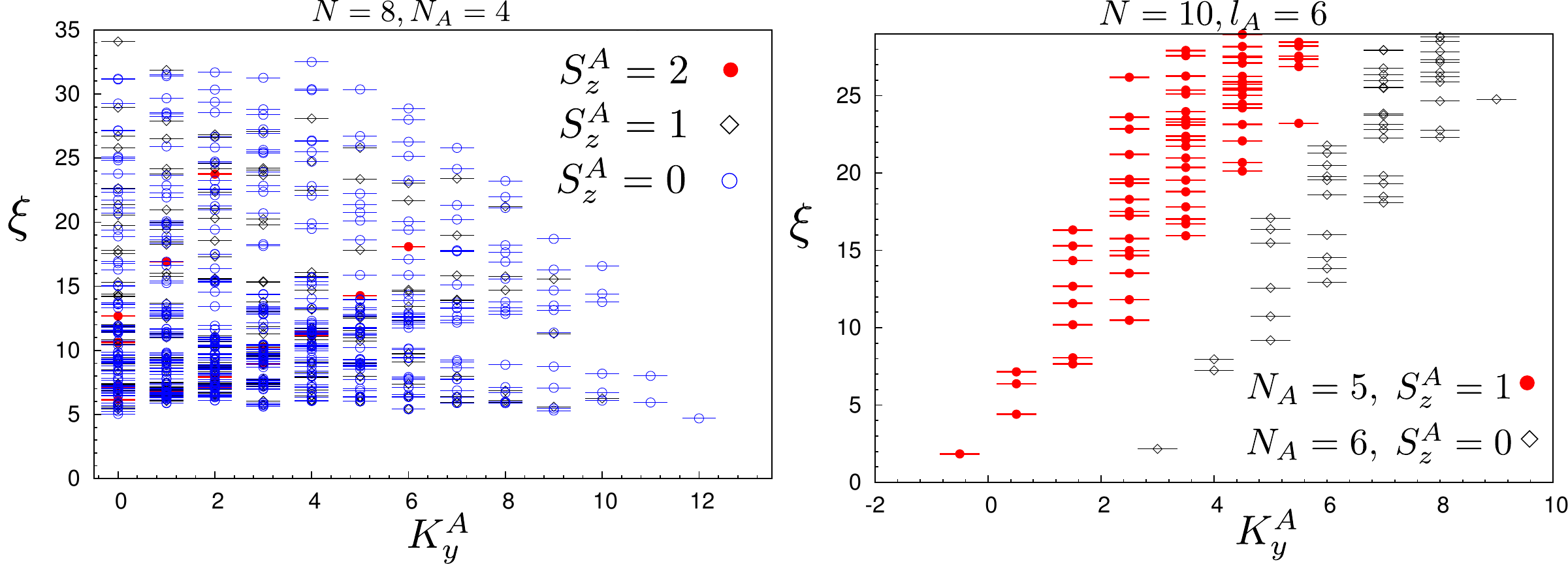}
  \end{minipage}
%	\captionsetup{justification=centerlast}
\caption{(Color online) PES (left) and OES (right) for the $\Psi_{221}$ state on a finite cylinder with aspect ratio equal to 1. For PES, we consider $N=8$ bosons and the subsystem $A$ contains $N_A=4$ particles. The PES is plotted as a function of momentum $K_y^A$ of subsystem $A$. Different values of $S_z^A$, the total $z$ component of spin in part $A$, are indicated in the inset. For OES, we consider $N=10$ bosons, the subsystem $A$ contains $l_A=6$ orbitals and the OES is plotted as a function of momentum $K_y^A$ of subsystem $A$, for two values of the total number of particles in subsystem $A$ ($N_A=5$ and $N_A=6$).}
\label{fig:perm221}
\end{figure*}

To begin with, we want to derive many-body projection Hamiltonians whose null space contains $\Phi_{lln}$. We first consider 2-body interactions. From Eq.~\ref{phidef}, $\Phi_{lln}$ vanishes as the $l^{th}$ power when $z_i\rightarrow z_j$ (or $w_i\rightarrow w_j$). Hence we need 2-body terms $U^2_{m,S=1}$ where $m<l$ and $|S|=1$, i.e. in the $\ket{\uparrow\uparrow}$ or $\ket{\downarrow\downarrow}$ channels. 
In the $S=0$ (or $\ket{\uparrow\downarrow}$) channel, however, one term in the permanent will cause $\Phi_{lln}$ not to vanish for $z_i\rightarrow w_k$. Hence we do not need any $U^2_{m,S=0}$.

For 3-body interactions, we find that the Halperin function parts contribute a degree of $l+2n$ for $S=\frac{1}{2}$ (or the $\ket{\uparrow\uparrow\downarrow}$, $\ket{\uparrow\downarrow\downarrow}$, etc.) channels. Also, the permanent will remove one degree. Hence we need $U^3_{m,1/2}$ for $m<l+2n-1$. An analogous analysis for $\ket{\uparrow\uparrow\uparrow}$ or $\ket{\downarrow\downarrow\downarrow}$ reveals that we also need $U^3_{m,3/2}$ for $m<3l$.
The matter simplifies a bit since not all of the requisite $U^{2,3}_{m,S}$ mentioned above actually exist. In Sec.~\ref{sec:spinful} and also Table 1 of Ref.~\onlinecite{davenport2012}, the PPs $U^2_{m,S}$ are nonzero for even $m$ when $l$ is even and $S=1$, or $l$ odd and $S=0$ (and vice versa for total anti-symmetry in spin-orbit space). For $N=3$, $U^3_{m,1/2}$ exists for $m\geq 1$, but $U^3_{m,3/2}$ does not exist for $m=0,1,2$ in the case of fermions (odd $l$). 
In summary, the above considerations for $n=1$ and $l=1,2$ imply that only $U^3_{1,1/2}$ is required to produce $\Phi_{111}$, consistent with Ref.~\onlinecite{Read-PhysRevB.54.16864}. On the other hand, $U^2_{0,1}$, $U^3_{1,1/2}$, $U^3_{2,1/2}$, and $U^3_{m,3/2}$ for $m=0,2,3,4,5$ are all required to produce $\Phi_{221}$. Therefore, we see that formulating the parent Hamiltonian for $\Phi_{221}$ is indeed rather involved, and crucially benefits from the systematic approach we follow here.

For completeness, we quote the explicit expressions for these two Hamiltonians in the form $U^{N}_{m,S}=\sum_{n}b_n^{m,S\dagger}b_n^{m,S} + (\ket{\uparrow}\leftrightarrow \ket{\downarrow})$.  
For the $\Phi_{111}$ state, we have 
\begin{eqnarray}
H_{111} &=& U^3_{1,1/2}\propto \sum_{n}b^{1,1/2\dagger}_{n}b^{1,1/2}_{n}, 
\end{eqnarray}
with the corresponding operator and polynomial amplitude
\begin{eqnarray}
b^{1,1/2}_n &=& \kappa^{3/2}\sum_{\sum n_i =n} p_{1,1/2}(\kappa; \{n\})e^{-\frac{\kappa^2}{3} W^2}c_{n_1}c_{n_2}c_{n_3}, \notag\\ \\
\nonumber p_{1,1/2} &=& (\bar n_1 - \bar n_2) \ket{\uparrow\uparrow\downarrow} +(\bar n_2 - \bar n_3) \ket{\downarrow\uparrow\uparrow} \\
&+& (\bar n_3 - \bar n_1) \ket{\uparrow\downarrow\uparrow},
\end{eqnarray}
where as before $\bar n_i= n_i-\frac{1}{3}\sum_j n_j$. 
For $\Phi_{221}$, we have 
\begin{eqnarray}\label{perm221ham}
\nonumber H_{221} &=& U^2_{0,1} + \sum_{m=1,2}\gamma_mU^3_{m,1/2} \\
&+&\sum_{m=0,2,3,4,5}\eta_mU^3_{m,3/2},
\end{eqnarray}
where again $\gamma_m$ and $\eta_m$ can be chosen to be arbitrary positive constants. 

We have numerically diagonalized the Hamiltonian (\ref{perm221ham}) and verified that it gives a unique zero energy ground state on finite cylinders when $N_\phi=3N/2-3$. By stretching the cylinder, we confirmed that the ground state reduces to the pattern $(\uparrow,\downarrow)00(\uparrow,\downarrow)00...$, which coincides with the root configuration given in Ref.~\onlinecite{ardonne2011}. In Fig.~\ref{fig:perm221} (left) we  compute the PES for $N=8$ bosons, which can be directly compared with Fig. 7 in Ref.~\onlinecite{ardonne2011}. As expected, we find an agreement in the counting of the PES levels between our results on the cylinder and the sphere data in Ref.~\onlinecite{ardonne2011}. Finally, we also provide a plot of the OES in Fig.~\ref{fig:perm221} (right). As shown by Li and Haldane~\cite{LiPhysRevLett.101.010504}, a non-Abelian state like the Moore-Read state will have different counting depending on the position of the orbital entanglement cut, which effectively probes different topological sectors of the theory. This is also the case in our example, as we see the counting changes depending on the number of particles in subsystem $A$ while the location of the cut remains fixed. Furthermore, the spectrum is fully chiral, which suggests that this state may be described by a simple CFT as conjectured in Ref.~\onlinecite{ardonne2011}. The CFT can in principle be identified using the counting of the OES in Fig.~\ref{fig:perm221}. For this purpose, larger systems may be required in order to verify that the counting we see in Fig.~\ref{fig:perm221} is saturated and indeed corresponds to the thermodynamic limit, which we defer to future work.

\section{Conclusion}\label{sec:conclusions}

We have provided a general method for constructing quantum Hall parent Hamiltonians with the desired type of clustering properties on cylinder and torus geometries. The approach developed here completes the program initiated in Ref.~\onlinecite{Simon-PhysRevB.75.075318} which is suited for the disk and sphere geometry that have a conserved $z$-component of angular momentum.
We have performed extensive analytic and numerical checks of the proposed model Hamiltonians, and generally find complete agreement with previous results if available. We have also demonstrated that it is possible to construct the parent Hamiltonians for some rather complex spinful non-Abelian states which previously have not been provided in the literature.

The method presented here opens up several future directions. One can use the Hamiltonians derived here to systematically scan for filling factors where the Hamiltonians have zero-energy ground states, in the spirit of Ref.~\onlinecite{Simon-PhysRevB.81.121301}. Performing this kind of search is much more natural in the torus geometry which does not suffer from the known shift bias~\cite{wenzee}. Furthermore, our approach straightforwardly generalizes to 4-body interactions whose possible incompressible ground states have not been systematically investigated before. Another interesting extension to pursue is the study of multicomponent states beyond the familiar SU(2) spin, such as the SU(4) valley/spin symmetric non-Abelian states. They may have relevance in the realization of FQHE in graphene, where the open accessibility of the electron gas may allow for further tunability of the interaction profile~\cite{grapheneFQHE,PhysRevLett.107.176602}.   

\begin{acknowledgments}
CH Lee is supported by an NSS scholarship from the Agency of Science, Technology and Research of Singapore. ZP acknowledges support by DOE grant DE-SC0002140. Research at Perimeter Institute
is supported by the Government of Canada through Industry
Canada and by the Province of Ontario through the Ministry
of Economic Development \& Innovation. RT is supported by the European Research Council through ERC-StG-TOPOLECTRICS-336012.
\end{acknowledgments}

\appendix
\section{Barycentric coordinates for many-body pseudopotentials}
\label{sec:barycentric}
We describe how to find a manifestly $S_N$-symmetric embedding of the tuple $(\bar n_1,...,\bar n_N)$ onto the $(N-1)$-dimensional simplex in $\mathbb{R}^{N-1}$. This can be achieved in barycentric coordinates, which is also useful for diverse applications involving permutation symmetry with a linear constraint~\cite{leeandy2014}. The most straightforward construction is to write 
\begin{equation}
\vec x = \kappa\sum_{k=1}^N \bar n_k \vec\beta_k,
\label{xbary}
\end{equation}where $\vec x\in \mathbb{R}^{N-1}$ and $\{\vec\beta_k\}$, $k=1,...,N$ forms a linearly dependent set of basis vectors, also in $\mathbb{R}^{N-1}$, normalized so that 
\begin{equation}
\vec\beta_j \cdot \vec\beta_k = \frac{N}{N-1}\delta_{jk} - \frac{1}{N-1}.
\label{betadot}\end{equation}
Geometrically, the $\beta_k$'s define the vertices of a simplex, and are at angle of $\cos^{-1}(-1/(N-1))$ from one another. The vertex $k$ corresponds to the least isotropic configuration with $\bar n_k\propto \frac{N-1}{N}$ and $\bar n_{j}\propto -\frac{1}{N}$, $j\neq k$. A basis consistent with the above requirements is \begin{subequations}\begin{align}
\vec\beta_1 &= (1,0,\ldots,0), \\
\vec\beta_2 &= (C_1, S_1,\ldots,0), \\
\vec\beta_3 &= (C_1, S_1C_2, S_1S_2,\ldots,0), \\
\vec\beta_4 &= (C_1, S_1C_2, S_1S_2C_3,S_1S_2S_3,\ldots,0), \\
&\vdots \notag \\
\vec\beta_{N-1} &= (C_1,S_1C_2,S_1S_2C_3,\ldots, S_1\cdots S_{N-2}), \\
\vec\beta_N &= (C_1,S_1C_2,S_1S_2C_3,\ldots, -S_1\cdots S_{N-2}),
\end{align}\label{barybasis}\end{subequations}
with $S^2_k+C^2_k=1$, $1\leq k\leq N-2$. Upon enforcing the scalar product constraint in Eq. \ref{betadot}, we require that $S_{k+1}=1-(1+1/N)/(\prod_{j=1}^k S_k)^2)$, which also implies that $C_k^2+S_k^2C_{k+1}=C_k$.  
A simple solution fortunately exists: 
\begin{equation}
C_k = -\frac{1}{N-k}, %\;\;\;\; S_k^2+C_k^2 = 1,
\label{recur2}\end{equation}
which implies that the $k$-th projected component of the relative angles between the position vectors of the vertices approaches $\pi$ as $k$ and $N$ increase. We proceed by substituting Eq.~\ref{recur2} into the explicit form of vertex positions $\vec\beta_k$, and expressing the latter in terms of the spherical coordinates.
From Eq. \ref{xbary}, we can easily check \begin{equation}
\bar n_k =  \frac{N-1}{N\kappa} \vec x \cdot \vec\beta_k,  \label{eqg10}
\end{equation}
and that 
\begin{equation}
W^2=|\vec x|^2 = \frac{N\kappa^2}{N-1}\sum_k^N\left(n_k-\frac{n}{N}\right)^2=\frac{N\kappa^2}{N-1}\sum_k^N \bar n_k^2.
\label{eqg11}
\end{equation}
For the purpose of orthogonalizing the PPs over the inner product measure in Eq.~\ref{innerproduct}, we will also need to express $\vec x$ explicitly in terms of angles in origin-centered spherical coordinates: \begin{subequations}\begin{align}
x_1 &= W \cos \varphi_1, \\
x_2 &= W \sin\varphi_1 \cos\varphi_2, \\
&\vdots \notag \\
x_{N-2} &= W  \cos \varphi_k \prod_{k=1}^{N-3}\sin\varphi_k, \\
x_{N-1} &= W \prod_{k=1}^{N-2}\sin\varphi_k.
\end{align}\label{spherical}\end{subequations}

Substituting the explicit expressions from Eqs. \ref{barybasis} and \ref{spherical} into Eq. \ref{eqg10}, we obtain
\begin{align}
\bar n_1&=  \frac{N-1}{N\kappa }W\cos\varphi_1,\\
\bar n_2&=  \frac{W}{N\kappa }\left(-\cos\varphi_1+\sqrt{N(N-2)}\sin\varphi_1\cos\varphi_2\right),\\
\bar n_k&= \frac{W}{N\kappa }[-\cos\varphi_1-\sqrt{\frac{N}{N-2}}\sin\varphi_1\cos\varphi_2 \notag \\
& \;\;\;\;\; - \sum_{j=2}^{k-2}\sqrt{\frac{N!}{(N-j-2)!}}\frac{\left(\prod_{i=1}^{j} \sin\varphi_i\right )\cos\varphi_{j+1}}{(N-j)(N-j-1)}\notag\\
& \;\;\;\;\; + \sqrt{\frac{N!}{(N-k-1)!}}\frac{\left(\prod_{i=1}^{k-1} \sin\varphi_i\right )\cos\varphi_{k}}{N-k+1}],\label{thirdline}\\
&\vdots \notag \\
\bar n_{N-1}&= \frac{W}{N\kappa }[-\cos\varphi_1-\sqrt{\frac{N}{N-2}}\sin\varphi_1\cos\varphi_2 \notag \\
& \;\;\;\;\; - \sum_{j=2}^{N-3}\sqrt{\frac{N!}{(N-j-2)!}}\frac{\left(\prod_{i=1}^{j} \sin\varphi_i\right )\cos\varphi_{j+1}}{(N-j)(N-j-1)}\notag\\
& \;\;\;\;\; + \frac{\sqrt{N!}}{2}\left(\prod_{i=1}^{N-2} \sin\varphi_i\right)],\\
\bar n_{N}&= \frac{W}{N\kappa }[-\cos\varphi_1-\sqrt{\frac{N}{N-2}}\sin\varphi_1\cos\varphi_2 \notag \\
& \;\;\;\;\; - \sum_{j=2}^{N-3}\sqrt{\frac{N!}{(N-j-2)!}}\frac{\left(\prod_{i=1}^{j} \sin\varphi_i\right )\cos\varphi_{j+1}}{(N-j)(N-j-1)}\notag\\
& \;\;\;\;\; - \frac{\sqrt{N!}}{2}\left(\prod_{i=1}^{N-2} \sin\varphi_i\right)].
\label{qs}
\end{align}
These are the explicit expressions for transcribing the $\bar n_j$, $1\leq j\leq N$ indices directly into spherical coordinates. In~\eqref{thirdline}, $k$ ranges from $3$ to $N-2$. 
\section{Second-quantized pseudopotentials vs. real-space projection Hamiltonians}\label{sec:TK}
We provide a brief comparison between the second-quantized PPs that have been derived in this paper and real-space projection Hamiltonians elsewhere in the literature, e.g. Trugman-Kivelson type Hamiltonians on the infinite plane. We show that our geometric PP construction avoids certain ambiguities that plague the latter approaches.

Neglecting internal DOFs for simplicity, a generic real-space PP living in total relative angular momentum sectors up to $m$ takes the form
\begin{equation} H(\vec r_1,...,\vec r_N)=\left(\prod^N_{j=1}\nabla_j^{2d_j}\right)\delta^2(\vec r_1-\vec r_2)...\delta^2(\vec r_{N-1}-\vec r_N)\end{equation}
such that $\sum_j d_j=m$. The various $d_j$'s refer to how the derivatives are distributed among the particles. In this form, there is \emph{no} simple one-to-one correspondence between the $d_j$ values and the relative weight of $H$ in the various $m$ sectors. To find the relative weights, one has to project $H$ onto the sector spanned by a state with angular momentum $m$. It is most convenient to adopt the symmetric gauge, where such a state takes the form $\ket{\Psi_m}\propto p(\tilde z_1,...,\tilde z_N)e^{-\frac{1}{4}\sum_j |z_j|^2}|z_1,...,z_N\rangle$ where $p$ is a symmetric or antisymmetric polynomial and $z_j=x_j+iy_j$ refers to the particle position ($\tilde z$ defined below):
\begin{eqnarray} 
&&\langle \Psi_m|H\ket{\Psi_m} = \nonumber \\
&& \left(\prod_{j=1}^N \int d^2\vec r_j\right)\Psi_m^*(\vec r_1,...,\vec r_N)H(\vec r_1,...,\vec r_N) \Psi_m(\vec r_1,...,\vec r_N). \nonumber \\
\end{eqnarray}
Upon substituting $\Psi_m$ and integrating by parts, we can express $\langle \Psi_m|H|\Psi_m\rangle$ as 
\begin{eqnarray}
&& \langle \Psi_m|H\ket{\Psi_m} = \notag \\
&&  \left(\prod_{j=1}^N \int d^2\vec r_j\right)\delta^2(\vec r_1-\vec r_2)...\delta^2(\vec r_{N-1}-\vec r_N) \notag \\
&\times&  \left(\prod^N_{j=1}\nabla_j^{2d_j}\right)[\Psi_m^*(\vec r_1,...,\vec r_N)\Psi_m(\vec r_1,...,\vec r_N)],
\end{eqnarray}
which can be evaluated as
\begin{eqnarray}
&& \langle \Psi_m|H\ket{\Psi_m} = \notag \\
&=& 4^m\left(\prod_{j=1}^N \int dz_jd z^*_j\right)\delta^2( z_1- z_2)...\delta^2( z_{N-1}- z_N)  \notag \\
&& \left(\prod^N_{j=1}\partial_{ z^*_j}^{2d_j}\Psi_m^*(  z_1^*,...,z_N^*)\right)\left(\prod^N_{j=1}\partial_{z_j}^{2d_j}\Psi_m( z_,...,z_N)\right),\notag\\
\label{realproject}
\end{eqnarray}
where $\nabla^2 = 4 \partial_z\partial_{\bar z}$ has been used. The delta functions, which enforce that all $z_j$ be equal, also enforce $\tilde z_j=z_j-\frac{1}{N}\sum_j z_j=0 \; \forall \; j$. Thus we only get a nonzero contribution from constant integrands. Specific examples were worked out in the appendices of Ref. \onlinecite{simon2007}; in general, an operator $H$ has an overlap with various $m$ sectors, and a PP that lies \emph{purely} in one $m$ sector must be a complicated linear combination of the Trugman-Kivelson terms with various sets of $d_j$'s. Some progress was made in Ref.~\onlinecite{lee2013}, where operators containing $L_m\left(\frac{N\kappa^2}{2(N-1)\nabla_1^2}\right)$ were shown to be contained purely in one $m$ sector. That, however, does not always work for fermions as will be explained below.

The mathematical complications above are further aggravated when the space of PPs in the sector $m$ is degenerate. In particular, the matrix $\langle \Psi_{ma}|H\ket{\Psi_{mb}}$ is not of full rank and some combinations of bona-fide PPs of degree $m$ will still not give a positive energy penalty to states in the $m$ sector~\cite{simon2007}. 
In particular, it is possible for certain Trugman-Kivelson Hamiltonians with $2m$ derivatives to evaluate to zero upon taking into account the fermionic antisymmetry. Consider for instance the $N=3$ fermionic state $\ket{\Psi_3} \propto  (\tilde z_1 - \tilde z_2)(\tilde z_2 - \tilde z_3)(\tilde z_3 - \tilde z_1)|z_1,z_2,z_3\rangle$. The following operators have a zero projection on it:

\[ \langle \Psi_3|\nabla^6_j\delta^2(r_1-r_2)\delta^2(r_2-r_3)|\Psi_3\rangle =0\] and
\[ \langle \Psi_3|\nabla^2_1\nabla^2_2\nabla^2_3\delta^2(r_1-r_2)\delta^2(r_2-r_3)|\Psi_3\rangle =0.\]  
Indeed, the only nonvanishing Trugman-Kivelson type fermionic PP in the $U^{N=3}_{m=3}$ sector is $H=\nabla^4_i\nabla^2_j\delta^2(r_1-r_2)\delta^2(r_2-r_3)$. As $|\Psi_3\rangle$ is of degree $\le 2$ in each variable, $H$ must contain a derivative of degrees $2d_j\leq 4$ in each variable for a nonzero projection. Since only constant terms in the integrand of Eq. \ref{realproject} contribute, we also need derivatives in all three particle coordinates. 

By contrast, our PP construction appeals to the orthogonality structure of the PPs {\it from the outset}, and avoids all of the above problems that generically arise when one attempts to use the Trugman-Kivelson approach in more complicated many-body cases. 

\section{Pseudopotential parent Hamiltonian from a given wave function}
\label{sec:power}
We briefly summarize the heuristic procedure how to determine the PPs whose null spaces contain a given QH state. These are the PPs that should be included in its parent Hamiltonian, since they penalize denser or ``simpler" states that will otherwise be realized at the same filling fraction. The method consists of two steps: (i) identify the thin-torus root pattern from the first-quantized wave function (Sec.~\ref{sec:root}), and (ii) use the root pattern and power counting to obtain the PPs (Sec.~\ref{sec:parent}).

\subsection{Wave function to root pattern}\label{sec:root}

\begin{figure}[htb]
\includegraphics[width=0.98\linewidth]{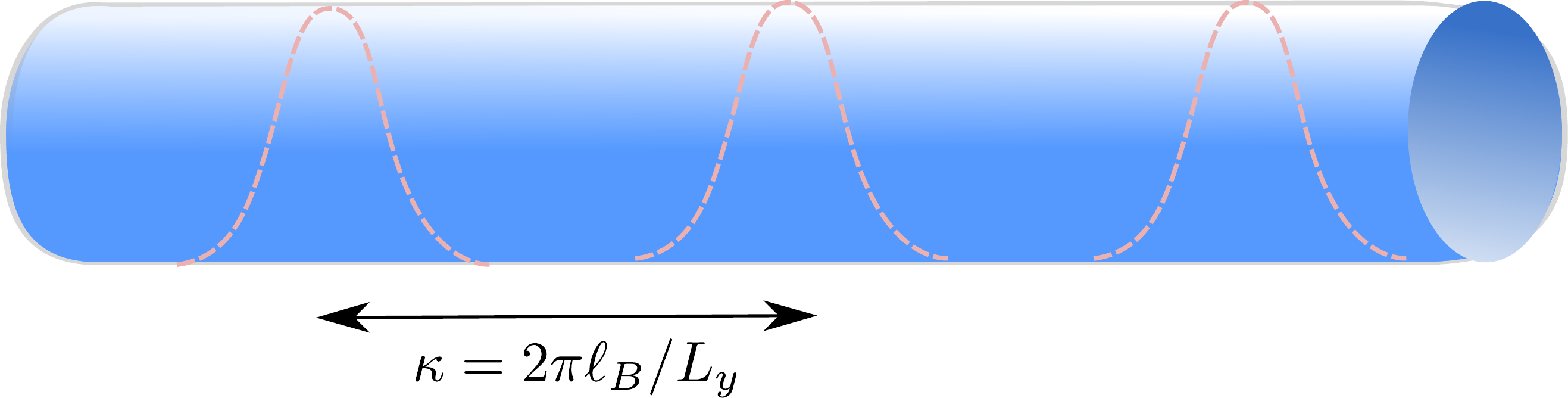}
%\captionsetup{justification=centerlast}
\caption{(Color online) In the limit of thin cylinder ($\kappa\to\infty$), the single-particle orbitals become well-separated from one another. As a consequence, the dominant configurations (``root patterns") in the expansion of a many-body state $\Psi$ are those Fock states where particles minimize their classical electrostatic energy.}
\label{fig:thincylinder}
\end{figure}

Suppose we are given a QH wave function with the polynomial part $\psi(z_1,z_2,...)$, where the $z_i$'s denote the positions of particles. In the infinite plane, $z_j$ is the usual complex coordinate $z_j=x_j+iy_j$. Since we will work on a cylinder geometry, however, we need to perform a stereographic mapping: $z_j \to e^{\kappa z_j}$, where $\kappa=2\pi/L_y$ is shown in Fig.~\ref{fig:thincylinder}. For simplicity, assume there are no internal degrees of freedom. 

To each wave function defined on a cylinder, we can assign one or several thin-cylinder \emph{root patterns}. Those are the Fock states with non-zero weight as the circumference of the cylinder is taken to zero ($\kappa\to\infty$). Typically, any given FQH state $\ket{\psi}$ in the thermodynamic limit decomposes onto many Fock states, as $\ket{\psi}$ represents a strongly correlated state. However, when the cylinder is stretched, the weights of most of the Fock states in the decomposition vanish. The only Fock states surviving in the limit $\kappa\to\infty$ are called ``root patterns" and are useful for finding the parent Hamiltonian (Laplacian) for $\ket{\psi}$ as its kernel.

In order to find the root pattern, we must examine the dominant terms in the decomposition of $\psi$ as $\kappa\to\infty$. Usually, this amounts to solving a classical electrostatic problem. Here, we consider the case of an open cylinder, where the root pattern will typically be unique. To get the remaining root patterns, that arise due to topological degeneracy or non-Abelian statistics, one must consider the system on a torus. Due to the complicated form of torus wave functions, it is more convenient to work on a cylinder but perform a flux insertion which allows one to access other topological sectors. Details of this are given e.g. in Ref.~\onlinecite{Seidel-PhysRevLett.101.036804}. 

Fixing the total number of particles, we study the dominating term in $\psi$. It is $z_1^{m_1}z_2^{m_2}z_3^{m_3}...$ such that \emph{no} other term (denoted by $'$) has $m'_1>m_1$ or, $m'_2>m_2$ if $m'_1=m_1$ or, $m'_3>m_3$ if $m'_1=m_1$ and $m'_2=m_2$, etc. The root pattern can then be written down as the partition defined by $m_1,m_2,...$:
\begin{equation}
(m_1,m_2,m_3,...)\rightarrow 10...010...01...
\end{equation}
where the $1$'s are at position $m_1,m_2,...$. If two $m_i$'s coincide, as can occur for bosons or multicomponent fermions, we shall label that position by '$2$' and so forth.
For instance, consider the $1/m$ Laughlin state with $2$ particles. We have $\psi(z_1,z_2)=(z_1-z_2)^m$ which, when expanded out, contains monomials
 \begin{equation}
\{ z_1^m, z_1^{m-1}z_2, z_1^{m-2}z_2^2,...\}
\end{equation}
Evidently, the term $z_1^mz_2^0$ is in the root pattern. Had we considered a droplet of $N$ particles, this term would have become $z_1^{m(N-1)}z_2^{m(N-2)}...z_{N-1}^mz_N^0$, with total degree $\frac{mN(n-1)}{2}$ and the root
\begin{equation}10^{m-1}10^{m-1}1...,\end{equation}
where $0^{m-1}$ represents a string of $m-1$ consecutive zeroes. For this simplest case, a droplet of $N=2$ particles is sufficient to determine the root configuration. Because we deal with translationally invariant states, a root configuration for larger $N$ is obtained by repeating the minimal pattern.

Let us now consider a more complicated example such as the generalized Pfaffian state introduced in Eq.~(\ref{Pfq}) in Sec. \ref{sec:examples}. 
By using the identity $[Pf(A)]^2=Det(A)$, one can show by induction starting from $n=4$, the most compact particle droplet, that the dominating term is $z_1^{3q-1}z_2^{2q}z_3^{q-1}$. This defines the root configuration in Eq.~(\ref{pfaffianroot}), $10^{q-2}10^{q}10^{q-2}10^q...$. It is also easy to read off the filling fraction from the above: we have $2$ particles per period, and each period has length $2q$. Hence the filling fraction is $\frac{2}{2q}=1/q$.

\subsection{Root pattern to parent Hamiltonian}\label{sec:parent}

Given a root pattern, we can easily infer what relative angular momentum states should be penalized to construct a projector onto the state. For instance, if the closest pair of particles are $...10^l1...$, no two particles have relative angular momentum less than $l+1$. Hence the state should lie in the null space of any two-body PPs with $U^2_m$, $m\leq l$. Consequently, the parent Hamiltonian should be a linear combination of these PPs, so that it penalizes all states denser than the given one. This is discussed at length in Ref. \onlinecite{Simon-PhysRevB.75.075318}.

To figure out the required many-body PPs, we apply the above procedure analogously to clusters of particles with minimal relative angular momentum.  Note that some many-body PPs need not be included if the interaction is already precluded by particle (anti)symmetry:
Consider the $1/q$ Pfaffian state described above, with a root given by Eq. \ref{pfaffianroot}. Obviously, no two particles have relative angular momentum less than $q-1$. Naively, we will then need to include all $U^2_{m}$ with $m\leq q-2$. The $1/q$ Pfaffian state, however, is fermionic (bosonic) when $q$ is even(odd), and $U^2_{q-2}$ exactly vanishes in both cases. Hence the parent Hamiltonian only contains $U^2_{m< q-2}$. 

Let us finally analyze 3-body interactions. From the root pattern, the minimal total relative angular momentum is $l=(q-1)+2q=3q-1$. However, configurations with $l=3q-2=2(q-1)+q$ automatically vanish since they require pairs of particles with relative angular momenta $q$ and $q-1$ simultaneously, which are forbidden by either fermionic or bosonic statistics. With lower 3-body relative angular momentum already excluded by the 2-body PPs, the only 3-body PP we need to include in the parent Hamiltonian is $U^3_{3q-3}$.  As the root pattern repeats from the third particle, there is no need to consider interactions with $4$ or more bodies.  

\bibliography{pseudopotentials}

\begin{thebibliography}{71}%
\makeatletter
\providecommand \@ifxundefined [1]{%
 \@ifx{#1\undefined}
}%
\providecommand \@ifnum [1]{%
 \ifnum #1\expandafter \@firstoftwo
 \else \expandafter \@secondoftwo
 \fi
}%
\providecommand \@ifx [1]{%
 \ifx #1\expandafter \@firstoftwo
 \else \expandafter \@secondoftwo
 \fi
}%
\providecommand \natexlab [1]{#1}%
\providecommand \enquote  [1]{``#1''}%
\providecommand \bibnamefont  [1]{#1}%
\providecommand \bibfnamefont [1]{#1}%
\providecommand \citenamefont [1]{#1}%
\providecommand \href@noop [0]{\@secondoftwo}%
\providecommand \href [0]{\begingroup \@sanitize@url \@href}%
\providecommand \@href[1]{\@@startlink{#1}\@@href}%
\providecommand \@@href[1]{\endgroup#1\@@endlink}%
\providecommand \@sanitize@url [0]{\catcode `\\12\catcode `\$12\catcode
  `\&12\catcode `\#12\catcode `\^12\catcode `\_12\catcode `\%12\relax}%
\providecommand \@@startlink[1]{}%
\providecommand \@@endlink[0]{}%
\providecommand \url  [0]{\begingroup\@sanitize@url \@url }%
\providecommand \@url [1]{\endgroup\@href {#1}{\urlprefix }}%
\providecommand \urlprefix  [0]{URL }%
\providecommand \Eprint [0]{\href }%
\providecommand \doibase [0]{http://dx.doi.org/}%
\providecommand \selectlanguage [0]{\@gobble}%
\providecommand \bibinfo  [0]{\@secondoftwo}%
\providecommand \bibfield  [0]{\@secondoftwo}%
\providecommand \translation [1]{[#1]}%
\providecommand \BibitemOpen [0]{}%
\providecommand \bibitemStop [0]{}%
\providecommand \bibitemNoStop [0]{.\EOS\space}%
\providecommand \EOS [0]{\spacefactor3000\relax}%
\providecommand \BibitemShut  [1]{\csname bibitem#1\endcsname}%
\let\auto@bib@innerbib\@empty
%</preamble>
\bibitem [{\citenamefont {Tsui}\ \emph {et~al.}(1982)\citenamefont {Tsui},
  \citenamefont {Stormer},\ and\ \citenamefont
  {Gossard}}]{Tsui-PhysRevLett.48.1559}%
  \BibitemOpen
  \bibfield  {author} {\bibinfo {author} {\bibfnamefont {D.~C.}\ \bibnamefont
  {Tsui}}, \bibinfo {author} {\bibfnamefont {H.~L.}\ \bibnamefont {Stormer}}, \
  and\ \bibinfo {author} {\bibfnamefont {A.~C.}\ \bibnamefont {Gossard}},\
  }\bibfield  {title} {\enquote {\bibinfo {title} {Two-dimensional
  magnetotransport in the extreme quantum limit},}\ }\href {\doibase
  10.1103/PhysRevLett.48.1559} {\bibfield  {journal} {\bibinfo  {journal}
  {Phys. Rev. Lett.}\ }\textbf {\bibinfo {volume} {48}},\ \bibinfo {pages}
  {1559--1562} (\bibinfo {year} {1982})}\BibitemShut {NoStop}%
\bibitem [{\citenamefont {Laughlin}(1983)}]{Laughlin-PhysRevLett.50.1395}%
  \BibitemOpen
  \bibfield  {author} {\bibinfo {author} {\bibfnamefont {R.~B.}\ \bibnamefont
  {Laughlin}},\ }\bibfield  {title} {\enquote {\bibinfo {title} {Anomalous
  quantum hall effect: An incompressible quantum fluid with fractionally
  charged excitations},}\ }\href {\doibase 10.1103/PhysRevLett.50.1395}
  {\bibfield  {journal} {\bibinfo  {journal} {Phys. Rev. Lett.}\ }\textbf
  {\bibinfo {volume} {50}},\ \bibinfo {pages} {1395--1398} (\bibinfo {year}
  {1983})}\BibitemShut {NoStop}%
\bibitem [{\citenamefont {Haldane}(1983)}]{Haldane1983}%
  \BibitemOpen
  \bibfield  {author} {\bibinfo {author} {\bibfnamefont {F~Duncan~M}\
  \bibnamefont {Haldane}},\ }\bibfield  {title} {\enquote {\bibinfo {title}
  {Fractional quantization of the hall effect-a hierarchy of incompressible
  quantum fluid states},}\ }\href
  {http://adsabs.harvard.edu/abs/1983PhRvL..51..605H} {\bibfield  {journal}
  {\bibinfo  {journal} {Physical Review Letters}\ }\textbf {\bibinfo {volume}
  {51}},\ \bibinfo {pages} {605--608} (\bibinfo {year} {1983})}\BibitemShut
  {NoStop}%
\bibitem [{\citenamefont {Haldane}(1990)}]{prangegirvin}%
  \BibitemOpen
  \bibfield  {author} {\bibinfo {author} {\bibfnamefont {F.~D.~M.}\
  \bibnamefont {Haldane}},\ }\href@noop {} {\emph {\bibinfo {title} {The
  Quantum Hall Effect}}},\ edited by\ \bibinfo {editor} {\bibfnamefont {R.~E.}\
  \bibnamefont {Prange}}\ and\ \bibinfo {editor} {\bibfnamefont {S.~M.}\
  \bibnamefont {Girvin}}\ (\bibinfo  {publisher} {Springer},\ \bibinfo
  {address} {New York},\ \bibinfo {year} {1990})\BibitemShut {NoStop}%
\bibitem [{\citenamefont {Fano}\ \emph {et~al.}(1986)\citenamefont {Fano},
  \citenamefont {Ortolani},\ and\ \citenamefont
  {Colombo}}]{FanoPhysRevB.34.2670}%
  \BibitemOpen
  \bibfield  {author} {\bibinfo {author} {\bibfnamefont {G.}~\bibnamefont
  {Fano}}, \bibinfo {author} {\bibfnamefont {F.}~\bibnamefont {Ortolani}}, \
  and\ \bibinfo {author} {\bibfnamefont {E.}~\bibnamefont {Colombo}},\
  }\bibfield  {title} {\enquote {\bibinfo {title} {Configuration-interaction
  calculations on the fractional quantum hall effect},}\ }\href {\doibase
  10.1103/PhysRevB.34.2670} {\bibfield  {journal} {\bibinfo  {journal} {Phys.
  Rev. B}\ }\textbf {\bibinfo {volume} {34}},\ \bibinfo {pages} {2670--2680}
  (\bibinfo {year} {1986})}\BibitemShut {NoStop}%
\bibitem [{\citenamefont {Nayak}\ \emph {et~al.}(2008)\citenamefont {Nayak},
  \citenamefont {Simon}, \citenamefont {Stern}, \citenamefont {Freedman},\ and\
  \citenamefont {Das~Sarma}}]{RevModPhys.80.1083}%
  \BibitemOpen
  \bibfield  {author} {\bibinfo {author} {\bibfnamefont {Chetan}\ \bibnamefont
  {Nayak}}, \bibinfo {author} {\bibfnamefont {Steven~H.}\ \bibnamefont
  {Simon}}, \bibinfo {author} {\bibfnamefont {Ady}\ \bibnamefont {Stern}},
  \bibinfo {author} {\bibfnamefont {Michael}\ \bibnamefont {Freedman}}, \ and\
  \bibinfo {author} {\bibfnamefont {Sankar}\ \bibnamefont {Das~Sarma}},\
  }\bibfield  {title} {\enquote {\bibinfo {title} {Non-abelian anyons and
  topological quantum computation},}\ }\href {\doibase
  10.1103/RevModPhys.80.1083} {\bibfield  {journal} {\bibinfo  {journal} {Rev.
  Mod. Phys.}\ }\textbf {\bibinfo {volume} {80}},\ \bibinfo {pages}
  {1083--1159} (\bibinfo {year} {2008})}\BibitemShut {NoStop}%
\bibitem [{\citenamefont {Pachos}(2012)}]{pachos2012introduction}%
  \BibitemOpen
  \bibfield  {author} {\bibinfo {author} {\bibfnamefont {J.K.}\ \bibnamefont
  {Pachos}},\ }\href {https://books.google.ca/books?id=XDciVh6bAE0C} {\emph
  {\bibinfo {title} {Introduction to Topological Quantum Computation}}},\
  Introduction to Topological Quantum Computation\ (\bibinfo  {publisher}
  {Cambridge University Press},\ \bibinfo {year} {2012})\BibitemShut {NoStop}%
\bibitem [{\citenamefont {Moore}\ and\ \citenamefont
  {Read}(1991)}]{Moore1991362}%
  \BibitemOpen
  \bibfield  {author} {\bibinfo {author} {\bibfnamefont {Gregory}\ \bibnamefont
  {Moore}}\ and\ \bibinfo {author} {\bibfnamefont {Nicholas}\ \bibnamefont
  {Read}},\ }\bibfield  {title} {\enquote {\bibinfo {title} {Nonabelions in the
  fractional quantum hall effect},}\ }\href {\doibase
  10.1016/0550-3213(91)90407-O} {\bibfield  {journal} {\bibinfo  {journal}
  {Nuclear Physics B}\ }\textbf {\bibinfo {volume} {360}},\ \bibinfo {pages}
  {362 -- 396} (\bibinfo {year} {1991})}\BibitemShut {NoStop}%
\bibitem [{\citenamefont {Greiter}\ \emph {et~al.}(1991)\citenamefont
  {Greiter}, \citenamefont {Wen},\ and\ \citenamefont {Wilczek}}]{Greiter91}%
  \BibitemOpen
  \bibfield  {author} {\bibinfo {author} {\bibfnamefont {Martin}\ \bibnamefont
  {Greiter}}, \bibinfo {author} {\bibfnamefont {Xiao-Gang}\ \bibnamefont
  {Wen}}, \ and\ \bibinfo {author} {\bibfnamefont {Frank}\ \bibnamefont
  {Wilczek}},\ }\bibfield  {title} {\enquote {\bibinfo {title} {Paired hall
  state at half filling},}\ }\href@noop {} {\bibfield  {journal} {\bibinfo
  {journal} {Physical review letters}\ }\textbf {\bibinfo {volume} {66}},\
  \bibinfo {pages} {3205} (\bibinfo {year} {1991})}\BibitemShut {NoStop}%
\bibitem [{\citenamefont {Greiter}\ \emph {et~al.}(1992)\citenamefont
  {Greiter}, \citenamefont {Wen},\ and\ \citenamefont {Wilczek}}]{Greiter92}%
  \BibitemOpen
  \bibfield  {author} {\bibinfo {author} {\bibfnamefont {Martin}\ \bibnamefont
  {Greiter}}, \bibinfo {author} {\bibfnamefont {XG}~\bibnamefont {Wen}}, \ and\
  \bibinfo {author} {\bibfnamefont {Frank}\ \bibnamefont {Wilczek}},\
  }\bibfield  {title} {\enquote {\bibinfo {title} {Paired hall states},}\
  }\href@noop {} {\bibfield  {journal} {\bibinfo  {journal} {Nuclear Physics
  B}\ }\textbf {\bibinfo {volume} {374}},\ \bibinfo {pages} {567--614}
  (\bibinfo {year} {1992})}\BibitemShut {NoStop}%
\bibitem [{\citenamefont {Read}\ and\ \citenamefont
  {Rezayi}(1996)}]{Read-PhysRevB.54.16864}%
  \BibitemOpen
  \bibfield  {author} {\bibinfo {author} {\bibfnamefont {N.}~\bibnamefont
  {Read}}\ and\ \bibinfo {author} {\bibfnamefont {E.}~\bibnamefont {Rezayi}},\
  }\bibfield  {title} {\enquote {\bibinfo {title} {Quasiholes and fermionic
  zero modes of paired fractional quantum hall states: The mechanism for
  non-abelian statistics},}\ }\href {\doibase 10.1103/PhysRevB.54.16864}
  {\bibfield  {journal} {\bibinfo  {journal} {Phys. Rev. B}\ }\textbf {\bibinfo
  {volume} {54}},\ \bibinfo {pages} {16864--16887} (\bibinfo {year}
  {1996})}\BibitemShut {NoStop}%
\bibitem [{\citenamefont {Read}(2001)}]{nick2001}%
  \BibitemOpen
  \bibfield  {author} {\bibinfo {author} {\bibfnamefont {N}~\bibnamefont
  {Read}},\ }\bibfield  {title} {\enquote {\bibinfo {title} {Paired fractional
  quantum hall states and the $\nu$= 5/2 puzzle},}\ }\href@noop {} {\bibfield
  {journal} {\bibinfo  {journal} {Physica B: Condensed Matter}\ }\textbf
  {\bibinfo {volume} {298}},\ \bibinfo {pages} {121--128} (\bibinfo {year}
  {2001})}\BibitemShut {NoStop}%
\bibitem [{\citenamefont {Read}\ and\ \citenamefont
  {Rezayi}(1999)}]{Read-PhysRevB.59.8084}%
  \BibitemOpen
  \bibfield  {author} {\bibinfo {author} {\bibfnamefont {N.}~\bibnamefont
  {Read}}\ and\ \bibinfo {author} {\bibfnamefont {E.}~\bibnamefont {Rezayi}},\
  }\bibfield  {title} {\enquote {\bibinfo {title} {Beyond paired quantum hall
  states: Parafermions and incompressible states in the first excited landau
  level},}\ }\href {\doibase 10.1103/PhysRevB.59.8084} {\bibfield  {journal}
  {\bibinfo  {journal} {Phys. Rev. B}\ }\textbf {\bibinfo {volume} {59}},\
  \bibinfo {pages} {8084--8092} (\bibinfo {year} {1999})}\BibitemShut {NoStop}%
\bibitem [{\citenamefont {Greiter}\ and\ \citenamefont
  {Thomale}(2009)}]{Ronny-CSLPhysRevLett.102.207203}%
  \BibitemOpen
  \bibfield  {author} {\bibinfo {author} {\bibfnamefont {Martin}\ \bibnamefont
  {Greiter}}\ and\ \bibinfo {author} {\bibfnamefont {Ronny}\ \bibnamefont
  {Thomale}},\ }\bibfield  {title} {\enquote {\bibinfo {title} {Non-abelian
  statistics in a quantum antiferromagnet},}\ }\href {\doibase
  10.1103/PhysRevLett.102.207203} {\bibfield  {journal} {\bibinfo  {journal}
  {Phys. Rev. Lett.}\ }\textbf {\bibinfo {volume} {102}},\ \bibinfo {pages}
  {207203} (\bibinfo {year} {2009})}\BibitemShut {NoStop}%
\bibitem [{\citenamefont {Jackson}\ \emph {et~al.}(2013)\citenamefont
  {Jackson}, \citenamefont {Read},\ and\ \citenamefont {Simon}}]{nick2013}%
  \BibitemOpen
  \bibfield  {author} {\bibinfo {author} {\bibfnamefont {T.~S.}\ \bibnamefont
  {Jackson}}, \bibinfo {author} {\bibfnamefont {N.}~\bibnamefont {Read}}, \
  and\ \bibinfo {author} {\bibfnamefont {S.~H.}\ \bibnamefont {Simon}},\
  }\bibfield  {title} {\enquote {\bibinfo {title} {Entanglement subspaces,
  trial wave functions, and special hamiltonians in the fractional quantum hall
  effect},}\ }\href {\doibase 10.1103/PhysRevB.88.075313} {\bibfield  {journal}
  {\bibinfo  {journal} {Phys. Rev. B}\ }\textbf {\bibinfo {volume} {88}},\
  \bibinfo {pages} {075313} (\bibinfo {year} {2013})}\BibitemShut {NoStop}%
\bibitem [{\citenamefont {Ortiz}\ \emph {et~al.}(2013)\citenamefont {Ortiz},
  \citenamefont {Nussinov}, \citenamefont {Dukelsky},\ and\ \citenamefont
  {Seidel}}]{PhysRevB.88.165303}%
  \BibitemOpen
  \bibfield  {author} {\bibinfo {author} {\bibfnamefont {G.}~\bibnamefont
  {Ortiz}}, \bibinfo {author} {\bibfnamefont {Z.}~\bibnamefont {Nussinov}},
  \bibinfo {author} {\bibfnamefont {J.}~\bibnamefont {Dukelsky}}, \ and\
  \bibinfo {author} {\bibfnamefont {A.}~\bibnamefont {Seidel}},\ }\bibfield
  {title} {\enquote {\bibinfo {title} {Repulsive interactions in quantum hall
  systems as a pairing problem},}\ }\href {\doibase 10.1103/PhysRevB.88.165303}
  {\bibfield  {journal} {\bibinfo  {journal} {Phys. Rev. B}\ }\textbf {\bibinfo
  {volume} {88}},\ \bibinfo {pages} {165303} (\bibinfo {year}
  {2013})}\BibitemShut {NoStop}%
\bibitem [{\citenamefont {Chen}\ and\ \citenamefont
  {Seidel}(2015)}]{PhysRevB.91.085103}%
  \BibitemOpen
  \bibfield  {author} {\bibinfo {author} {\bibfnamefont {Li}~\bibnamefont
  {Chen}}\ and\ \bibinfo {author} {\bibfnamefont {Alexander}\ \bibnamefont
  {Seidel}},\ }\bibfield  {title} {\enquote {\bibinfo {title} {Algebraic
  approach to the study of zero modes of haldane pseudopotentials},}\ }\href
  {\doibase 10.1103/PhysRevB.91.085103} {\bibfield  {journal} {\bibinfo
  {journal} {Phys. Rev. B}\ }\textbf {\bibinfo {volume} {91}},\ \bibinfo
  {pages} {085103} (\bibinfo {year} {2015})}\BibitemShut {NoStop}%
\bibitem [{\citenamefont {Mazaheri}\ \emph {et~al.}(2015)\citenamefont
  {Mazaheri}, \citenamefont {Ortiz}, \citenamefont {Nussinov},\ and\
  \citenamefont {Seidel}}]{PhysRevB.91.085115}%
  \BibitemOpen
  \bibfield  {author} {\bibinfo {author} {\bibfnamefont {Tahereh}\ \bibnamefont
  {Mazaheri}}, \bibinfo {author} {\bibfnamefont {Gerardo}\ \bibnamefont
  {Ortiz}}, \bibinfo {author} {\bibfnamefont {Zohar}\ \bibnamefont {Nussinov}},
  \ and\ \bibinfo {author} {\bibfnamefont {Alexander}\ \bibnamefont {Seidel}},\
  }\bibfield  {title} {\enquote {\bibinfo {title} {Zero modes, bosonization,
  and topological quantum order: The laughlin state in second quantization},}\
  }\href {\doibase 10.1103/PhysRevB.91.085115} {\bibfield  {journal} {\bibinfo
  {journal} {Phys. Rev. B}\ }\textbf {\bibinfo {volume} {91}},\ \bibinfo
  {pages} {085115} (\bibinfo {year} {2015})}\BibitemShut {NoStop}%
\bibitem [{\citenamefont {Hansson}\ \emph {et~al.}(2009)\citenamefont
  {Hansson}, \citenamefont {Hermanns},\ and\ \citenamefont
  {Viefers}}]{PhysRevB.80.165330}%
  \BibitemOpen
  \bibfield  {author} {\bibinfo {author} {\bibfnamefont {T.~H.}\ \bibnamefont
  {Hansson}}, \bibinfo {author} {\bibfnamefont {M.}~\bibnamefont {Hermanns}}, \
  and\ \bibinfo {author} {\bibfnamefont {S.}~\bibnamefont {Viefers}},\
  }\bibfield  {title} {\enquote {\bibinfo {title} {Quantum hall quasielectron
  operators in conformal field theory},}\ }\href {\doibase
  10.1103/PhysRevB.80.165330} {\bibfield  {journal} {\bibinfo  {journal} {Phys.
  Rev. B}\ }\textbf {\bibinfo {volume} {80}},\ \bibinfo {pages} {165330}
  (\bibinfo {year} {2009})}\BibitemShut {NoStop}%
\bibitem [{\citenamefont {Greiter}\ \emph {et~al.}()\citenamefont {Greiter},
  \citenamefont {Schnells},\ and\ \citenamefont {Thomale}}]{schnells}%
  \BibitemOpen
  \bibfield  {author} {\bibinfo {author} {\bibfnamefont {M.}~\bibnamefont
  {Greiter}}, \bibinfo {author} {\bibfnamefont {V.}~\bibnamefont {Schnells}}, \
  and\ \bibinfo {author} {\bibfnamefont {R.}~\bibnamefont {Thomale}},\
  }\href@noop {} {\enquote {\bibinfo {title} {Laughlin states and their
  quasi-particle excitations on the torus},}\ }\bibinfo {note}
  {ArXiv:1405.0742}\BibitemShut {NoStop}%
\bibitem [{\citenamefont {Girvin}\ \emph {et~al.}(1985)\citenamefont {Girvin},
  \citenamefont {MacDonald},\ and\ \citenamefont {Platzman}}]{GMP85}%
  \BibitemOpen
  \bibfield  {author} {\bibinfo {author} {\bibfnamefont {SM}~\bibnamefont
  {Girvin}}, \bibinfo {author} {\bibfnamefont {AH}~\bibnamefont {MacDonald}}, \
  and\ \bibinfo {author} {\bibfnamefont {PM}~\bibnamefont {Platzman}},\
  }\bibfield  {title} {\enquote {\bibinfo {title} {Collective-excitation gap in
  the fractional quantum hall effect},}\ }\href@noop {} {\bibfield  {journal}
  {\bibinfo  {journal} {Physical review letters}\ }\textbf {\bibinfo {volume}
  {54}},\ \bibinfo {pages} {581} (\bibinfo {year} {1985})}\BibitemShut
  {NoStop}%
\bibitem [{\citenamefont {Girvin}\ \emph {et~al.}(1986)\citenamefont {Girvin},
  \citenamefont {MacDonald},\ and\ \citenamefont {Platzman}}]{GMP86}%
  \BibitemOpen
  \bibfield  {author} {\bibinfo {author} {\bibfnamefont {SM}~\bibnamefont
  {Girvin}}, \bibinfo {author} {\bibfnamefont {AH}~\bibnamefont {MacDonald}}, \
  and\ \bibinfo {author} {\bibfnamefont {PM}~\bibnamefont {Platzman}},\
  }\bibfield  {title} {\enquote {\bibinfo {title} {Magneto-roton theory of
  collective excitations in the fractional quantum hall effect},}\ }\href@noop
  {} {\bibfield  {journal} {\bibinfo  {journal} {Physical Review B}\ }\textbf
  {\bibinfo {volume} {33}},\ \bibinfo {pages} {2481} (\bibinfo {year}
  {1986})}\BibitemShut {NoStop}%
\bibitem [{\citenamefont {Repellin}\ \emph {et~al.}(2014)\citenamefont
  {Repellin}, \citenamefont {Neupert}, \citenamefont {Papi{\'c}},\ and\
  \citenamefont {Regnault}}]{SMA}%
  \BibitemOpen
  \bibfield  {author} {\bibinfo {author} {\bibfnamefont {C{\'e}cile}\
  \bibnamefont {Repellin}}, \bibinfo {author} {\bibfnamefont {Titus}\
  \bibnamefont {Neupert}}, \bibinfo {author} {\bibfnamefont {Zlatko}\
  \bibnamefont {Papi{\'c}}}, \ and\ \bibinfo {author} {\bibfnamefont {Nicolas}\
  \bibnamefont {Regnault}},\ }\bibfield  {title} {\enquote {\bibinfo {title}
  {Single-mode approximation for fractional chern insulators and the fractional
  quantum hall effect on the torus},}\ }\href@noop {} {\bibfield  {journal}
  {\bibinfo  {journal} {Physical Review B}\ }\textbf {\bibinfo {volume} {90}},\
  \bibinfo {pages} {045114} (\bibinfo {year} {2014})}\BibitemShut {NoStop}%
\bibitem [{\citenamefont {Repellin}\ \emph {et~al.}(2015)\citenamefont
  {Repellin}, \citenamefont {Neupert}, \citenamefont {Bernevig},\ and\
  \citenamefont {Regnault}}]{projective}%
  \BibitemOpen
  \bibfield  {author} {\bibinfo {author} {\bibfnamefont {C{\'e}cile}\
  \bibnamefont {Repellin}}, \bibinfo {author} {\bibfnamefont {Titus}\
  \bibnamefont {Neupert}}, \bibinfo {author} {\bibfnamefont {B~Andrei}\
  \bibnamefont {Bernevig}}, \ and\ \bibinfo {author} {\bibfnamefont {Nicolas}\
  \bibnamefont {Regnault}},\ }\href@noop {} {\enquote {\bibinfo {title}
  {Projective construction of the $\mathbb{Z}_k $ read-rezayi fractional
  quantum hall states and their excitations on the torus geometry},}\ }
  (\bibinfo {year} {2015})\BibitemShut {NoStop}%
\bibitem [{\citenamefont {Li}\ and\ \citenamefont
  {Haldane}(2008)}]{LiPhysRevLett.101.010504}%
  \BibitemOpen
  \bibfield  {author} {\bibinfo {author} {\bibfnamefont {Hui}\ \bibnamefont
  {Li}}\ and\ \bibinfo {author} {\bibfnamefont {F.~D.~M.}\ \bibnamefont
  {Haldane}},\ }\bibfield  {title} {\enquote {\bibinfo {title} {Entanglement
  spectrum as a generalization of entanglement entropy: Identification of
  topological order in non-abelian fractional quantum hall effect states},}\
  }\href {\doibase 10.1103/PhysRevLett.101.010504} {\bibfield  {journal}
  {\bibinfo  {journal} {Phys. Rev. Lett.}\ }\textbf {\bibinfo {volume} {101}},\
  \bibinfo {pages} {010504} (\bibinfo {year} {2008})}\BibitemShut {NoStop}%
\bibitem [{\citenamefont {Thomale}\ \emph {et~al.}(2010)\citenamefont
  {Thomale}, \citenamefont {Sterdyniak}, \citenamefont {Regnault},\ and\
  \citenamefont {Bernevig}}]{ThomalePhysRevLett.104.180502}%
  \BibitemOpen
  \bibfield  {author} {\bibinfo {author} {\bibfnamefont {R.}~\bibnamefont
  {Thomale}}, \bibinfo {author} {\bibfnamefont {A.}~\bibnamefont {Sterdyniak}},
  \bibinfo {author} {\bibfnamefont {N.}~\bibnamefont {Regnault}}, \ and\
  \bibinfo {author} {\bibfnamefont {B.~Andrei}\ \bibnamefont {Bernevig}},\
  }\bibfield  {title} {\enquote {\bibinfo {title} {Entanglement gap and a new
  principle of adiabatic continuity},}\ }\href {\doibase
  10.1103/PhysRevLett.104.180502} {\bibfield  {journal} {\bibinfo  {journal}
  {Phys. Rev. Lett.}\ }\textbf {\bibinfo {volume} {104}},\ \bibinfo {pages}
  {180502} (\bibinfo {year} {2010})}\BibitemShut {NoStop}%
\bibitem [{\citenamefont {Sterdyniak}\ \emph {et~al.}(2011)\citenamefont
  {Sterdyniak}, \citenamefont {Regnault},\ and\ \citenamefont
  {Bernevig}}]{PES}%
  \BibitemOpen
  \bibfield  {author} {\bibinfo {author} {\bibfnamefont {A}~\bibnamefont
  {Sterdyniak}}, \bibinfo {author} {\bibfnamefont {N}~\bibnamefont {Regnault}},
  \ and\ \bibinfo {author} {\bibfnamefont {BA}~\bibnamefont {Bernevig}},\
  }\bibfield  {title} {\enquote {\bibinfo {title} {Extracting excitations from
  model state entanglement},}\ }\href@noop {} {\bibfield  {journal} {\bibinfo
  {journal} {Physical review letters}\ }\textbf {\bibinfo {volume} {106}},\
  \bibinfo {pages} {100405} (\bibinfo {year} {2011})}\BibitemShut {NoStop}%
\bibitem [{\citenamefont {Haldane}\ and\ \citenamefont
  {Rezayi}(1988)}]{Haldane-PhysRevLett.60.956}%
  \BibitemOpen
  \bibfield  {author} {\bibinfo {author} {\bibfnamefont {F.~D.~M.}\
  \bibnamefont {Haldane}}\ and\ \bibinfo {author} {\bibfnamefont {E.~H.}\
  \bibnamefont {Rezayi}},\ }\bibfield  {title} {\enquote {\bibinfo {title}
  {Spin-singlet wave function for the half-integral quantum hall effect},}\
  }\href {\doibase 10.1103/PhysRevLett.60.956} {\bibfield  {journal} {\bibinfo
  {journal} {Phys. Rev. Lett.}\ }\textbf {\bibinfo {volume} {60}},\ \bibinfo
  {pages} {956--959} (\bibinfo {year} {1988})}\BibitemShut {NoStop}%
\bibitem [{\citenamefont {Greiter}(1993)}]{greiterGaff}%
  \BibitemOpen
  \bibfield  {author} {\bibinfo {author} {\bibfnamefont {M.}~\bibnamefont
  {Greiter}},\ }\href@noop {} {\bibfield  {journal} {\bibinfo  {journal} {Bull.
  Am. Phys. Soc.}\ }\textbf {\bibinfo {volume} {38}},\ \bibinfo {pages} {137}
  (\bibinfo {year} {1993})}\BibitemShut {NoStop}%
\bibitem [{\citenamefont {Simon}\ \emph
  {et~al.}(2007{\natexlab{a}})\citenamefont {Simon}, \citenamefont {Rezayi},\
  and\ \citenamefont {Cooper}}]{Simon-PhysRevB.75.075318}%
  \BibitemOpen
  \bibfield  {author} {\bibinfo {author} {\bibfnamefont {Steven~H.}\
  \bibnamefont {Simon}}, \bibinfo {author} {\bibfnamefont {E.~H.}\ \bibnamefont
  {Rezayi}}, \ and\ \bibinfo {author} {\bibfnamefont {Nigel~R.}\ \bibnamefont
  {Cooper}},\ }\bibfield  {title} {\enquote {\bibinfo {title} {Generalized
  quantum hall projection hamiltonians},}\ }\href {\doibase
  10.1103/PhysRevB.75.075318} {\bibfield  {journal} {\bibinfo  {journal} {Phys.
  Rev. B}\ }\textbf {\bibinfo {volume} {75}},\ \bibinfo {pages} {075318}
  (\bibinfo {year} {2007}{\natexlab{a}})}\BibitemShut {NoStop}%
\bibitem [{\citenamefont {Wen}\ and\ \citenamefont
  {Niu}(1990)}]{WenPhysRevB.41.9377}%
  \BibitemOpen
  \bibfield  {author} {\bibinfo {author} {\bibfnamefont {X.~G.}\ \bibnamefont
  {Wen}}\ and\ \bibinfo {author} {\bibfnamefont {Q.}~\bibnamefont {Niu}},\
  }\bibfield  {title} {\enquote {\bibinfo {title} {Ground-state degeneracy of
  the fractional quantum hall states in the presence of a random potential and
  on high-genus riemann surfaces},}\ }\href {\doibase 10.1103/PhysRevB.41.9377}
  {\bibfield  {journal} {\bibinfo  {journal} {Phys. Rev. B}\ }\textbf {\bibinfo
  {volume} {41}},\ \bibinfo {pages} {9377--9396} (\bibinfo {year}
  {1990})}\BibitemShut {NoStop}%
\bibitem [{\citenamefont {Wen}(1990)}]{Wen-modular}%
  \BibitemOpen
  \bibfield  {author} {\bibinfo {author} {\bibfnamefont {Xiao-Gang}\
  \bibnamefont {Wen}},\ }\bibfield  {title} {\enquote {\bibinfo {title}
  {Topological orders in rigid states},}\ }\href@noop {} {\bibfield  {journal}
  {\bibinfo  {journal} {International Journal of Modern Physics B}\ }\textbf
  {\bibinfo {volume} {4}},\ \bibinfo {pages} {239--271} (\bibinfo {year}
  {1990})}\BibitemShut {NoStop}%
\bibitem [{\citenamefont {Zhang}\ \emph {et~al.}(2012)\citenamefont {Zhang},
  \citenamefont {Grover}, \citenamefont {Turner}, \citenamefont {Oshikawa},\
  and\ \citenamefont {Vishwanath}}]{Zhang-modular}%
  \BibitemOpen
  \bibfield  {author} {\bibinfo {author} {\bibfnamefont {Yi}~\bibnamefont
  {Zhang}}, \bibinfo {author} {\bibfnamefont {Tarun}\ \bibnamefont {Grover}},
  \bibinfo {author} {\bibfnamefont {Ari}\ \bibnamefont {Turner}}, \bibinfo
  {author} {\bibfnamefont {Masaki}\ \bibnamefont {Oshikawa}}, \ and\ \bibinfo
  {author} {\bibfnamefont {Ashvin}\ \bibnamefont {Vishwanath}},\ }\bibfield
  {title} {\enquote {\bibinfo {title} {Quasiparticle statistics and braiding
  from ground-state entanglement},}\ }\href {\doibase
  10.1103/PhysRevB.85.235151} {\bibfield  {journal} {\bibinfo  {journal} {Phys.
  Rev. B}\ }\textbf {\bibinfo {volume} {85}},\ \bibinfo {pages} {235151}
  (\bibinfo {year} {2012})}\BibitemShut {NoStop}%
\bibitem [{\citenamefont {Lee}\ and\ \citenamefont
  {Leinaas}(2004)}]{LeePhysRevLett.92.096401}%
  \BibitemOpen
  \bibfield  {author} {\bibinfo {author} {\bibfnamefont {Dung-Hai}\
  \bibnamefont {Lee}}\ and\ \bibinfo {author} {\bibfnamefont {Jon~Magne}\
  \bibnamefont {Leinaas}},\ }\bibfield  {title} {\enquote {\bibinfo {title}
  {Mott insulators without symmetry breaking},}\ }\href {\doibase
  10.1103/PhysRevLett.92.096401} {\bibfield  {journal} {\bibinfo  {journal}
  {Phys. Rev. Lett.}\ }\textbf {\bibinfo {volume} {92}},\ \bibinfo {pages}
  {096401} (\bibinfo {year} {2004})}\BibitemShut {NoStop}%
\bibitem [{\citenamefont {Seidel}\ \emph {et~al.}(2005)\citenamefont {Seidel},
  \citenamefont {Fu}, \citenamefont {Lee}, \citenamefont {Leinaas},\ and\
  \citenamefont {Moore}}]{seidel2005}%
  \BibitemOpen
  \bibfield  {author} {\bibinfo {author} {\bibfnamefont {Alexander}\
  \bibnamefont {Seidel}}, \bibinfo {author} {\bibfnamefont {Henry}\
  \bibnamefont {Fu}}, \bibinfo {author} {\bibfnamefont {Dung-Hai}\ \bibnamefont
  {Lee}}, \bibinfo {author} {\bibfnamefont {Jon~Magne}\ \bibnamefont
  {Leinaas}}, \ and\ \bibinfo {author} {\bibfnamefont {Joel}\ \bibnamefont
  {Moore}},\ }\bibfield  {title} {\enquote {\bibinfo {title} {Incompressible
  quantum liquids and new conservation laws},}\ }\href
  {http://prl.aps.org/abstract/PRL/v95/i26/e266405} {\bibfield  {journal}
  {\bibinfo  {journal} {Physical review letters}\ }\textbf {\bibinfo {volume}
  {95}},\ \bibinfo {pages} {266405} (\bibinfo {year} {2005})}\BibitemShut
  {NoStop}%
\bibitem [{\citenamefont {Jansen}(2012)}]{Jansen-2012JMP....53l3306J}%
  \BibitemOpen
  \bibfield  {author} {\bibinfo {author} {\bibfnamefont {Sabine}\ \bibnamefont
  {Jansen}},\ }\bibfield  {title} {\enquote {\bibinfo {title} {Fermionic and
  bosonic laughlin state on thick cylinders},}\ }\href@noop {} {\bibfield
  {journal} {\bibinfo  {journal} {Journal of Mathematical Physics}\ }\textbf
  {\bibinfo {volume} {53}},\ \bibinfo {pages} {123306} (\bibinfo {year}
  {2012})}\BibitemShut {NoStop}%
\bibitem [{\citenamefont {Wang}\ and\ \citenamefont
  {Nakamura}(2013)}]{Wang-PhysRevB.87.245119}%
  \BibitemOpen
  \bibfield  {author} {\bibinfo {author} {\bibfnamefont {Zheng-Yuan}\
  \bibnamefont {Wang}}\ and\ \bibinfo {author} {\bibfnamefont {Masaaki}\
  \bibnamefont {Nakamura}},\ }\bibfield  {title} {\enquote {\bibinfo {title}
  {One-dimensional lattice model with an exact matrix-product ground state
  describing the laughlin wave function},}\ }\href {\doibase
  10.1103/PhysRevB.87.245119} {\bibfield  {journal} {\bibinfo  {journal} {Phys.
  Rev. B}\ }\textbf {\bibinfo {volume} {87}},\ \bibinfo {pages} {245119}
  (\bibinfo {year} {2013})}\BibitemShut {NoStop}%
\bibitem [{\citenamefont {Soul\'e}\ and\ \citenamefont
  {Jolicoeur}(2012)}]{Soule-PhysRevB.85.155116}%
  \BibitemOpen
  \bibfield  {author} {\bibinfo {author} {\bibfnamefont {Paul}\ \bibnamefont
  {Soul\'e}}\ and\ \bibinfo {author} {\bibfnamefont {Thierry}\ \bibnamefont
  {Jolicoeur}},\ }\bibfield  {title} {\enquote {\bibinfo {title} {Exact wave
  functions for excitations of the $\nu=\frac{1}{3}$ fractional quantum hall
  state from a model hamiltonian},}\ }\href {\doibase
  10.1103/PhysRevB.85.155116} {\bibfield  {journal} {\bibinfo  {journal} {Phys.
  Rev. B}\ }\textbf {\bibinfo {volume} {85}},\ \bibinfo {pages} {155116}
  (\bibinfo {year} {2012})}\BibitemShut {NoStop}%
\bibitem [{\citenamefont {Papi{\'c}}(2014)}]{Papic14}%
  \BibitemOpen
  \bibfield  {author} {\bibinfo {author} {\bibfnamefont {Z}~\bibnamefont
  {Papi{\'c}}},\ }\bibfield  {title} {\enquote {\bibinfo {title} {Solvable
  models for unitary and nonunitary topological phases},}\ }\href@noop {}
  {\bibfield  {journal} {\bibinfo  {journal} {Physical Review B}\ }\textbf
  {\bibinfo {volume} {90}},\ \bibinfo {pages} {075304} (\bibinfo {year}
  {2014})}\BibitemShut {NoStop}%
\bibitem [{\citenamefont {Seidel}\ and\ \citenamefont
  {Yang}(2011)}]{Seidel-PhysRevB.84.085122}%
  \BibitemOpen
  \bibfield  {author} {\bibinfo {author} {\bibfnamefont {Alexander}\
  \bibnamefont {Seidel}}\ and\ \bibinfo {author} {\bibfnamefont {Kun}\
  \bibnamefont {Yang}},\ }\bibfield  {title} {\enquote {\bibinfo {title}
  {Gapless excitations in the haldane-rezayi state: The thin-torus limit},}\
  }\href {\doibase 10.1103/PhysRevB.84.085122} {\bibfield  {journal} {\bibinfo
  {journal} {Phys. Rev. B}\ }\textbf {\bibinfo {volume} {84}},\ \bibinfo
  {pages} {085122} (\bibinfo {year} {2011})}\BibitemShut {NoStop}%
\bibitem [{\citenamefont {Weerasinghe}\ and\ \citenamefont
  {Seidel}(2014)}]{Weerasinghe14}%
  \BibitemOpen
  \bibfield  {author} {\bibinfo {author} {\bibfnamefont {Amila}\ \bibnamefont
  {Weerasinghe}}\ and\ \bibinfo {author} {\bibfnamefont {Alexander}\
  \bibnamefont {Seidel}},\ }\bibfield  {title} {\enquote {\bibinfo {title}
  {Thin torus perturbative analysis of elementary excitations in the gaffnian
  and haldane-rezayi quantum hall states},}\ }\href@noop {} {\bibfield
  {journal} {\bibinfo  {journal} {Physical Review B}\ }\textbf {\bibinfo
  {volume} {90}},\ \bibinfo {pages} {125146} (\bibinfo {year}
  {2014})}\BibitemShut {NoStop}%
\bibitem [{\citenamefont {Simon}\ \emph
  {et~al.}(2007{\natexlab{b}})\citenamefont {Simon}, \citenamefont {Rezayi},\
  and\ \citenamefont {Cooper}}]{simon2007}%
  \BibitemOpen
  \bibfield  {author} {\bibinfo {author} {\bibfnamefont {Steven~H}\
  \bibnamefont {Simon}}, \bibinfo {author} {\bibfnamefont {EH}~\bibnamefont
  {Rezayi}}, \ and\ \bibinfo {author} {\bibfnamefont {Nigel~R}\ \bibnamefont
  {Cooper}},\ }\bibfield  {title} {\enquote {\bibinfo {title} {Pseudopotentials
  for multiparticle interactions in the quantum hall regime},}\ }\href@noop {}
  {\bibfield  {journal} {\bibinfo  {journal} {Physical Review B}\ }\textbf
  {\bibinfo {volume} {75}},\ \bibinfo {pages} {195306} (\bibinfo {year}
  {2007}{\natexlab{b}})}\BibitemShut {NoStop}%
\bibitem [{\citenamefont {Davenport}\ and\ \citenamefont
  {Simon}(2012)}]{davenport2012}%
  \BibitemOpen
  \bibfield  {author} {\bibinfo {author} {\bibfnamefont {Simon~C}\ \bibnamefont
  {Davenport}}\ and\ \bibinfo {author} {\bibfnamefont {Steven~H}\ \bibnamefont
  {Simon}},\ }\bibfield  {title} {\enquote {\bibinfo {title} {Multiparticle
  pseudopotentials for multicomponent quantum hall systems},}\ }\href@noop {}
  {\bibfield  {journal} {\bibinfo  {journal} {Physical Review B}\ }\textbf
  {\bibinfo {volume} {85}},\ \bibinfo {pages} {075430} (\bibinfo {year}
  {2012})}\BibitemShut {NoStop}%
\bibitem [{\citenamefont {Lee}\ \emph {et~al.}(2013)\citenamefont {Lee},
  \citenamefont {Thomale},\ and\ \citenamefont {Qi}}]{lee2013}%
  \BibitemOpen
  \bibfield  {author} {\bibinfo {author} {\bibfnamefont {Ching~Hua}\
  \bibnamefont {Lee}}, \bibinfo {author} {\bibfnamefont {Ronny}\ \bibnamefont
  {Thomale}}, \ and\ \bibinfo {author} {\bibfnamefont {Xiao-Liang}\
  \bibnamefont {Qi}},\ }\bibfield  {title} {\enquote {\bibinfo {title}
  {Pseudopotential formalism for fractional chern insulators},}\ }\href
  {http://prb.aps.org/abstract/PRB/v88/i3/e035101} {\bibfield  {journal}
  {\bibinfo  {journal} {Physical Review B}\ }\textbf {\bibinfo {volume} {88}},\
  \bibinfo {pages} {035101} (\bibinfo {year} {2013})}\BibitemShut {NoStop}%
\bibitem [{\citenamefont {Wu}\ \emph {et~al.}(2013)\citenamefont {Wu},
  \citenamefont {Regnault},\ and\ \citenamefont {Bernevig}}]{YangleWu}%
  \BibitemOpen
  \bibfield  {author} {\bibinfo {author} {\bibfnamefont {Yang-Le}\ \bibnamefont
  {Wu}}, \bibinfo {author} {\bibfnamefont {Nicolas}\ \bibnamefont {Regnault}},
  \ and\ \bibinfo {author} {\bibfnamefont {B~Andrei}\ \bibnamefont
  {Bernevig}},\ }\bibfield  {title} {\enquote {\bibinfo {title} {Bloch model
  wave functions and pseudopotentials for all fractional chern insulators},}\
  }\href@noop {} {\bibfield  {journal} {\bibinfo  {journal} {Physical review
  letters}\ }\textbf {\bibinfo {volume} {110}},\ \bibinfo {pages} {106802}
  (\bibinfo {year} {2013})}\BibitemShut {NoStop}%
\bibitem [{\citenamefont {Lee}\ and\ \citenamefont {Qi}(2014)}]{lee2014}%
  \BibitemOpen
  \bibfield  {author} {\bibinfo {author} {\bibfnamefont {Ching~Hua}\
  \bibnamefont {Lee}}\ and\ \bibinfo {author} {\bibfnamefont {Xiao-Liang}\
  \bibnamefont {Qi}},\ }\bibfield  {title} {\enquote {\bibinfo {title} {Lattice
  construction of pseudopotential hamiltonians for fractional chern
  insulators},}\ }\href@noop {} {\bibfield  {journal} {\bibinfo  {journal}
  {Physical Review B}\ }\textbf {\bibinfo {volume} {90}},\ \bibinfo {pages}
  {085103} (\bibinfo {year} {2014})}\BibitemShut {NoStop}%
\bibitem [{\citenamefont {Claassen}\ \emph {et~al.}(2015)\citenamefont
  {Claassen}, \citenamefont {Lee}, \citenamefont {Thomale}, \citenamefont
  {Qi},\ and\ \citenamefont {Devereaux}}]{claassen2015position}%
  \BibitemOpen
  \bibfield  {author} {\bibinfo {author} {\bibfnamefont {Martin}\ \bibnamefont
  {Claassen}}, \bibinfo {author} {\bibfnamefont {Ching~Hua}\ \bibnamefont
  {Lee}}, \bibinfo {author} {\bibfnamefont {Ronny}\ \bibnamefont {Thomale}},
  \bibinfo {author} {\bibfnamefont {Xiao-Liang}\ \bibnamefont {Qi}}, \ and\
  \bibinfo {author} {\bibfnamefont {Thomas~P.}\ \bibnamefont {Devereaux}},\
  }\href {\doibase 10.1103/PhysRevLett.114.236802} {\enquote {\bibinfo {title}
  {Position-momentum duality and fractional quantum hall effect in chern
  insulators},}\ } (\bibinfo {year} {2015})\BibitemShut {NoStop}%
\bibitem [{\citenamefont {Haldane}(2011)}]{HaldaneGeometry}%
  \BibitemOpen
  \bibfield  {author} {\bibinfo {author} {\bibfnamefont {F.~D.~M.}\
  \bibnamefont {Haldane}},\ }\bibfield  {title} {\enquote {\bibinfo {title}
  {Geometrical description of the fractional quantum hall effect},}\ }\href
  {\doibase 10.1103/PhysRevLett.107.116801} {\bibfield  {journal} {\bibinfo
  {journal} {Phys. Rev. Lett.}\ }\textbf {\bibinfo {volume} {107}},\ \bibinfo
  {pages} {116801} (\bibinfo {year} {2011})}\BibitemShut {NoStop}%
\bibitem [{\citenamefont {Bernevig}\ and\ \citenamefont
  {Haldane}(2008)}]{jack}%
  \BibitemOpen
  \bibfield  {author} {\bibinfo {author} {\bibfnamefont {B~Andrei}\
  \bibnamefont {Bernevig}}\ and\ \bibinfo {author} {\bibfnamefont {FDM}\
  \bibnamefont {Haldane}},\ }\bibfield  {title} {\enquote {\bibinfo {title}
  {Model fractional quantum hall states and jack polynomials},}\ }\href@noop {}
  {\bibfield  {journal} {\bibinfo  {journal} {Physical review letters}\
  }\textbf {\bibinfo {volume} {100}},\ \bibinfo {pages} {246802} (\bibinfo
  {year} {2008})}\BibitemShut {NoStop}%
\bibitem [{\citenamefont {Wen}\ and\ \citenamefont
  {Wang}(2008)}]{PhysRevB.77.235108}%
  \BibitemOpen
  \bibfield  {author} {\bibinfo {author} {\bibfnamefont {Xiao-Gang}\
  \bibnamefont {Wen}}\ and\ \bibinfo {author} {\bibfnamefont {Zhenghan}\
  \bibnamefont {Wang}},\ }\bibfield  {title} {\enquote {\bibinfo {title}
  {Classification of symmetric polynomials of infinite variables: Construction
  of abelian and non-abelian quantum hall states},}\ }\href {\doibase
  10.1103/PhysRevB.77.235108} {\bibfield  {journal} {\bibinfo  {journal} {Phys.
  Rev. B}\ }\textbf {\bibinfo {volume} {77}},\ \bibinfo {pages} {235108}
  (\bibinfo {year} {2008})}\BibitemShut {NoStop}%
\bibitem [{\citenamefont {Haldane}\ and\ \citenamefont
  {Rezayi}(1985)}]{HalRezTorusPhysRevB.31.2529}%
  \BibitemOpen
  \bibfield  {author} {\bibinfo {author} {\bibfnamefont {F.~D.~M.}\
  \bibnamefont {Haldane}}\ and\ \bibinfo {author} {\bibfnamefont {E.~H.}\
  \bibnamefont {Rezayi}},\ }\bibfield  {title} {\enquote {\bibinfo {title}
  {Periodic laughlin-jastrow wave functions for the fractional quantized hall
  effect},}\ }\href {\doibase 10.1103/PhysRevB.31.2529} {\bibfield  {journal}
  {\bibinfo  {journal} {Phys. Rev. B}\ }\textbf {\bibinfo {volume} {31}},\
  \bibinfo {pages} {2529--2531} (\bibinfo {year} {1985})}\BibitemShut {NoStop}%
\bibitem [{\citenamefont {Haldane}(1985)}]{Haldane-PhysRevLett.55.2095}%
  \BibitemOpen
  \bibfield  {author} {\bibinfo {author} {\bibfnamefont {F.~D.~M.}\
  \bibnamefont {Haldane}},\ }\bibfield  {title} {\enquote {\bibinfo {title}
  {Many-particle translational symmetries of two-dimensional electrons at
  rational landau-level filling},}\ }\href {\doibase
  10.1103/PhysRevLett.55.2095} {\bibfield  {journal} {\bibinfo  {journal}
  {Phys. Rev. Lett.}\ }\textbf {\bibinfo {volume} {55}},\ \bibinfo {pages}
  {2095--2098} (\bibinfo {year} {1985})}\BibitemShut {NoStop}%
\bibitem [{\citenamefont {Trugman}\ and\ \citenamefont
  {Kivelson}(1985)}]{trugman1985}%
  \BibitemOpen
  \bibfield  {author} {\bibinfo {author} {\bibfnamefont {SA}~\bibnamefont
  {Trugman}}\ and\ \bibinfo {author} {\bibfnamefont {S}~\bibnamefont
  {Kivelson}},\ }\bibfield  {title} {\enquote {\bibinfo {title} {Exact results
  for the fractional quantum hall effect with general interactions},}\ }\href
  {http://prb.aps.org/abstract/PRB/v31/i8/p5280_1} {\bibfield  {journal}
  {\bibinfo  {journal} {Physical Review B}\ }\textbf {\bibinfo {volume} {31}},\
  \bibinfo {pages} {5280} (\bibinfo {year} {1985})}\BibitemShut {NoStop}%
\bibitem [{\citenamefont {Simon}\ \emph
  {et~al.}(2007{\natexlab{c}})\citenamefont {Simon}, \citenamefont {Rezayi},
  \citenamefont {Cooper},\ and\ \citenamefont
  {Berdnikov}}]{Simon-PhysRevB.75.075317}%
  \BibitemOpen
  \bibfield  {author} {\bibinfo {author} {\bibfnamefont {Steven~H.}\
  \bibnamefont {Simon}}, \bibinfo {author} {\bibfnamefont {E.~H.}\ \bibnamefont
  {Rezayi}}, \bibinfo {author} {\bibfnamefont {N.~R.}\ \bibnamefont {Cooper}},
  \ and\ \bibinfo {author} {\bibfnamefont {I.}~\bibnamefont {Berdnikov}},\
  }\bibfield  {title} {\enquote {\bibinfo {title} {Construction of a paired
  wave function for spinless electrons at filling fraction $\nu=2/5$},}\ }\href
  {\doibase 10.1103/PhysRevB.75.075317} {\bibfield  {journal} {\bibinfo
  {journal} {Phys. Rev. B}\ }\textbf {\bibinfo {volume} {75}},\ \bibinfo
  {pages} {075317} (\bibinfo {year} {2007}{\natexlab{c}})}\BibitemShut
  {NoStop}%
\bibitem [{\citenamefont {Flavin}\ \emph {et~al.}(2012)\citenamefont {Flavin},
  \citenamefont {Thomale},\ and\ \citenamefont {Seidel}}]{PhysRevB.86.125316}%
  \BibitemOpen
  \bibfield  {author} {\bibinfo {author} {\bibfnamefont {John}\ \bibnamefont
  {Flavin}}, \bibinfo {author} {\bibfnamefont {Ronny}\ \bibnamefont {Thomale}},
  \ and\ \bibinfo {author} {\bibfnamefont {Alexander}\ \bibnamefont {Seidel}},\
  }\bibfield  {title} {\enquote {\bibinfo {title} {Gaffnian holonomy through
  the coherent state method},}\ }\href {\doibase 10.1103/PhysRevB.86.125316}
  {\bibfield  {journal} {\bibinfo  {journal} {Phys. Rev. B}\ }\textbf {\bibinfo
  {volume} {86}},\ \bibinfo {pages} {125316} (\bibinfo {year}
  {2012})}\BibitemShut {NoStop}%
\bibitem [{\citenamefont {Seidel}\ and\ \citenamefont
  {Yang}(2008)}]{Seidel-PhysRevLett.101.036804}%
  \BibitemOpen
  \bibfield  {author} {\bibinfo {author} {\bibfnamefont {Alexander}\
  \bibnamefont {Seidel}}\ and\ \bibinfo {author} {\bibfnamefont {Kun}\
  \bibnamefont {Yang}},\ }\bibfield  {title} {\enquote {\bibinfo {title}
  {Halperin $(m,m',n)$ bilayer quantum hall states on thin cylinders},}\ }\href
  {\doibase 10.1103/PhysRevLett.101.036804} {\bibfield  {journal} {\bibinfo
  {journal} {Phys. Rev. Lett.}\ }\textbf {\bibinfo {volume} {101}},\ \bibinfo
  {pages} {036804} (\bibinfo {year} {2008})}\BibitemShut {NoStop}%
\bibitem [{\citenamefont {Bernevig}\ and\ \citenamefont
  {Regnault}(2009)}]{jack2}%
  \BibitemOpen
  \bibfield  {author} {\bibinfo {author} {\bibfnamefont {B~Andrei}\
  \bibnamefont {Bernevig}}\ and\ \bibinfo {author} {\bibfnamefont
  {N}~\bibnamefont {Regnault}},\ }\bibfield  {title} {\enquote {\bibinfo
  {title} {Anatomy of abelian and non-abelian fractional quantum hall
  states},}\ }\href@noop {} {\bibfield  {journal} {\bibinfo  {journal}
  {Physical review letters}\ }\textbf {\bibinfo {volume} {103}},\ \bibinfo
  {pages} {206801} (\bibinfo {year} {2009})}\BibitemShut {NoStop}%
\bibitem [{\citenamefont {Luhman}\ \emph {et~al.}(2008)\citenamefont {Luhman},
  \citenamefont {Pan}, \citenamefont {Tsui}, \citenamefont {Pfeiffer},
  \citenamefont {Baldwin},\ and\ \citenamefont {West}}]{luhman}%
  \BibitemOpen
  \bibfield  {author} {\bibinfo {author} {\bibfnamefont {DR}~\bibnamefont
  {Luhman}}, \bibinfo {author} {\bibfnamefont {W}~\bibnamefont {Pan}}, \bibinfo
  {author} {\bibfnamefont {DC}~\bibnamefont {Tsui}}, \bibinfo {author}
  {\bibfnamefont {LN}~\bibnamefont {Pfeiffer}}, \bibinfo {author}
  {\bibfnamefont {KW}~\bibnamefont {Baldwin}}, \ and\ \bibinfo {author}
  {\bibfnamefont {KW}~\bibnamefont {West}},\ }\bibfield  {title} {\enquote
  {\bibinfo {title} {Observation of a fractional quantum hall state at $\nu$=
  1/4 in a wide gaas quantum well},}\ }\href@noop {} {\bibfield  {journal}
  {\bibinfo  {journal} {Physical review letters}\ }\textbf {\bibinfo {volume}
  {101}},\ \bibinfo {pages} {266804} (\bibinfo {year} {2008})}\BibitemShut
  {NoStop}%
\bibitem [{\citenamefont {Papi{\'c}}\ \emph {et~al.}(2009)\citenamefont
  {Papi{\'c}}, \citenamefont {M{\"o}ller}, \citenamefont {Milovanovi{\'c}},
  \citenamefont {Regnault},\ and\ \citenamefont {Goerbig}}]{Papic09}%
  \BibitemOpen
  \bibfield  {author} {\bibinfo {author} {\bibfnamefont {Z}~\bibnamefont
  {Papi{\'c}}}, \bibinfo {author} {\bibfnamefont {G}~\bibnamefont
  {M{\"o}ller}}, \bibinfo {author} {\bibfnamefont {MV}~\bibnamefont
  {Milovanovi{\'c}}}, \bibinfo {author} {\bibfnamefont {N}~\bibnamefont
  {Regnault}}, \ and\ \bibinfo {author} {\bibfnamefont {MO}~\bibnamefont
  {Goerbig}},\ }\bibfield  {title} {\enquote {\bibinfo {title} {Fractional
  quantum hall state at $\nu= 1/4$ in a wide quantum well},}\ }\href@noop {}
  {\bibfield  {journal} {\bibinfo  {journal} {Physical Review B}\ }\textbf
  {\bibinfo {volume} {79}},\ \bibinfo {pages} {245325} (\bibinfo {year}
  {2009})}\BibitemShut {NoStop}%
\bibitem [{\citenamefont {Green}(2001)}]{green-10thesis}%
  \BibitemOpen
  \bibfield  {author} {\bibinfo {author} {\bibfnamefont {D.}~\bibnamefont
  {Green}},\ }\href@noop {} {\bibfield  {journal} {\bibinfo  {journal} {Ph.D.
  thesis, Yale University, New Haven}\ } (\bibinfo {year} {2001})},\ \bibinfo
  {note} {arXiv:cond-mat/0202455}\BibitemShut {NoStop}%
\bibitem [{\citenamefont {Hermanns}\ \emph {et~al.}(2011)\citenamefont
  {Hermanns}, \citenamefont {Regnault}, \citenamefont {Bernevig},\ and\
  \citenamefont {Ardonne}}]{Hermanns-PhysRevB.83.241302}%
  \BibitemOpen
  \bibfield  {author} {\bibinfo {author} {\bibfnamefont {M.}~\bibnamefont
  {Hermanns}}, \bibinfo {author} {\bibfnamefont {N.}~\bibnamefont {Regnault}},
  \bibinfo {author} {\bibfnamefont {B.~A.}\ \bibnamefont {Bernevig}}, \ and\
  \bibinfo {author} {\bibfnamefont {E.}~\bibnamefont {Ardonne}},\ }\bibfield
  {title} {\enquote {\bibinfo {title} {From irrational to nonunitary: Haffnian
  and haldane-rezayi wave functions},}\ }\href {\doibase
  10.1103/PhysRevB.83.241302} {\bibfield  {journal} {\bibinfo  {journal} {Phys.
  Rev. B}\ }\textbf {\bibinfo {volume} {83}},\ \bibinfo {pages} {241302}
  (\bibinfo {year} {2011})}\BibitemShut {NoStop}%
\bibitem [{\citenamefont {Davenport}\ \emph {et~al.}(2013)\citenamefont
  {Davenport}, \citenamefont {Ardonne}, \citenamefont {Regnault},\ and\
  \citenamefont {Simon}}]{davenport2013}%
  \BibitemOpen
  \bibfield  {author} {\bibinfo {author} {\bibfnamefont {Simon~C}\ \bibnamefont
  {Davenport}}, \bibinfo {author} {\bibfnamefont {Eddy}\ \bibnamefont
  {Ardonne}}, \bibinfo {author} {\bibfnamefont {Nicolas}\ \bibnamefont
  {Regnault}}, \ and\ \bibinfo {author} {\bibfnamefont {Steven~H}\ \bibnamefont
  {Simon}},\ }\bibfield  {title} {\enquote {\bibinfo {title} {Spin-singlet
  gaffnian wave function for fractional quantum hall systems},}\ }\href@noop {}
  {\bibfield  {journal} {\bibinfo  {journal} {Phys. Rev. B}\ }\textbf {\bibinfo
  {volume} {87}},\ \bibinfo {pages} {045310} (\bibinfo {year}
  {2013})}\BibitemShut {NoStop}%
\bibitem [{\citenamefont {Ardonne}\ \emph {et~al.}(2001)\citenamefont
  {Ardonne}, \citenamefont {Read}, \citenamefont {Rezayi},\ and\ \citenamefont
  {Schoutens}}]{ardonne2001}%
  \BibitemOpen
  \bibfield  {author} {\bibinfo {author} {\bibfnamefont {Eddy}\ \bibnamefont
  {Ardonne}}, \bibinfo {author} {\bibfnamefont {N}~\bibnamefont {Read}},
  \bibinfo {author} {\bibfnamefont {Edward}\ \bibnamefont {Rezayi}}, \ and\
  \bibinfo {author} {\bibfnamefont {Kareljan}\ \bibnamefont {Schoutens}},\
  }\bibfield  {title} {\enquote {\bibinfo {title} {Non-abelian spin-singlet
  quantum hall states: wave functions and quasihole state counting},}\
  }\href@noop {} {\bibfield  {journal} {\bibinfo  {journal} {Nucl. Phys. B}\
  }\textbf {\bibinfo {volume} {607}},\ \bibinfo {pages} {549--576} (\bibinfo
  {year} {2001})}\BibitemShut {NoStop}%
\bibitem [{\citenamefont {Ardonne}\ and\ \citenamefont
  {Regnault}(2011)}]{ardonne2011}%
  \BibitemOpen
  \bibfield  {author} {\bibinfo {author} {\bibfnamefont {E}~\bibnamefont
  {Ardonne}}\ and\ \bibinfo {author} {\bibfnamefont {N}~\bibnamefont
  {Regnault}},\ }\bibfield  {title} {\enquote {\bibinfo {title} {Structure of
  spinful quantum hall states: A squeezing perspective},}\ }\href@noop {}
  {\bibfield  {journal} {\bibinfo  {journal} {Phys. Rev. B}\ }\textbf {\bibinfo
  {volume} {84}},\ \bibinfo {pages} {205134} (\bibinfo {year}
  {2011})}\BibitemShut {NoStop}%
\bibitem [{\citenamefont {Thomale}\ \emph {et~al.}(2011)\citenamefont
  {Thomale}, \citenamefont {Estienne}, \citenamefont {Regnault},\ and\
  \citenamefont {Bernevig}}]{PhysRevB.84.045127}%
  \BibitemOpen
  \bibfield  {author} {\bibinfo {author} {\bibfnamefont {Ronny}\ \bibnamefont
  {Thomale}}, \bibinfo {author} {\bibfnamefont {Benoit}\ \bibnamefont
  {Estienne}}, \bibinfo {author} {\bibfnamefont {Nicolas}\ \bibnamefont
  {Regnault}}, \ and\ \bibinfo {author} {\bibfnamefont {B.~Andrei}\
  \bibnamefont {Bernevig}},\ }\bibfield  {title} {\enquote {\bibinfo {title}
  {Decomposition of fractional quantum hall model states: Product rule
  symmetries and approximations},}\ }\href {\doibase
  10.1103/PhysRevB.84.045127} {\bibfield  {journal} {\bibinfo  {journal} {Phys.
  Rev. B}\ }\textbf {\bibinfo {volume} {84}},\ \bibinfo {pages} {045127}
  (\bibinfo {year} {2011})}\BibitemShut {NoStop}%
\bibitem [{\citenamefont {Halperin}(1983)}]{Halperin83}%
  \BibitemOpen
  \bibfield  {author} {\bibinfo {author} {\bibfnamefont {Bertrand~I}\
  \bibnamefont {Halperin}},\ }\bibfield  {title} {\enquote {\bibinfo {title}
  {Theory of the quantized hall conductance},}\ }\href@noop {} {\bibfield
  {journal} {\bibinfo  {journal} {Helv. Phys. Acta}\ }\textbf {\bibinfo
  {volume} {56}},\ \bibinfo {pages} {75} (\bibinfo {year} {1983})}\BibitemShut
  {NoStop}%
\bibitem [{\citenamefont {Simon}\ \emph {et~al.}(2010)\citenamefont {Simon},
  \citenamefont {Rezayi},\ and\ \citenamefont
  {Regnault}}]{Simon-PhysRevB.81.121301}%
  \BibitemOpen
  \bibfield  {author} {\bibinfo {author} {\bibfnamefont {Steven~H.}\
  \bibnamefont {Simon}}, \bibinfo {author} {\bibfnamefont {Edward~H.}\
  \bibnamefont {Rezayi}}, \ and\ \bibinfo {author} {\bibfnamefont {Nicolas}\
  \bibnamefont {Regnault}},\ }\bibfield  {title} {\enquote {\bibinfo {title}
  {Quantum hall wave functions based on ${S}_{3}$ conformal field theories},}\
  }\href {\doibase 10.1103/PhysRevB.81.121301} {\bibfield  {journal} {\bibinfo
  {journal} {Phys. Rev. B}\ }\textbf {\bibinfo {volume} {81}},\ \bibinfo
  {pages} {121301} (\bibinfo {year} {2010})}\BibitemShut {NoStop}%
\bibitem [{\citenamefont {Wen}\ and\ \citenamefont {Zee}(1992)}]{wenzee}%
  \BibitemOpen
  \bibfield  {author} {\bibinfo {author} {\bibfnamefont {XG}~\bibnamefont
  {Wen}}\ and\ \bibinfo {author} {\bibfnamefont {A}~\bibnamefont {Zee}},\
  }\bibfield  {title} {\enquote {\bibinfo {title} {Shift and spin vector: New
  topological quantum numbers for the hall fluids},}\ }\href@noop {} {\bibfield
   {journal} {\bibinfo  {journal} {Physical review letters}\ }\textbf {\bibinfo
  {volume} {69}},\ \bibinfo {pages} {953} (\bibinfo {year} {1992})}\BibitemShut
  {NoStop}%
\bibitem [{\citenamefont {Dean}\ \emph {et~al.}(2011)\citenamefont {Dean},
  \citenamefont {Young}, \citenamefont {Cadden-Zimansky}, \citenamefont {Wang},
  \citenamefont {Ren}, \citenamefont {Watanabe}, \citenamefont {Taniguchi},
  \citenamefont {Kim}, \citenamefont {Hone},\ and\ \citenamefont
  {Shepard}}]{grapheneFQHE}%
  \BibitemOpen
  \bibfield  {author} {\bibinfo {author} {\bibfnamefont {CR}~\bibnamefont
  {Dean}}, \bibinfo {author} {\bibfnamefont {AF}~\bibnamefont {Young}},
  \bibinfo {author} {\bibfnamefont {Pet}\ \bibnamefont {Cadden-Zimansky}},
  \bibinfo {author} {\bibfnamefont {L}~\bibnamefont {Wang}}, \bibinfo {author}
  {\bibfnamefont {H}~\bibnamefont {Ren}}, \bibinfo {author} {\bibfnamefont
  {K}~\bibnamefont {Watanabe}}, \bibinfo {author} {\bibfnamefont
  {T}~\bibnamefont {Taniguchi}}, \bibinfo {author} {\bibfnamefont
  {P}~\bibnamefont {Kim}}, \bibinfo {author} {\bibfnamefont {J}~\bibnamefont
  {Hone}}, \ and\ \bibinfo {author} {\bibfnamefont {KL}~\bibnamefont
  {Shepard}},\ }\bibfield  {title} {\enquote {\bibinfo {title} {Multicomponent
  fractional quantum hall effect in graphene},}\ }\href@noop {} {\bibfield
  {journal} {\bibinfo  {journal} {Nature Physics}\ }\textbf {\bibinfo {volume}
  {7}},\ \bibinfo {pages} {693--696} (\bibinfo {year} {2011})}\BibitemShut
  {NoStop}%
\bibitem [{\citenamefont {Papi\ifmmode~\acute{c}\else \'{c}\fi{}}\ \emph
  {et~al.}(2011)\citenamefont {Papi\ifmmode~\acute{c}\else \'{c}\fi{}},
  \citenamefont {Thomale},\ and\ \citenamefont
  {Abanin}}]{PhysRevLett.107.176602}%
  \BibitemOpen
  \bibfield  {author} {\bibinfo {author} {\bibfnamefont {Z.}~\bibnamefont
  {Papi\ifmmode~\acute{c}\else \'{c}\fi{}}}, \bibinfo {author} {\bibfnamefont
  {R.}~\bibnamefont {Thomale}}, \ and\ \bibinfo {author} {\bibfnamefont
  {D.~A.}\ \bibnamefont {Abanin}},\ }\bibfield  {title} {\enquote {\bibinfo
  {title} {Tunable electron interactions and fractional quantum hall states in
  graphene},}\ }\href {\doibase 10.1103/PhysRevLett.107.176602} {\bibfield
  {journal} {\bibinfo  {journal} {Phys. Rev. Lett.}\ }\textbf {\bibinfo
  {volume} {107}},\ \bibinfo {pages} {176602} (\bibinfo {year}
  {2011})}\BibitemShut {NoStop}%
\bibitem [{\citenamefont {Lee}\ and\ \citenamefont
  {Lucas}(2014)}]{leeandy2014}%
  \BibitemOpen
  \bibfield  {author} {\bibinfo {author} {\bibfnamefont {Ching~Hua}\
  \bibnamefont {Lee}}\ and\ \bibinfo {author} {\bibfnamefont {Andrew}\
  \bibnamefont {Lucas}},\ }\bibfield  {title} {\enquote {\bibinfo {title}
  {Simple model for multiple-choice collective decision making},}\ }\href@noop
  {} {\bibfield  {journal} {\bibinfo  {journal} {Physical Review E}\ }\textbf
  {\bibinfo {volume} {90}},\ \bibinfo {pages} {052804} (\bibinfo {year}
  {2014})}\BibitemShut {NoStop}%
\end{thebibliography}%

\end{document}